%% file: GHM-4mar2021.tex
\documentclass[a4paper,12pt]{article}
\pdfoutput=1
\usepackage{geometry}
\usepackage{amsmath,amssymb,amsfonts}
\usepackage{xcolor,graphicx,cite,soul}
\usepackage{array,booktabs,longtable}
\usepackage{caption,microtype}
\usepackage{subcaption}
\usepackage{hyperref}
\usepackage{ulem}
\usepackage{float}

\usepackage{datetime}
\usepackage{appendix}
\usepackage{bigstrut}
\usepackage{multirow}
\definecolor{OliveGreen}{rgb}{0,0.6,0}

\newcommand{\bear}{\begin{eqnarray}}
\newcommand{\eear}{\end{eqnarray}}
\newcommand{\mL}{\mathcal{L}}

\def\gY{g'}
\def\gYc{g^{'2}}
\input{paperdef}

\graphicspath{{figs/}}
\allowdisplaybreaks
\begin{document}
\thispagestyle{empty}

\def\thefootnote{\fnsymbol{footnote}}

%\begin{flushleft}
%\mbox{Compiled on \today, at \currenttime}
%\end{flushleft}
\begin{flushright}
\mbox{}
IFT--UAM/CSIC-20-147\\
FTUAM-20-22
\end{flushright}

\vspace{0.5cm}

\begin{center}

{\large\sc 
{\bf Testing anomalous $H-W$ couplings and Higgs self-couplings via double and triple Higgs production at $e^+e^-$ colliders}}\\

\vspace{1cm}

{\sc
M.~Gonzalez-Lopez$^{1,2}$%
\footnote{email: manuel.gonzalezl@uam.es}% 
, M.J.~Herrero$^{1,2}$%
\footnote{email: maria.herrero@uam.es}%
~and P.~Martinez-Suarez$^{1,2,3}$%
\footnote{email: paula.martinez.suarez@cern.ch}%
{
\def\thefootnote{\arabic{footnote}}
\footnotetext[3]{Currently at the Institut de F\'isica d'Altes Energies (IFAE) and the Departament de F\'isica from the Universitat Aut\`onoma de Barcelona (UAB) in Bellaterra, 08193, Barcelona, Spain.}
}

%\footnote{former address}%
}

\vspace*{.7cm}

{\sl
$^1$Departamento de F\'isica Te\'orica, 
Universidad Aut\'onoma de Madrid, \\ 
Cantoblanco, 28049, Madrid, Spain

\vspace*{0.1cm}

$^2$Instituto de F\'isica Te\'orica (UAM/CSIC), 
Universidad Aut\'onoma de Madrid, \\ 
Cantoblanco, 28049, Madrid, Spain

}

\end{center}

\vspace*{0.1cm}

\begin{abstract}
\noindent

In the present work we study the implications at the future $e^+e^-$ colliders of the modified interaction vertices $WWH$, $WWHH$, $HHH$ and $HHHH$ within the context of the non-linear effective field theory given by the Electroweak Chiral Lagrangian. These vertices are given by four parameters, $a$, $b$, $\kappa_3$ and $\kappa_4$, respectively, that are independent and without any constraint from symmetry considerations in this non-linear effective Lagrangian context, given the fact the Higgs field is a singlet. This is in contrast to the Standard Model, where the vertices are related by $V_{WWH}^{\rm SM}=v V_{WWHH}^{\rm SM}$ and $V_{HHH}^{\rm SM}=v V_{HHHH}^{\rm SM}$, with $v=246$ GeV. We investigate the implications of the absence of these relations in the Electroweak Chiral Lagrangian case. We explore the sensitivity to these Higgs anomalous couplings in the two main channels at these colliders: double and triple Higgs production (plus neutrinos). Concretely, we study the access to $a$ and $b$ in $e^+e^- \to HH \nu \bar{\nu}$ and the access to $\kappa_3$ and $\kappa_4$ in $e^+e^- \to HHH \nu \bar{\nu}$.  Our study of the beyond the Standard Model couplings  via triple Higgs boson production at $e^+e^-$ colliders is novel and shows for the first time the possible accessibility to the quartic Higgs self-coupling.

\end{abstract}

%\pac
\def\thefootnote{\arabic{footnote}}
\setcounter{page}{0}
\setcounter{footnote}{0}

\newpage

%%%%%%%%%%%%%%%%%%%%%%%%%%%%%%%%%%%%%%%%%%%%%%%%%%%%%%%%%%%%%%%%%%%%%%%%%%%%%%%
%%%%%%%%%%%%%%%%%%%%%%%%%%%%%%%%%%%%%%%%%%%%%%%%%%%%%%%%%%%%%%%%%%%%%%%%%%%%%%%

\section{Introduction}
\label{sec:intro}
One of the not yet fully explored pieces of the Standard Model (SM) of Particle Physics is the Higgs boson potential, involving the triple and quartic  Higgs self-couplings, which are both given in terms of the $\lambda$ parameter of the potential by $\lambda_{HHH}^{\rm SM}=\lambda_{HHHH}^{\rm SM}=\lambda $, and whose strength is related to the Higgs boson mass by  $m_H^2=2\lambda v^2$, with $v=246$ GeV. The implementation of all the Higgs boson interactions within the SM is done usually by setting the Higgs boson as a component of a doublet, the simplest linear realisation of the $SU(2)\times U(1)$ electroweak (EW) symmetry. This linear realisation leads to some correlations among the Higgs interaction vertices within the SM. In particular, the triple and quartic self-interaction vertices are related by $V_{HHH}^{\rm SM}=v V_{HHHH}^{\rm SM}$.  Similarly, there are also correlations among  other SM couplings in the bosonic sector, like the Higgs interaction vertices with $W$ gauge bosons, which are also related by $V_{WWH}^{\rm SM}=v V_{WWHH}^{\rm SM}$.  Deviations from these correlations may lead to new signals of beyond the Standard Model (BSM) physics. Therefore,  an important task at future colliders will be the improvement in the measurement of all these bosonic couplings and the test of their correlations (for a recent review on measurements and constraints on Higgs boson couplings at present and future colliders, see for instance \cite{Aad:2019mbh,Aad:2020kub,ATLAS:2019pbo,DiMicco:2019ngk,Cepeda:2019klc,Strube:2016eje,Roloff:2019crr,Fuks:2017zkg,Belyaev:2018fky}). 

In the present paper we study the sensitivity to these particular four BSM Higgs interactions, $WWH$, $WWHH$, $HHH$ and $HHHH$, and focus on future $e^+e^-$ colliders. To explore the BSM predictions we work within the context of an Effective Field Theory (EFT) for EW interactions  
(for a recent review on EFTs for the SM see, for instance, \cite{Brivio:2017vri}). In particular, we choose the non-linear effective field theory given by the Electroweak Chiral Lagrangian (EChL). This EChL, also called Higgs Effective Field Theory (HEFT) in the literature, was prosed long ago~\cite{Appelquist:1980vg,Longhitano:1980iz,Chanowitz:1985hj,Cheyette:1987jf, Dobado:1989ax,Dobado:1989ue,Dobado:1990zh,Espriu:1991vm,Feruglio:1992wf,Herrero:1993nc,Herrero:1994iu} to describe the BSM, $SU(2)\times U(1)$ gauge invariant interactions without the explicit Higgs field and using a non-linear realisation of the EW chiral symmetry.  In the last years, after the discovery of the Higgs boson particle, the EChL has been renewed with the incorporation of the Higgs field. Consequently, the new version of the EChL contains more effective operators that now include also the Higgs boson~\cite{Alonso:2012px,Brivio:2013pma,Espriu:2012ih,Espriu:2013fia,Delgado:2013loa,Delgado:2013hxa,Delgado:2014jda,Buchalla:2012qq,Buchalla:2013rka,Buchalla:2013eza,Buchalla:2015wfa,Buchalla:2015qju,deBlas:2018tjm}.  

In the EChL,  the Higgs field has the peculiarity of being a singlet under $SU(2)\times U(1)$, in clear contrast with the SM case and other EFTs like the SM effective field theory (SMEFT). As a consequence, the previously mentioned SM correlations between couplings involving two and three Higgs bosons are not present, leading to singular signals in multiple Higgs boson production. Thus, double and triple Higgs productions may exhibit very different rates with respect to the SM. In particular, we focus our studies here on two production channels at $e^+e^-$ colliders, $e^+e^- \to HH \nu \bar{\nu}$ and $e^+e^- \to HHH \nu  \bar{\nu}$, where the subprocesses initiated by $W^-W^+$ scattering (WWS, also called Vector Boson Fusion (VBF)  in the literature) play a very important role. Indeed, at high energies, in the TeV range,  these WWS subprocesses are known to be the dominant ones (for a recent review on VBF, see for instance \cite{Rauch:2016pai,Green:2016trm}.  

Within the context of the EChL, the BSM Higgs couplings to bosons we are interested in are parametrised by four independent EChL coefficients, $a$, $b$, $\kappa_3$ and $\kappa_4$ (also called anomalous couplings), one for each of the four BSM interactions: $WWH$, $WWHH$, $HHH$ and $HHHH$.   Some of the phenomenological consequences of these couplings via WWS have been examined. The effects of $a$ and $b$ in double Higgs production via WWS have been studied mostly for the case of proton-proton collisions, focusing mainly on the LHC and its future upgrades, both in energy and luminosity \cite{Bishara:2016kjn}, although the effects of $b$ at $e^+e^-$ future colliders have also been explored in \cite{Contino:2013gna}. The BSM effects of the triple Higgs coupling, $\lambda_{HHH}=\kappa_3 \lambda$,  at  LHC  via double Higgs production from WWS have been studied in \cite{Arganda:2018ftn}. The highest sensitivity to $\kappa_3$ in hadron colliders is reached, however, via gluon gluon fusion (for a review, see for instance \cite{DiMicco:2019ngk}). The quartic Higgs self-coupling, $\lambda_{HHHH}=\kappa_4 \lambda$,  and the consequences of a BSM $\kappa_4$, have barely been explored in the literature. Its effects via radiative corrections have been studied in \cite{Bizon:2018syu,Liu:2018peg,Borowka:2018pxx,Maltoni:2018ttu}, whereas the sensitivity to $\kappa_4$ in multi-TeV muon colliders has been considered in \cite{Chiesa:2020awd}. The combined effects of anomalous triple and quartic Higgs self-couplings via radiative corrections in multiple Higgs boson production at $e^+e^-$ colliders have been studied in \cite{Maltoni:2018ttu} within the context of the linear EFT approach. In that context, the correlations among the anomalous couplings, the implementation of gauge invariance and the dimensional counting and renormalisation program for the relevant operators are different from that in the non-linear EFT context that we follow here. The phenomenological implications in the two approaches, linear and non-linear, are consequently also different.  In the present study we  explore the sensitivity to the four EChL coefficients $a$, $b$, $\kappa_3$ and $\kappa_4$ via  these two $e^+e^- \to HH(H) \nu \bar{\nu}$ processes in the context of the non-linear EFT approach.  In particular, we present a novel study of triple Higgs production at $e^+e^-$ colliders, that we compare with the better known double Higgs production channel, which gives no access to the quartic Higgs couplings, and introduce a new proposal to test both the triple and quartic Higgs self-couplings at once via the $e^+e^- \to HHH   \nu \bar{\nu}$ process. 

\begin{table}[htb!]
%\begin{table}[h]
\begin{center}
\begin{tabular}{|lcc|}
\hline
Collider     & $\sqrt{s}$ (GeV) & $\mathcal{L}_\text{int}$ (ab$^{-1}$) \\ \hline\hline
ILC 
& 250
& $2$
\\
& 350
& $0.2$
\\
& 500
& $4$
\\
& 1000
& $8$
\\ \hline\hline
CLIC
& $380$
& $1$
\\
& $1500$
& $2.5$
\\
& $3000$
&  $5$
\\ \hline
\end{tabular}
\end{center}
\caption{Expected center of mass energies and integrated luminosities in the different stages of both the ILC and CLIC.}
\label{eecolliders}
\end{table}
But the measurement of the Higgs self-couplings is not an easy task, as the processes that depend on them involve multiple Higgs production, which is difficult to study and generally suffers of low rates. In the current LHC, with proton-proton collisions at a center of mass (CM) energy of 13 TeV, the SM production cross section of two or more Higgs bosons lies at the picobarn (pb) scale or below \cite{DiMicco:2019ngk}, and the luminosity is not high enough to produce good statistics. Nevertheless, the triple Higgs self-coupling has been constrained by the ATLAS Collaboration \cite{ATLAS:2019pbo} to $-2.3 < \lambda_{HHH}/\lambda_{HHH}^\text{SM} < 10.3$ at the 95\% CL, under the assumption that the possible new physics affects only the triple Higgs coupling. The quartic Higgs self-coupling $\lambda_{HHHH}$ remains experimentally unaccessible at the moment. A more precise determination of the triple Higgs self-coupling via double Higgs production is one of the aims of several future projects, such as the High Luminosity LHC (HL-LHC) and its high energy upgrade (HE-LHC) \cite{Cepeda:2019klc}, the International Linear Collider (ILC) \cite{Strube:2016eje} and the Compact Linear Collider (CLIC) \cite{Roloff:2019crr}. Much research on this topic has already been carried out. The study of triple Higgs production as a way to test new couplings of the Higgs boson to SM particles has already been proposed for future proton-proton colliders, such as the Future Circular Collider (FCC) \cite{Fuks:2017zkg,Belyaev:2018fky}. In this work we will focus on the two electron-positron colliders in the list: the ILC and CLIC. They will operate in several stages with different values of the CM energy and integrated luminosities, as it can be seen in \refta{eecolliders}.

This work is organized as follows: in \refse{EChLsec} we review the relevant interactions within the EChL for the present work ($WWH$, $WWHH$, $HHH$ and $HHHH$) and write them in terms of the four relevant EChL coefficients $a$, $b$, $\kappa_3$ and $\kappa_4$. In \refse{WWS} we analyze the role of WWS in multiple Higgs production at $e^+e^-$ colliders. In \refse{TestWWS} we present the computation and results of the cross sections for the two relevant WWS subprocesses, $W^-W^+\rightarrow HH(H)$, in the BSM case. We also compare these BSM results with the SM predictions, and check how sensitive they are to variations of the Higgs boson anomalous couplings. In \refse{TestProcess}, the corresponding results for  $e^+e^-$ collisions are presented for the various colliders. Finally, \refse{b-jet} explores the sensitivities to BSM couplings in multiple $b$-jet events that are expected considering the final Higgs decays. The parameter space region accessible at the future $e^+e^-$ colliders is also derived. The conclusions are summarized in \refse{conclu}. 
% --------------------------------------------------------------------------

%%%%%%%%%%%%%%%%%%%%%%%%%

\section{BSM Higgs couplings to bosons within the EChL}
\label{EChLsec}
%%%
%\subsection{The Electroweak Chiral Lagrangian}
To explore the BSM predictions in this work we use the non-linear EFT given by EChL,  where the Higgs field is a singlet under both the EW gauge (local) symmetry, $SU(2)_L \times U(1)_Y$, and the EW chiral (global) symmetry, $SU(2)_L \times SU(2)_R$. This is in contrast to the usual linear realisations, like the SM itself or other EFTs like the SMEFT, where the Higgs field is one component of an $SU(2)$ doublet. The Higgs singlet is usually introduced in the EChL via polynomial functions, generically leading to uncorrelated BSM Higgs couplings  involving one or more Higgs particles, while the three EW Goldstone bosons (GB) that arise from the EW symmetry breaking are usually included in an exponential representation and transform non-linearly under the EW symmetry. We will assume here that the dynamics of the Higgs, Goldstone and EW gauge bosons are given by the EChL, while those describing fermions will be the same as in the SM. The EChL follows a pattern and counting rules \cite{Buchalla:2013eza} similar to those in the chiral Lagrangian for low energy QCD \cite{Weinberg:1978kz}. It contains a  tower of effective operators ordered by their chiral dimension, providing predictions ordered in even powers of momentum, $p^n$. It also provides well defined predictions of observables beyond the tree level approximation, i.e. including loop corrections which scale as $(p/(4\pi v))^n$, such that the convergence of this momentum expansion is ensured for low energies, say below $4 \pi v \sim 3$ TeV. As in any other EFT of EW interactions, the predictions from the EChL are model independent. The information on the particular ultraviolet theory and the corresponding cut-off that lead to this EChL at low energies is encoded in the coefficients of the effective operators.  Therefore, to access the BSM physics via the EChL, the challenge is to find sensitivity to the EChL coefficients at present and future experiments.   
 
For our purposes, in this study of the BSM Higgs couplings to bosons we work at  the tree level approximation, so it is enough to use the leading order EChL, which reads:

\bear
\mathcal{L}_\text{EChL} & =  &
    \frac{v^2}{4}\left[1+2a\left(\frac{H}{v}\right)+b\left(\frac{H}{v}\right)^2+...\right]
    \text{Tr}\big[D_\mu U^\dagger D^\mu U\big]+  \nonumber\\[5pt]
    && +\frac{1}{2}(\partial_\mu H)(\partial^\mu H)
    -\frac{1}{2}m_H^2 H^2 - \kappa_3\lambda vH^3 - \kappa_4\frac{1}{4}\lambda H^4 
     \nonumber\\[5pt]
    && -\frac{1}{2 \gYc}
{\rm Tr} \Big[\hat{B}_{\mu\nu} \hat{B}^{\mu\nu}\Big] -\frac{1}{2 g^2} {\rm Tr}\Big[\hat{W}_{\mu\nu}\hat{W}^{\mu\nu}\Big] + \mL_{\rm GF} + \mL_{\rm FP}. 
    \label{EChL}
\eear

Here $v=246$ GeV, $g$ and $g'$ are the EW gauge couplings, $H$ is the Higgs boson field, and $U$ is the $2\times2$ matrix that contains the three GBs, $\omega_{1,2,3}$:
\begin{align}
    U  = \exp\left(\frac{i\vec{\omega}\cdot \vec{\tau}}{v}\right),
\end{align}
with $\tau^a$ being the Pauli matrices. The field strengths and covariant derivative are defined as: 
\bear
\hat{B}_{\mu\nu} &=& \partial_\mu \hat{B}_\nu -\partial_\nu \hat{B}_\mu \,,\quad
\hat{W}_{\mu\nu} = \partial_\mu \hat{W}_\nu - \partial_\nu \hat{W}_\mu + i  [\hat{W}_\mu,\hat{W}_\nu ] \,,  \nonumber\\
D_\mu U &=& \partial_\mu U + i\hat{W}_\mu U - i U\hat{B}_\mu \,,
\eear
where $\hat{B}_\mu = \gY B_\mu \tau^3/2$ and $\hat{W}_\mu = g W^a_\mu \tau^a/2$. The rotation from the interaction basis to the physical gauge bosons given by the weak angle $\theta_W$ is the same as in the SM, so the usual tree level relations among gauge couplings and gauge boson masses still hold. Similarly, the tree level relation $m_H^2=2\lambda v^2$ also holds here. The gauge fixing Lagrangian,  $\mL_{\rm GF}$, and the Fadeev-Popov Lagrangian, $\mL_{\rm FP}$, do not enter in our computation of cross-sections, since we work at the tree level and choosing the unitary gauge.

\begin{figure}[h]
\centering
\includegraphics[scale=0.95]{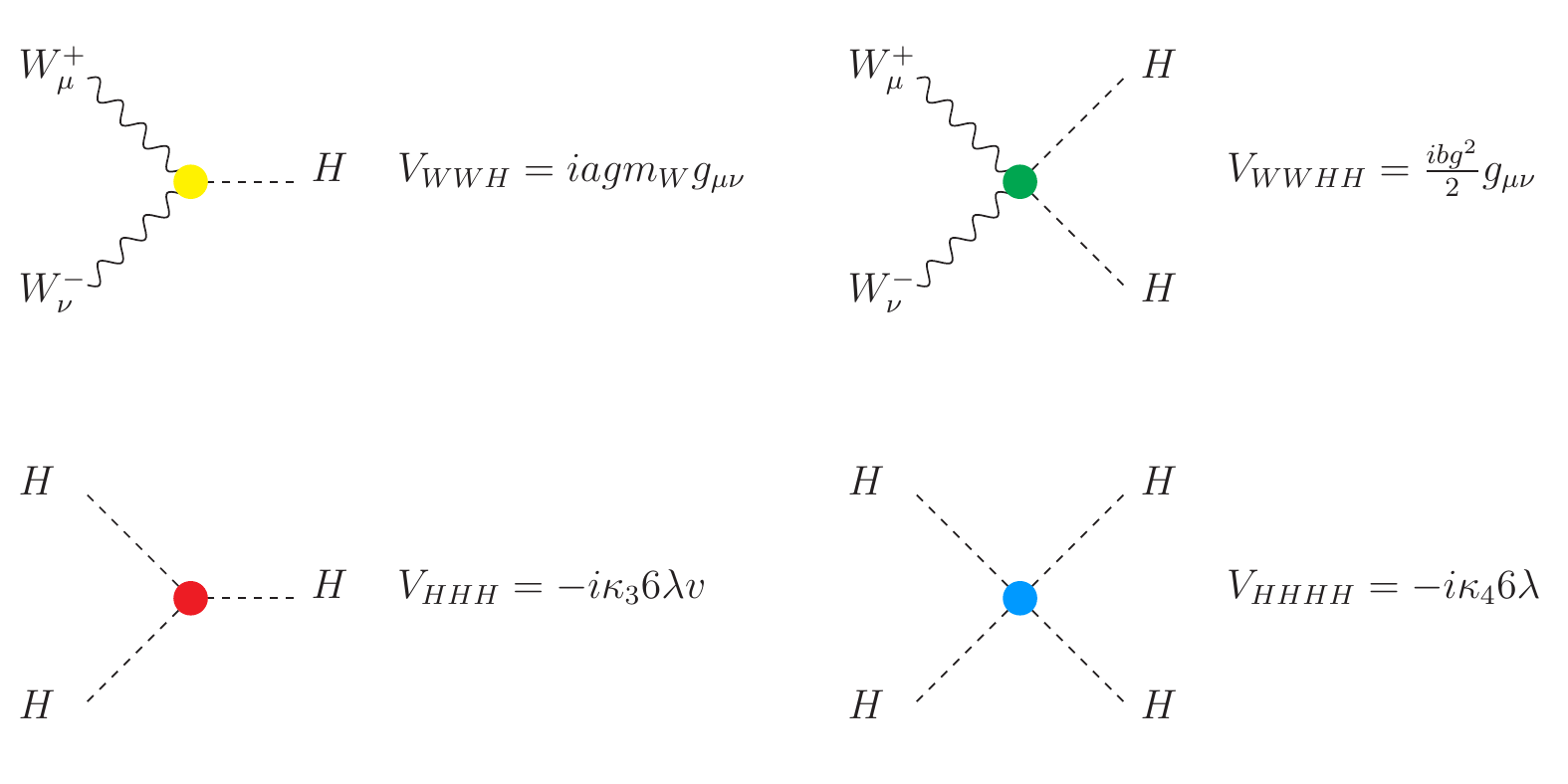}
\caption{Feynman rules for Higgs couplings to bosons within the EChL.}
\label{FR}
\end{figure}

In summary, the relevant interactions in this EChL for the present study, working at the tree level and in the unitary gauge, are given by the four vertices $V_{WWH}$, $V_{WWHH}$, $V_{HHH}$ and $V_{HHHH}$ collected in \reffi{FR}.
We see in this figure that the coefficients  $a$ and $b$ (yellow and green dots respectively) control the so-called anomalous couplings of the Higgs boson to $W$ and $Z$ bosons, while the coefficients $\kappa_3$ and $\kappa_4$ (red and blue dots respectively) control the so-called anomalous triple and quartic Higgs self-couplings, respectively.   The SM vertices are recovered when $a=b=\kappa_3=\kappa_4=1$.

It is worth noticing that the correlations among these vertices within the SM, $V_{WWH}^{\rm SM}= v V_{WWHH}^{\rm SM} $ and $V_{HHH}^{\rm SM}= v V_{HHHH}^{\rm SM} $, are not present within the EChL. Thus, exploring deviations with respect to these correlations (which occur if any of the EChL coefficients differs from 1) is an interesting way to access the BSM physics at colliders.
%%%%%%%%%%%%%%%%
Regarding the experimental bounds on these EChL coefficients from present colliders, there are available constraints from the LHC for $a$, $b$ and  $\kappa_3$,  but not for $\kappa_4$. The ATLAS Collaboration has set the following 95\% C.L. bounds on each of the Higgs couplings to gauge bosons using single and double Higgs production:
\begin{align}
    a\in[0.97,1.13] \;\hbox{\cite{Aad:2019mbh}}\qquad b\in[-0.76,2.90] \;\hbox{\cite{Aad:2020kub}}.
\end{align}
The value of $\kappa_3$ has also been constrained by the ATLAS Collaboration, using combined information from single and double Higgs production, but the sensitivity to this parameter is lower in comparison, which leads to a much less restrictive bound than the ones for $a$ and $b$:
\begin{align}
    \kappa_3\in [-2.3,10.3] \;\hbox{\cite{ATLAS:2019pbo}}
    \label{ATLASconstraintEq}
\end{align}
Again, we would like to emphasize that currently there are no bounds for $\kappa_4$. 

For the forthcoming numerical analysis in the present work we will explore the effects of BSM Higgs couplings to gauge bosons and Higgs self-couplings by varying $a$, $b$,  $\kappa_3$ and $\kappa_4$ within the following theoretical intervals:
\begin{align}
a \in [0.9,1.1] \,\,\, ,\,
b \in [-2,2] \,\,\, ,\,
    \kappa_3\in[-10,10]  \,\,\,,\,
    \kappa_4\in[-10,10].
\end{align}
 Notice that although some of these values are outside the experimental bounds, we believe that it is illustrative to study the behaviour with these parameters when they vary in a wider range. As we will show in the next section, there are indeed further constraints on some of these parameters from the potential violation of unitarity that may occur at high energies for too large values of these coefficients.  Concretely, 
$a$ and $b$ will be further restricted by unitarity, but $\kappa_3$ and $\kappa_4$ will not. In the last part of this work, all these constraints are taken into account.

\section{The role of WWS in multiple Higgs production at $e^+e^-$ colliders}
\label{WWS}
These BSM Higgs couplings modify the predictions of  double and triple Higgs production at future $e^+e^-$ colliders with respect to the SM predictions and can lead to testable departures. A first indication of the size of such departures in the particular $e^+e^- \to HH (H) \nu {\bar \nu}$ channels  is provided by the study of the subprocesses $W^-W^+ \to HH(H)$ for the case of two (three) Higgs bosons, respectively. These WWS  channels are known to be the dominant subprocesses at $e^+e^-$ colliders with energies at the TeV scale.  In this section we will illustrate the WWS dominance in the SM by performing cross sections computations in \textsc{MadGraph5} (MG5) \cite{Alwall:2014hca}. We will then compare these results to those predicted by the simplified effective $W$ approximation (EWA). This approximation is very useful, but as we will see it does not provide a sufficiently accurate prediction in all cases.
\subsection{The dominance of WWS within the SM}
\label{WWSSM}
\begin{figure}[h!]
\centering
\includegraphics[width=0.85\textwidth]{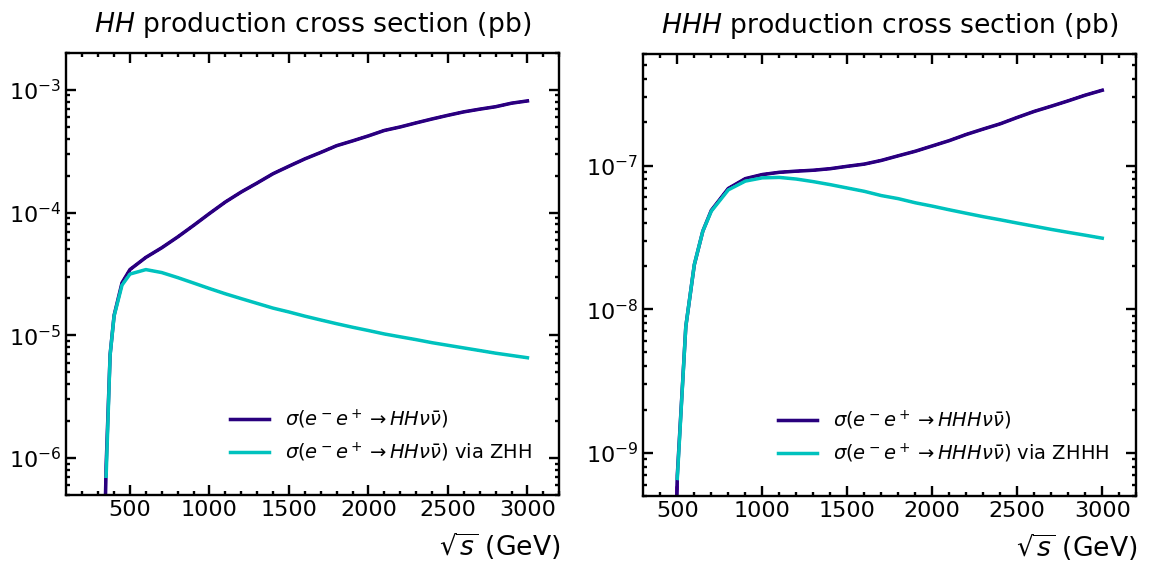}
\caption{Predictions for $\sigma(e^+e^-\rightarrow HH\nu\bar{\nu})$ (left) and $\sigma(e^+e^-\rightarrow HHH\nu\bar{\nu})$ (right) in the SM as a function of the CM energy $\sqrt{s}$. The corresponding cross sections coming from just the $Z$-mediated subprocesses, $\sigma(e^+e^-\rightarrow ZHH\rightarrow HH\nu\bar{\nu})$ and $\sigma(e^+e^-\rightarrow ZHHH\rightarrow HHH\nu\bar{\nu})$, are also shown for comparison.}
\label{prodXS1}
\end{figure}

Within the SM and neglecting the Yukawa couplings, there are
8 diagrams that mediate the $e^+e^- \to HH \nu {\bar \nu}$ process: 4 of them are of the WWS type, while in the other 4 the neutrino pair is produced via the $Z$ boson. In the $e^+e^- \to HHH \nu {\bar \nu}$ process there are 50 diagrams in total, 25 of which are WWS and 25 $Z$-mediated. We do not draw all these diagrams here, for shortness, but the subset of diagrams that are mediated by WWS are easily extracted from those of the corresponding subprocess, which have been collected in the appendices. MG5 takes into account all these diagrams, allowing to distinguish the different contributions when computing total cross sections. In \reffi{prodXS1} we show these cross sections, taking into account all diagrams (dark blue lines) and only the $Z$-mediated ones (light blue lines).

From \reffi{prodXS1} we learn that the dominant contribution to $\sigma(e^+e^-\rightarrow HH(H)\nu\bar{\nu})$ at low energies comes from the associated production of two (three) $H$'s with a $Z$ boson which then goes to $\nu \bar \nu$, denoted here as "$\sigma(e^+e^-\rightarrow HH(H)\nu\bar{\nu})$ via $ZHH(H)$". The cross section for this particular production mechanism can be computed as the cross section of the process $e^+e^-\rightarrow ZHH(H)$ times the branching ratio BR$(Z\rightarrow\text{invisible})=20\%$. As the energy increases, the associated production of two (three) $H$’s with a $Z$ boson becomes subdominant, which suggests that $HH(H)\nu\bar{\nu}$ production at the TeV scale is dominated by the other diagrams (WWS). However, to estimate the size of the WWS contributions one cannot naively extract these diagrams, since they generically do not form a gauge invariant subset. Thus, a different method must be used.
\begin{figure}[h!]
\centering
\includegraphics[width=0.85\textwidth]{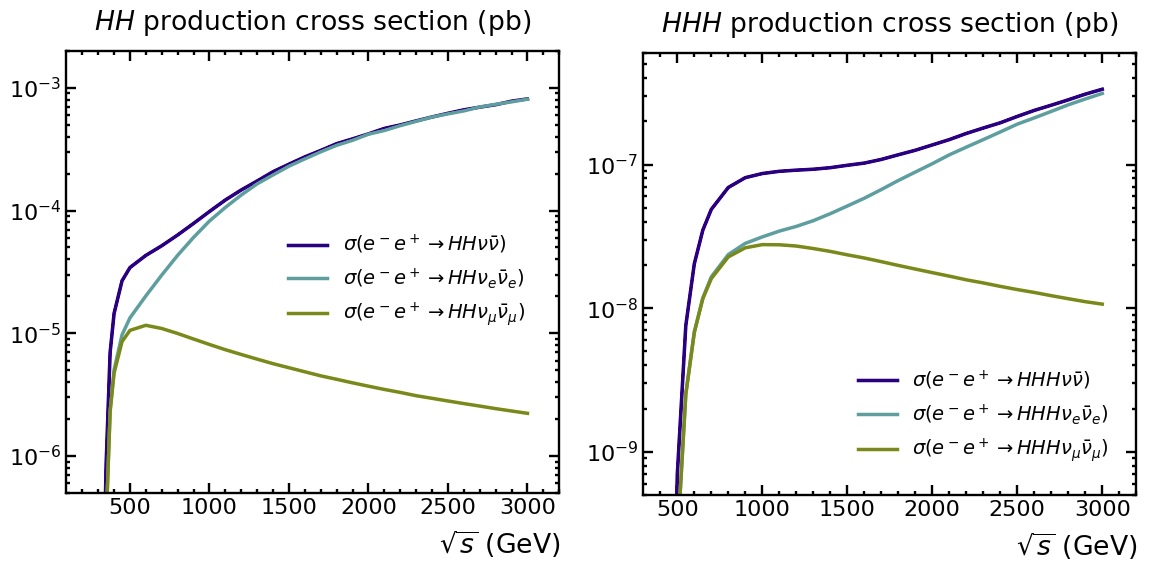}
\caption{Predictions for $\sigma(e^+e^-\rightarrow HH\nu_e\bar{\nu}_e)$ and $\sigma(e^+e^-\rightarrow HH\nu_\mu\bar{\nu}_\mu)$ (left) and $\sigma(e^+e^-\rightarrow HHH\nu_e\bar{\nu}_e)$ and $\sigma(e^+e^-\rightarrow HHH\nu_\mu\bar{\nu}_\mu)$ (right) in the SM as a function of the CM energy $\sqrt{s}$. The cross sections for $e^+e^-\rightarrow HH\nu\bar{\nu}$ and $e^+e^-\rightarrow HHH\nu\bar{\nu}$ from the previous figure are also shown for comparison.}
\label{prodXS2}
\end{figure}

\reffi{prodXS2} shows that the enhancement in $\sigma(e^+e^-\rightarrow HH(H)\nu\bar{\nu})$ at high energies actually comes from $\sigma(e^+e^-\rightarrow HH(H)\nu_e\bar{\nu}_e)$, that is, processes with electron neutrinos in the final state. 
This fact becomes clearer when we also represent the contribution of $\sigma(e^+e^-\rightarrow HH(H)\nu_\mu\bar{\nu}_\mu)$, which decreases at high energies.\footnote{For simplicity, we omit $\sigma(e^+e^-\rightarrow HH(H)\nu_\tau\bar{\nu}_\tau)$, which is the same as $\sigma(e^+e^-\rightarrow HH(H)\nu_\mu\bar{\nu}_\mu)$.}. 
\begin{figure}[h!]
\centering
\includegraphics[width=0.85\textwidth]{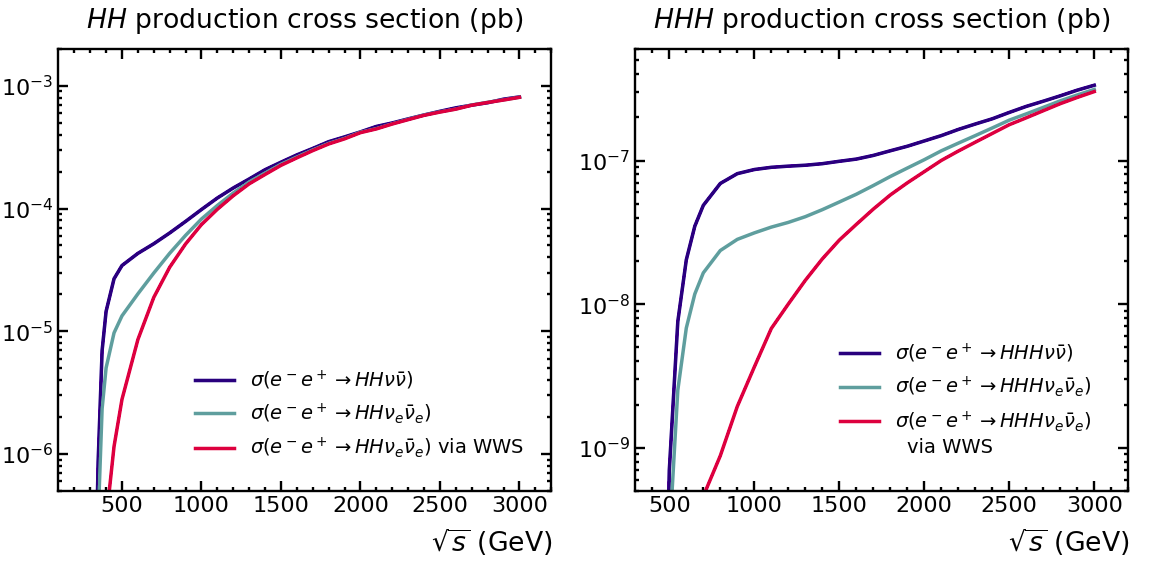}
\caption{Double (left) and triple (right) Higgs production cross section in $e^+e^-$ collisions with neutrinos in the final state and the corresponding contributions coming from $HH(H)\nu_e\bar{\nu}_e$ and WWS. The prediction for $\sigma(e^+e^-\rightarrow HH(H)\nu_e\bar{\nu}_e)$ via WWS is  defined in \refeq{WWSmadgraph}.}
\label{prodXS3}
\end{figure}
The main difference between both processes is that whereas $e^+e^-\rightarrow HH(H)\nu_e\bar{\nu}_e$ includes both types of diagrams, WWS and $ZHH(H)$, this is not the case in $e^+e^-\rightarrow HH(H)\nu_\mu\bar{\nu}_\mu$ where all diagrams are of $ZHH(H)$ type.
 Then, the contribution to $\sigma(e^+e^-\rightarrow HH(H)\nu_e\bar{\nu}_e)$ coming from WWS can be isolated using the following ``theoretical observable'':
\begin{align}
    \sigma_\text{WWS} = \sigma(e^+e^-\rightarrow HH(H)\nu_e\bar{\nu}_e) -
    \sigma(e^+e^-\rightarrow HH(H)\nu_\mu\bar{\nu}_\mu).
    \label{WWSmadgraph}
\end{align}
Therefore, this difference between the channels with electron and muon neutrinos quantifies the weight of the WWS diagrams in the full process without the need of separating diagrams which, as mentioned above, is not the proper way if one wishes to preserve gauge invariance. We also define the related quantity $R_\text{WWS}$,
\begin{equation}\label{rvbs}
R_\text{WWS}=\frac{\sigma(e^+e^-\rightarrow HH(H)\nu_e\bar{\nu_e})-\sigma(e^+e^-\rightarrow HH(H)\nu_\mu\bar{\nu_\mu})}{\sigma(e^+e^-\rightarrow HH(H)\nu_e\bar{\nu_e})}.
\end{equation}
This is an adimensional quantity that clearly determines how large is the contribution of WWS in the $e^+e^-\rightarrow HH(H)\nu_e\bar{\nu_e}$ process. If $R_\text{WWS}$ is close to 1, the cross section will be mostly dominated by WWS. The isolation of these WWS contributions is of relevance for  the present study, since, as we will see next, they are indeed the most sensitive ones to the anomalous Higgs couplings.

The importance of the WWS contribution in the two processes, $e^+e^-\rightarrow HH\nu_e\bar{\nu}_e$ and $e^+e^-\rightarrow HHH\nu_e\bar{\nu}_e$, is clearly illustrated in  \reffi{prodXS3}, where we have included the predictions for 
$\sigma(e^+e^-\rightarrow HH(H)\nu\bar{\nu})$, $\sigma(e^+e^-\rightarrow HH(H)\nu_e\bar{\nu}_e)$ and $\sigma_\text{WWS}$ as defined in \refeq{WWSmadgraph}. Above the ${\cal O}(1\,{\rm TeV})$  energies, $\sigma_\text{WWS}$ clearly approaches $\sigma(e^+e^-\rightarrow HH(H)\nu_e\bar{\nu}_e)$, which in turn approaches the total $\sigma(e^+e^-\rightarrow HH(H)\nu\bar{\nu})$. Thus, we conclude on the dominance of WWS in $\sigma(e^+e^-\rightarrow HH(H)\nu\bar{\nu})$, via the particular channel with electron neutrinos.

It is also important to notice that $ZZ$ scattering (ZZS) contributes to the final state $HH(H)e^+e^-$, which has a much lower production cross section (the probability of producing two or three Higgs bosons via ZZS is approximately ten times lower). Therefore, this ZZS will not be relevant in this work. 

Although the previous figures show separately the various contributions from the different neutrino species, $\sigma(e^+e^-\rightarrow HH(H)\nu_i\bar{\nu_i})$, one must realize that in the experiment it is only possible to measure their sum, $\sigma(e^+e^-\rightarrow HH(H)\nu\bar{\nu})$. However, the key point on studying such theoretical observables is to show, in a gauge invariant way, that WWS  is indeed the largest contribution at high energies.  As a consequence, the predictions for $\sigma(e^+e^-\rightarrow HH(H)\nu\bar{\nu})$ will approximately follow the same behaviour at these large energies, specifically that of $\sigma(e^+e^-\rightarrow HH(H)\nu_e\bar{\nu}_e)$.

Comparing triple with double Higgs production in Figs. \ref{prodXS1}, \ref{prodXS2} and \ref{prodXS3}, we find that they are roughly similar in shape, being the WWS enhancement in triple Higgs production displaced to higher energies, since one extra particle is produced. This causes $ZHHH$ to be more relevant compared to $HHH\nu\bar{\nu}$ than $ZHH$ compared to $HH\nu\bar{\nu}$, especially at energies below 2000 GeV. It is also important to note that the SM cross sections for triple Higgs production at the TeV energy scale are typically three orders of magnitude below those for double Higgs production, which is why it is not expected to measure SM-like $HHH$ production in future linear colliders. For this reason, we will focus our attention on BSM scenarios.
\subsection{The effective $W$ approximation versus full MG5 simulations}
\label{EWAvMG}
Before exploring the effects of BSM Higgs couplings, it is instructive to first evaluate the validity of the so-called effective $W$ approximation (EWA) \cite{Dawson:1984gx} by comparing its predictions to the full MG5 simulations. The EWA is a generalisation of the well known effective photon approximation, and treats vector bosons, $W$'s and $Z$'s, that are radiated from the initial fermions, as if they were partons inside the fermions. This allows to define distribution functions for these vector bosons in a similar way as the PDFs of quarks and gluons inside a proton. The EWA also assumes that the vector bosons are radiated colinearly by the fermions, and then they scatter on-shell in the subprocess. 
\begin{figure}[h!]
\centering
\includegraphics[width=0.33\textwidth]{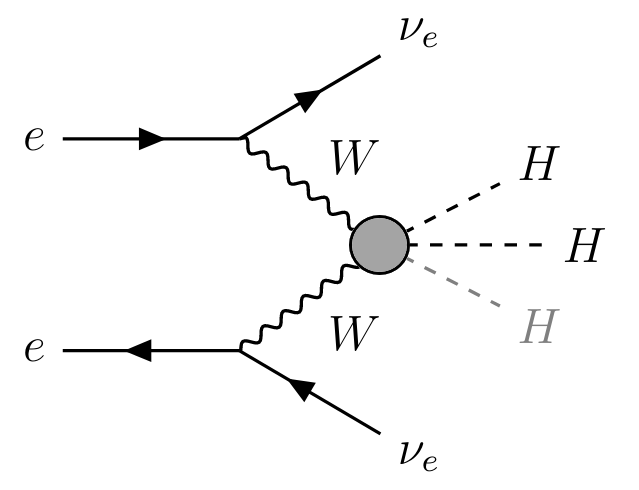}
\caption{Multiple Higgs production via WWS.}
\label{figWWS}
\end{figure}

This allows to employ factorisation, obtaining the total cross section of the whole process by convoluting that of the subprocess with the distribution functions of the vector bosons, providing a more analytical and intuitive approach to the computation than with MG5. The key point then is that the EWA assumes that the WWS is the dominant subprocess and, therefore, the validity of this approximation will depend on how much accurate this assumption is. This procedure is much simpler than computing the full cross section by a Monte Carlo method like MG5. In particular, the EWA approximation  has been used in the literature very often to simplify the estimates of BSM physics in  colliders, both for $e^+e^-$ and $pp$ collisions,  via WWS.  For the present case of $e^+e^-$ colliders the generic representation of the WWS participating in the collision and producing multiple (two or three in our case) Higgs bosons is drawn in \reffi{figWWS}.
The simple formula displaying the mentioned factorisation is:
\begin{align}
    \sigma(s) = \int dx_1\int dx_2 \sum_{i,j} f_i(x_1) f_j(x_2)\; \hat{\sigma}_{ij}(\hat{s}).
\end{align}
Here, $\sigma(s) = \sigma(e^+e^-\rightarrow HH(H)\nu_e\bar{\nu}_e)$ is the total cross section of the process of interest at a CM energy of $\sqrt{s}$, and $\hat{\sigma}_{ij}(\hat{s}) = \hat{\sigma}(W_iW_j\rightarrow HH(H))$ is the cross section of the WWS subprocess at a CM energy of $\sqrt{\hat{s}}$.  $x_1$ and $x_2$ are the momentum fractions carried by each $W$ boson and define the CM energy of the subprocess by $\hat{s} =x_1x_2 s$. The subindices $i,j$ refer to the polarization of the $W$ bosons (longitudinal or transverse). Different polarizations must be taken into account separately, as the probability of radiating a $W$ boson depends on whether it is longitudinally or transversely polarized. Consequently, each polarized cross section is convoluted with the corresponding combination of distribution functions $f_i(x)$. Note that this formula assumes that WWS is the dominant contribution to $\sigma(e^+e^-\rightarrow HH(H)\nu_e\bar{\nu}_e)$, so it is expected to work better at high energies. To compute this cross section we write $\hat{\sigma}_{ij}(\hat{s})$ in terms of the polarized amplitudes $\mathcal{M}_{ij}$, which we generate using \textsc{FeynArts-3.10} \cite{Hahn:2000kx} and \textsc{FormCalc-9.6} \cite{Hahn:1998yk}, and then perform the integration using VEGAS \cite{Lepage:1977sw} and a private PYTHON code. We do it for both the $HH\nu_e\bar{\nu}_e$ and $HHH\nu_e\bar{\nu}_e$ channels, with the corresponding phase space factors, which we omit here for shortness. The analytical expressions we use for the $W$ distribution functions are taken from \citere{Dawson:1984gx} and correspond to the so-called improved EWA, that keeps corrections of order $m_W^2/E^2$, with $E$ being the energy of the parent fermion  radiating the $W$. This improved EWA works better than the most frequently used Leading Log Approximation (LLA) EWA, which is only valid in the very high energy limit, $E\gg m_W$.  The formulas for the LLA-EWA can also be found in  \citere{Dawson:1984gx}. The Feynman diagrams contributing to the scattering amplitudes of the $W^-W^+\to HH(H)$ subprocesses are collected in the appendices. We omit the corresponding analytical expressions for shortness. 
\begin{figure}[h!]
\centering
\includegraphics[width=0.9\textwidth]{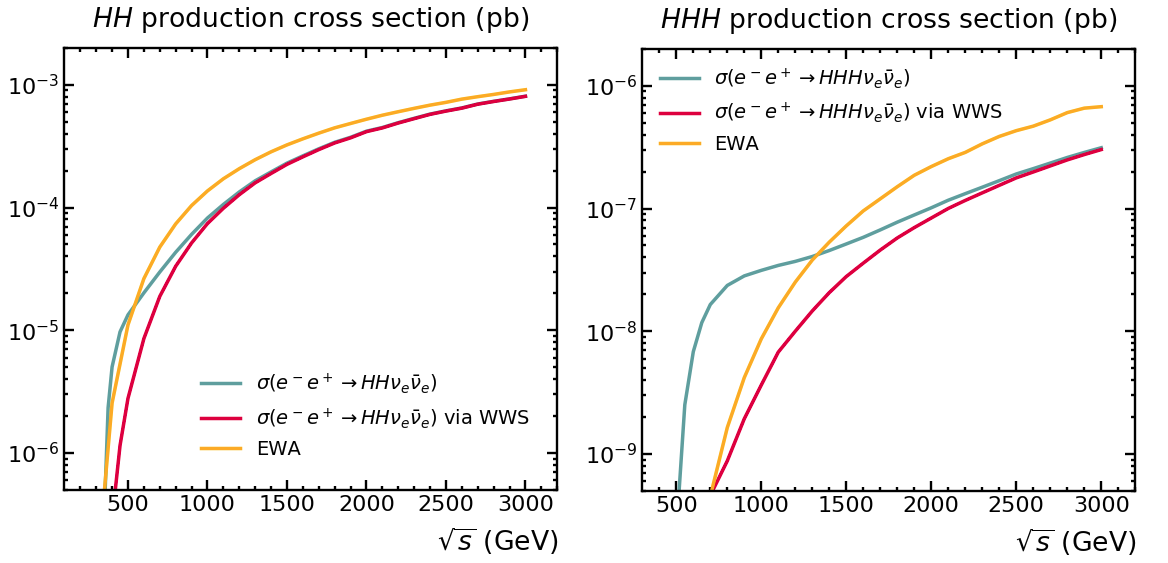}
\caption{MG5 and EWA predictions of the cross sections $\sigma(e^+e^-\rightarrow HH(H)\nu_e\bar{\nu}_e)$ as a function of energy for double Higgs (left panel) and triple Higgs (right panel) production. The predictions for $\sigma(e^+e^-\rightarrow HH(H)\nu_e\bar{\nu}_e)$ via WWS as defined in \refeq{WWSmadgraph} are also included.}
\label{MGvsEWAtotal}
\end{figure}

The numerical  results of our estimates of the cross sections with the improved EWA and their comparison with the MG5 results are presented next. In \reffi{MGvsEWAtotal} we show  the total cross section of the two processes, $e^+e^-\rightarrow HH\nu_e\bar{\nu}_e$ (left) and $e^+e^-\rightarrow HHH\nu_e\bar{\nu}_e$ (right,) as a function of the $e^+e^-$ CM energy. We include the predictions from the improved EWA and from MG5.  We also include the predictions of the contributions to the total cross section from WWS, $\sigma_{WWS}$,  as defined in \refeq{WWSmadgraph}. We see in this figure that at high energies the EWA predicts the $HH$ production cross section with good accuracy, while it fails in the predicted cross section in more than a factor of 2 for the $HHH$ case. The energies that we are considering in this work seem to be too low for the EWA to be a good approximation of triple Higgs production. 
We have also checked that the dominant contribution to the total cross section comes from longitudinally polarized $W$ bosons in both cases.
\begin{figure}[h!]
\centering
\includegraphics[width=0.999\textwidth]{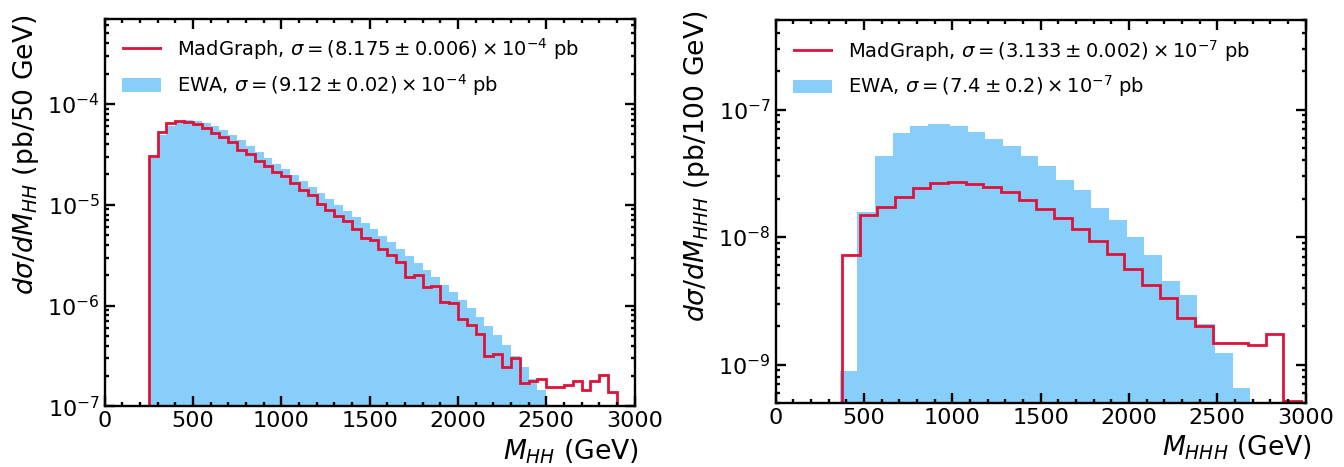}
\caption{Differential cross sections for $e^+e^- \to HH \nu \bar{\nu}$ (left) and $e^+e^- \to HHH \nu \bar{\nu}$ (right) at $\sqrt{s}=3000$ GeV with respect to the invariant mass of the final Higgs bosons, $M_{HH}$ (left) and $M_{HHH}$ (right). The predictions from both MG5 and the EWA are included for comparison.}
\label{MGvsEWAdiff}
\end{figure}

Another condition for the EWA to be a useful approximation is that it can reproduce the differential cross section distributions. As an example,  \reffi{MGvsEWAdiff} shows the differential cross section with respect to the invariant mass of the final state Higgs bosons, $M_{HH}$ and $M_{HHH}$ respectively, which is equivalent to $\sqrt{\hat{s}}$ due to 4-momentum conservation. The results for the SM case are again accurate for double Higgs production, but not so satisfactory in the case of three Higgs bosons. The good agreement of the EWA with the MG5 computation in the double Higgs production channel also occurs in the case of the BSM predictions. Indeed, the EWA works even better for BSM than for the SM. This will be shown in the next section.

From this brief analysis we conclude that the EWA is an interesting alternative to compute our results in the double Higgs case, but it is not accurate enough to study triple Higgs production in the range of energies available at the $e^+e^-$ colliders under consideration. For the final analysis at $e^+e^-$ colliders in sections 5 and 6, we will not use the EWA anymore, and our results on total and differential cross sections will be extracted from the full Monte Carlo simulation with  MG5.

\section{Deviations from BSM Higgs couplings in WWS}
\label{TestWWS}
In this section we study the effects of the anomalous Higgs couplings in double and triple Higgs production via WWS. In particular, we consider the $WWH$, $WWHH$, $HHH$ and $HHHH$ interactions, given by the EChL coefficients $a$, $b$, $\kappa_3$ and $\kappa_4$, and explore the departures with respect to the SM predictions both in total cross sections and in some differential distributions. First, we study the effects of $a$ and $b$ in $W^-W^+ \to HH$ and next we analyse the effects of $\kappa_3$ and $\kappa_4$ in $W^-W^+ \to HH(H)$. To simplify the computations, we take $\kappa_{3,4}=1$ when exploring the sensitivity to $a$ and $b$, and $a,b=1$ when studying the effects of $\kappa_3$ and $\kappa_4$.
 The computations presented in this section have been performed using \textsc{FeynArts-3.10} for the generation of diagrams in the unitary gauge and \textsc{FormCalc-9.6} to perform the analytical calculation of the corresponding amplitudes. In addition, we use VEGAS to integrate numerically over the corresponding phase space in order to obtain the corresponding cross sections. All these computations have been additionally checked with MG5.
 %%%
\subsection{Effects of $a$ and $b$ in $W^-W^+ \to HH$}
\label{abinWWS}
\begin{figure}[t!]
\centering
\includegraphics[width=0.49\textwidth]{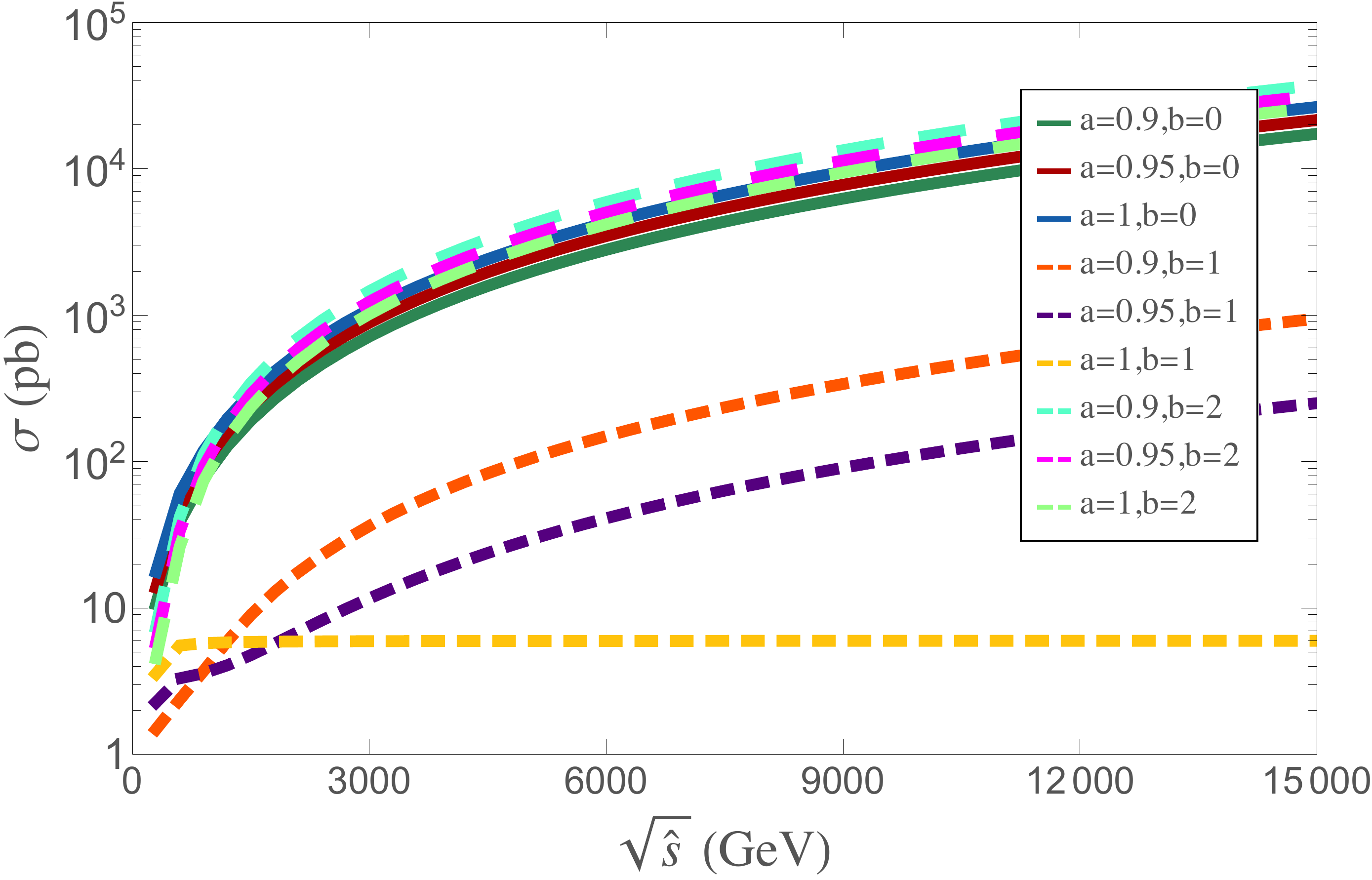}
\includegraphics[width=0.49\textwidth]{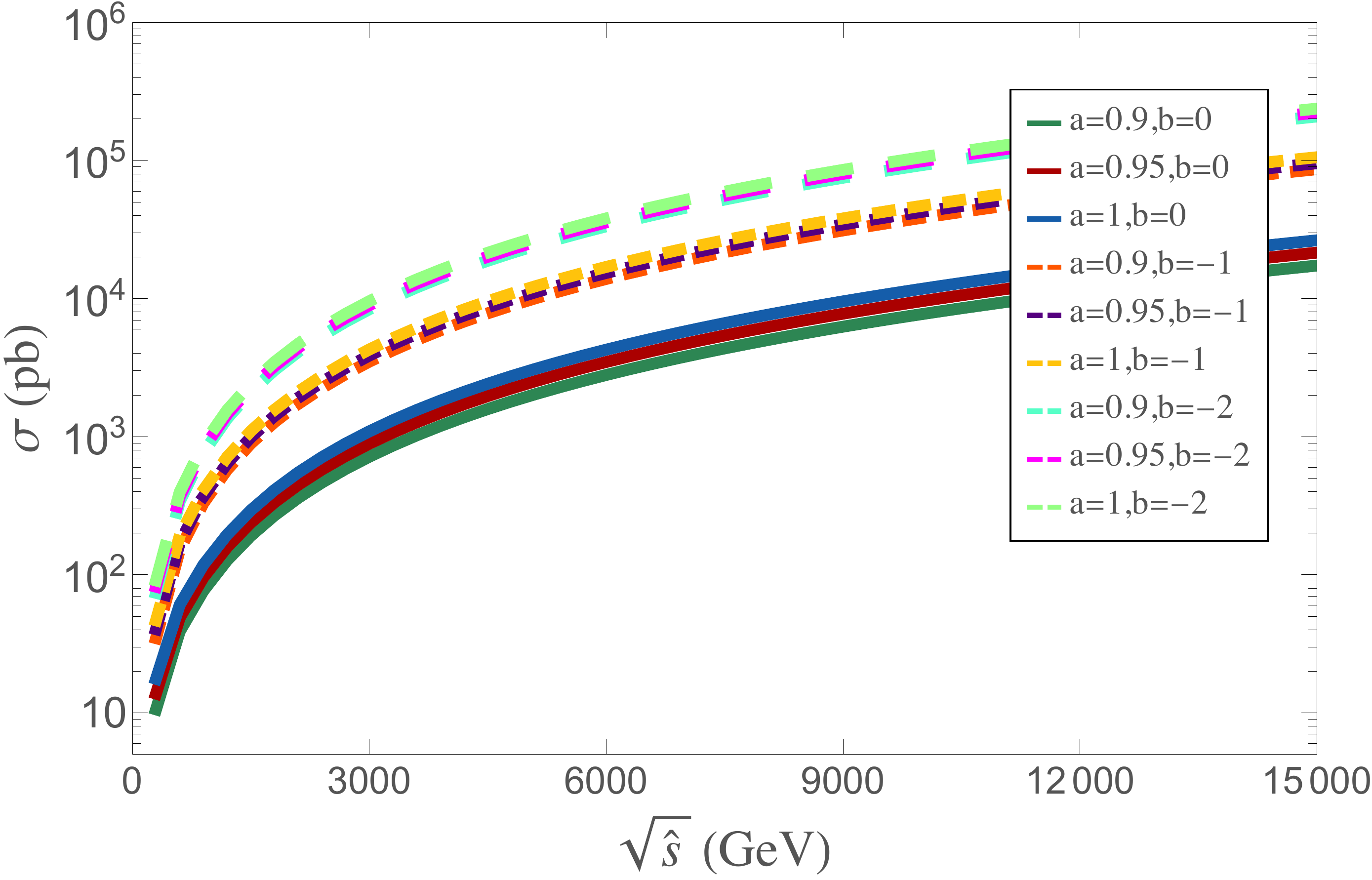}
\includegraphics[width=0.49\textwidth]{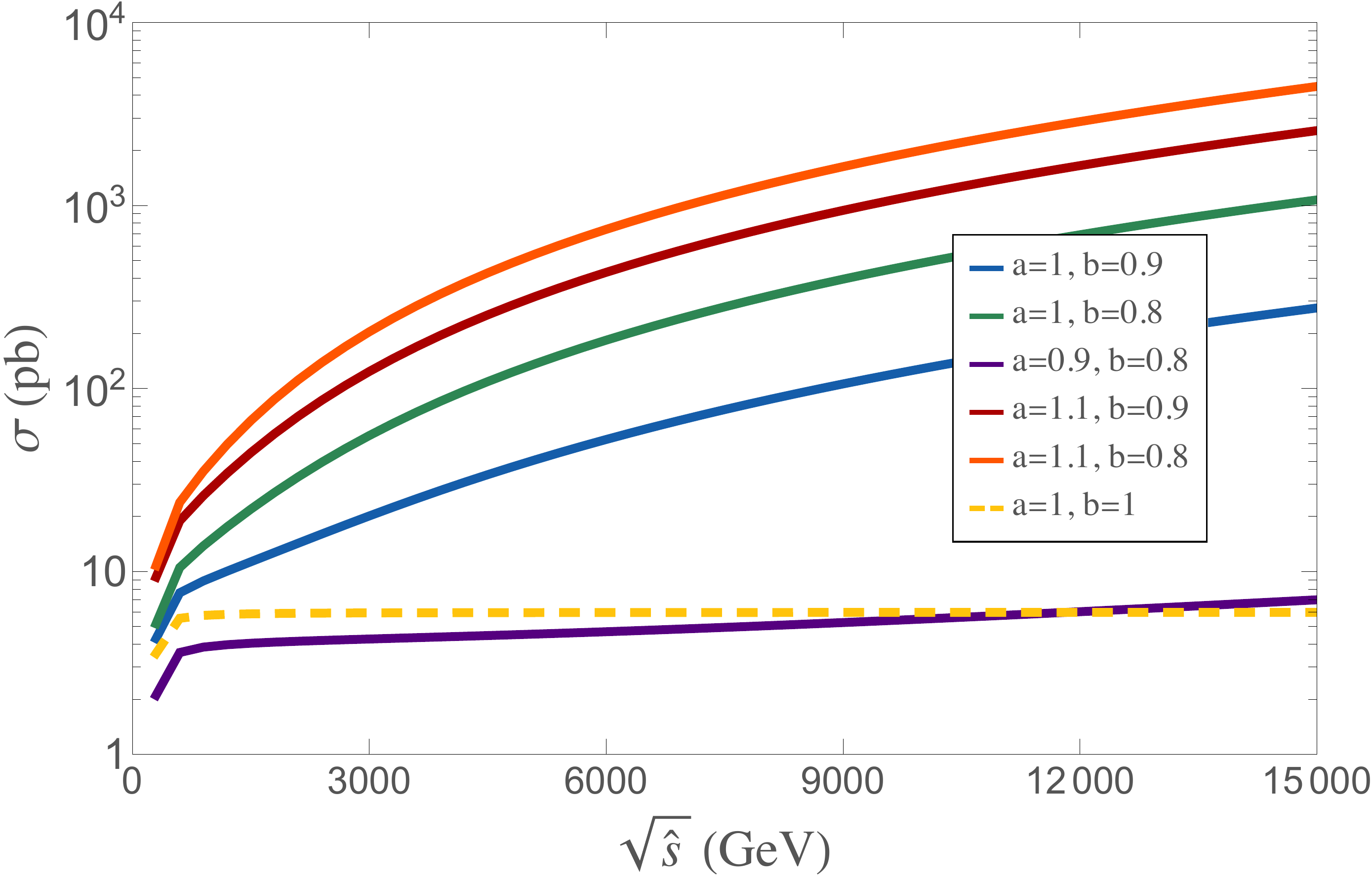}
\hspace{-0.25cm}
\includegraphics[width=0.495\textwidth]{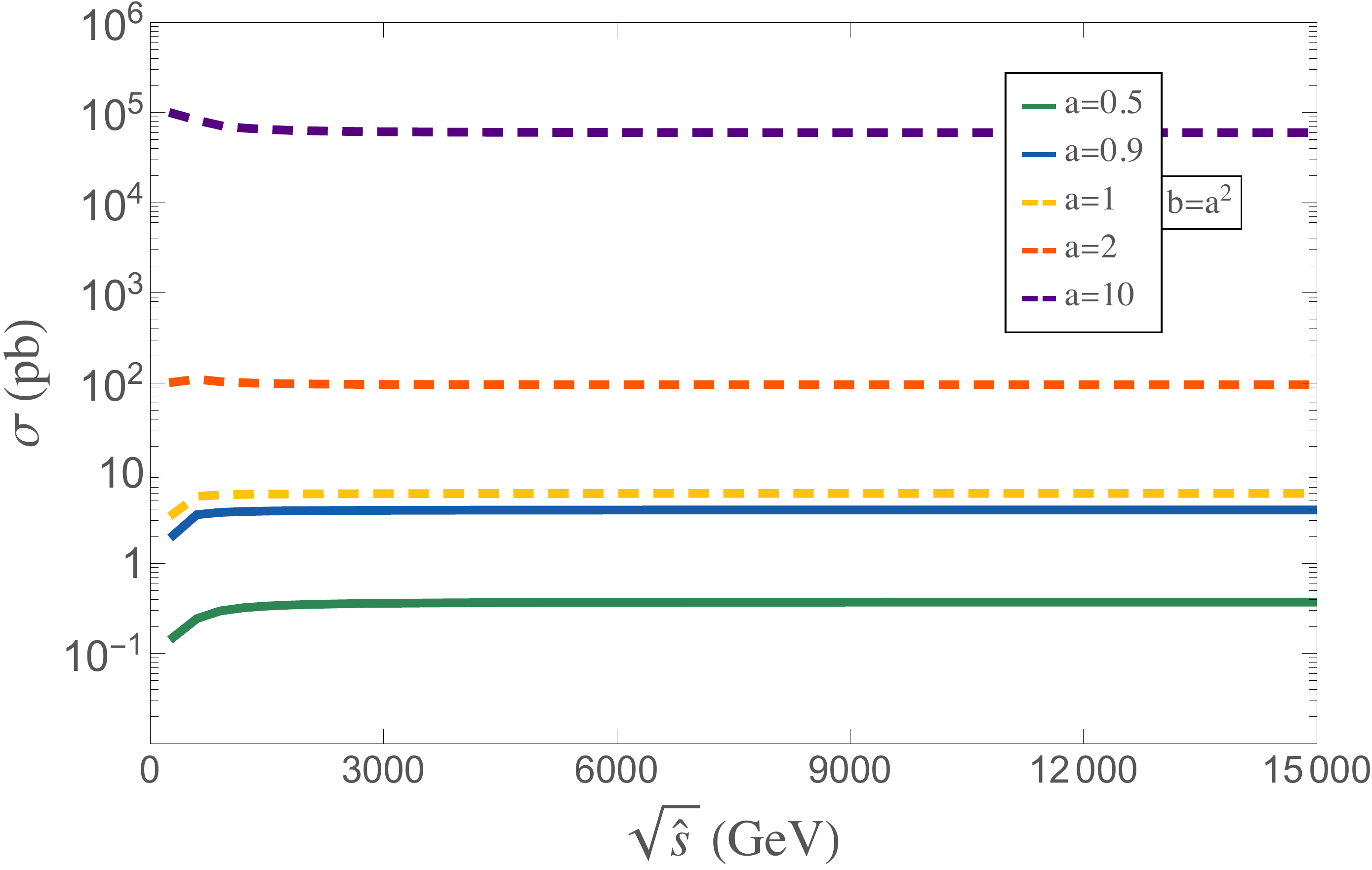}
\caption{Total cross sections of the $W^-W^+\rightarrow HH$ subprocess in the EChL, as a function of the CM energy, for different values of the parameters $a$ and $b$.  The upper left (right) panel shows positive (negative) values of $b$.  Smaller deviations of this parameter with respect to $b=1$ are displayed on the lower left panel.  In the lower right panel,  the values of $a$ and $b$ are determined by the relation $b=a^2$.  The dashed yellow lines correspond to the SM prediction ($a=b=1$),  except in the upper right panel,  where the SM case is not displayed.}
\label{xsechl1}
\end{figure}

%%%
Here we present the effects in the $W^-W^+ \to HH$ subprocess of the two anomalous Higgs couplings parametrised by $a$ and $b,$ and compare them with the SM case, i.e, with $a=b=1$. 
\reffi{xsechl1} shows how the total cross section of this subprocess depends on the CM energy when varying the EChL parameters $a$ and $b$.  We restrict our study to positive values of $a$ but consider both positive and negative values of $b$. Although some values are outside the experimental bounds, we believe that at this intial stage it is illustrative to show the effects of these parameters when they vary in a wider range. 

Some first conclusions can be extracted from \reffi{xsechl1}.  It is clear that the behaviour of the cross section with the energy in the EChL is,  in general,  very different than that of the SM prediction.  While in the SM case (dashed yellow line in the left plot), the behaviour is flat with the energy at the TeV region and above,  in most cases where $a$ and/or $b$ are different from 1 the cross section grows very steeply. This growth with the energy occurs typically in predictions from EFTs, due to the ultraviolet incompleteness of these theories. In particular, the EChL provides predictions, as  previously said, that generically grow with powers of the external momenta.  

In the SM case, the flatness with the energy occurs because there is a strong cancellation of the terms that grow with the energy among the contributions from the various diagrams. If we consider the dominant contribution to this SM cross section, that comes from the longitudinally polarized $W$ gauge bosons,  it turns out that, at high energies (for $\sqrt{\hat s} \gg m_W, m_H$), the contributions from the contact, $t$ and $u$ channels to the scattering amplitude ${\cal M}(W_L W_L \to HH)$ are proportional to ${\hat s}$, whereas the contribution from the $s$ channel has a constant behaviour with ${\hat s}$ \cite{Arganda:2018ftn}. In this case, i.e for $a=b=1$, there is  an exact cancellation of the terms growing linearly with ${\hat s}$  among the contact, $t$ and $u$ channels, and what remain  are just the terms having a constant behaviour with ${\hat s}$. However, this strong cancellation among diagrams does not happen in the case of the EChL for arbitrary $a$ and $b$ values. More concretely, this cancellation of the terms growing with energy occurs in the SM because of the relation mentioned previously among the two vertices, $V_{WWH}=v V_{WWHH}$, which is not fulfilled in the EChL case. To get a similar cancellation of the terms growing linearly with ${\hat s}$ in the EChL amplitude one should restrict the parameters to the particular setting given by $b=a^2$, which is obviously not necessarily true in the general case.  This can be seen in the lower right panel of \reffi{xsechl1},  which shows how the cross section for the $W^-W^+\to HH$ subprocess is practically constant with the energy when enforcing the relation $b=a^2$.  Notice also that this restriction on $a$ and $b$ is not required  by any symmetry argument (remember that the EChL is gauge and chiral invariant for all $a$ and $b$). Within the SM, the relation $V_{WWH}=v V_{WWHH}$ occurs as a consequence of the symmetry being linearly realised, i.e.  because the Higgs field is placed into a doublet. But in the EChL, the Higgs field is a singlet under the gauge and chiral symmetries and $a$ and $b$ can be arbitrary.    

%The dominant contribution to this cross section is the scattering amplitude of polarized $W$ bosons, ${\cal M}(W_L W_L \to HH)$. At high energies (for $\sqrt{\hat s} \gg m_W, m_H$), the contact, $t$- and $u$- channels are proportional to ${\hat s}$, whereas the contribution from the $s$-channel does not depend on the energy \cite{Arganda:2018ftn}. In the SM case, i.e. for $a=b=1$, there is an exact cancellation of the terms which grow linearly with ${\hat s}$ among the contact, $t$- and $u$-channels, and only the constants terms remain. This is due to the previously mentioned relation between the $WWH$ and $WWHH$ vertices, $V_{WWH}=v V_{WWHH}$, which is a consequence of the EW symmetry being linearly realised (i.e. of the Higgs field being placed in a doublet). In the EChL, the Higgs field is a singlet under the gauge and chiral symmetries; $a$ and $b$ can be arbitrary, so both vertices are not related anymore and the strong cancellation among diagrams will not occur for arbitrary values of $a$ and $b$. To get a similar cancellation of the terms growing linearly with ${\hat s}$ in the EChL amplitude, one should restrict the parameters to the particular setting given by $b=a^2$, which is obviously not fulfilled generically.  Notice also that this restriction on $a$ and $b$ is not required  by any symmetry argument (remember that the EChL is gauge and chiral invariant for all $a$ and $b$).

\begin{figure}[t!]
\centering
\includegraphics[scale=0.4]{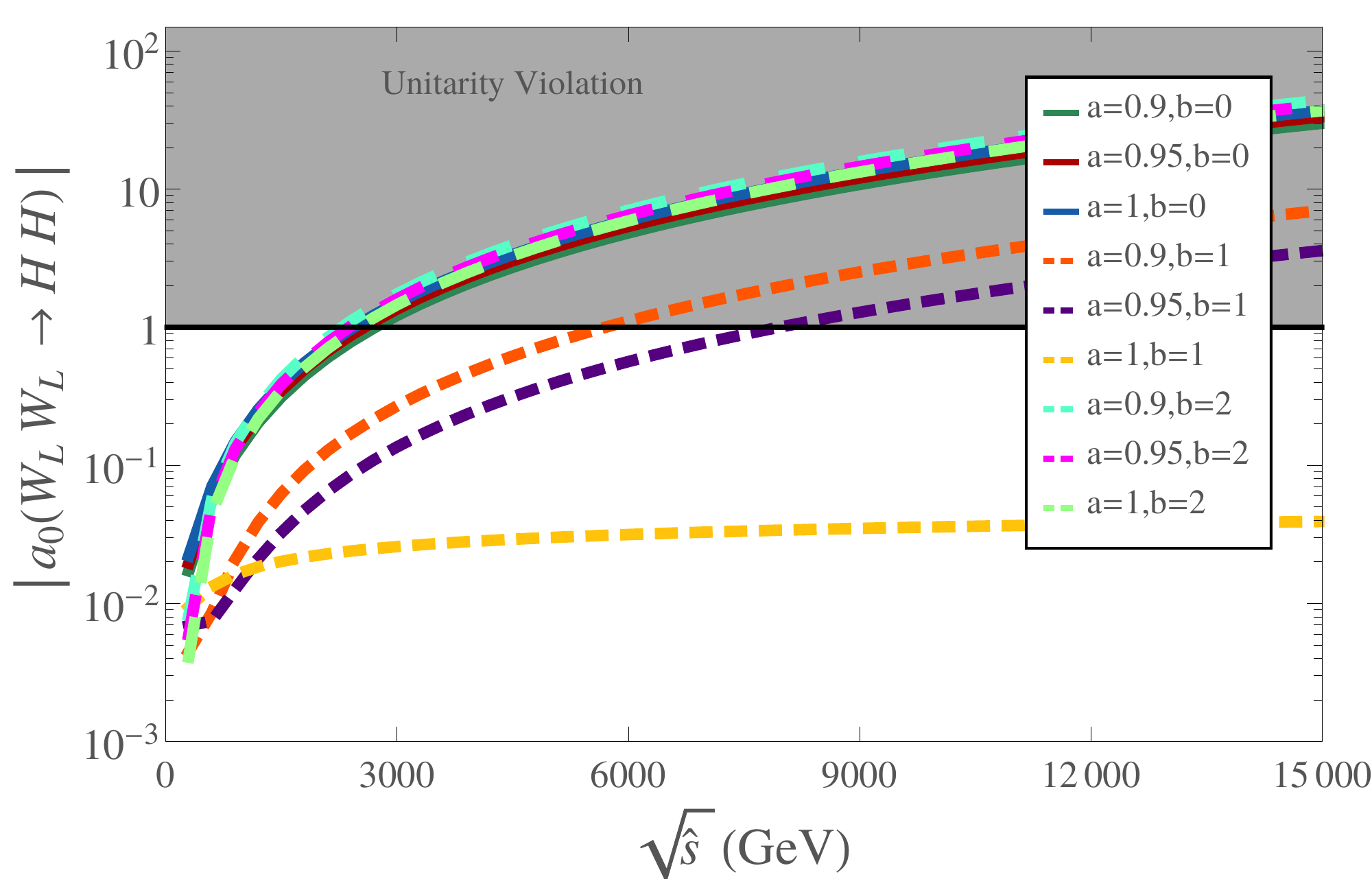}
\includegraphics[scale=0.4]{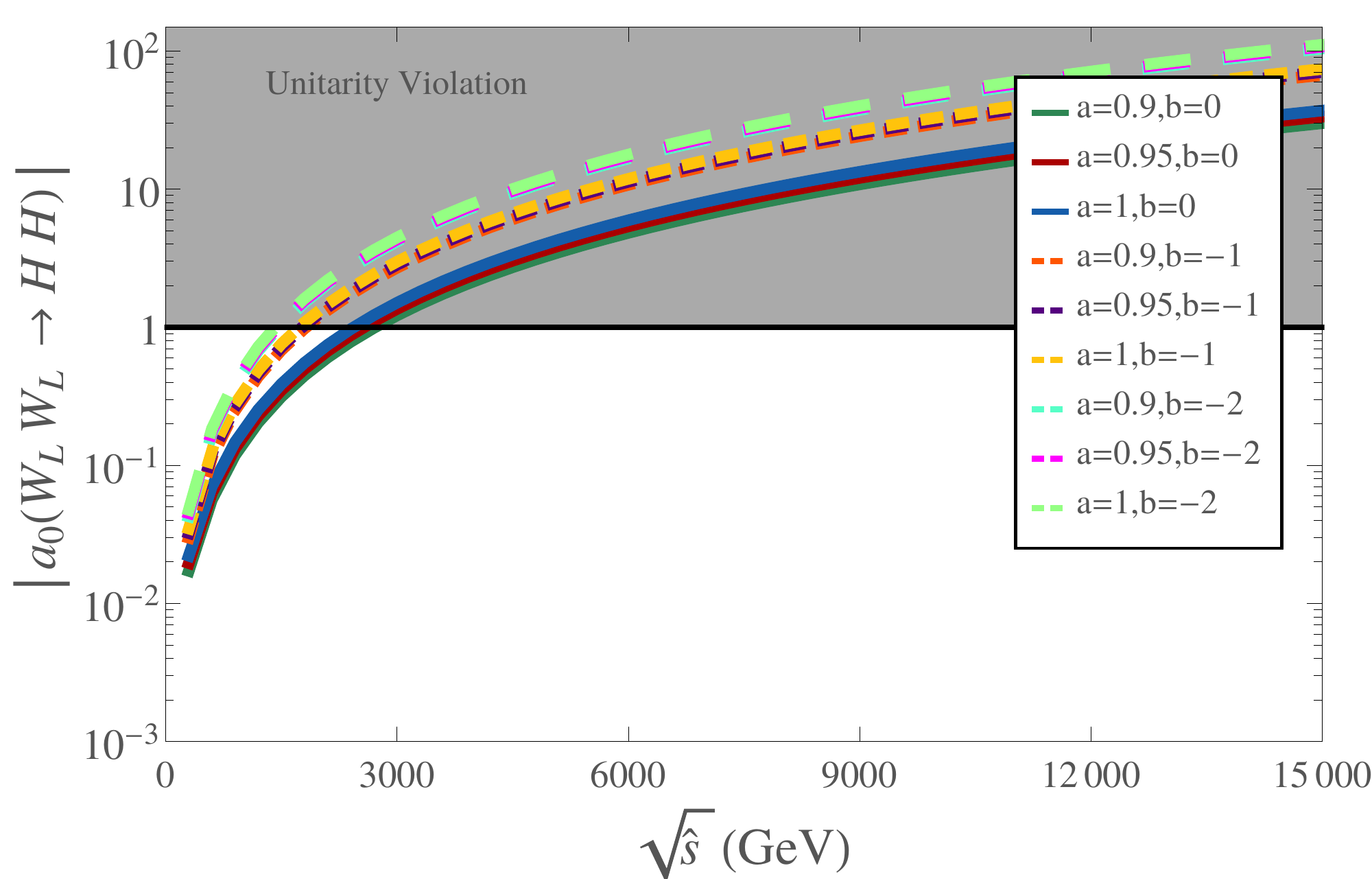}
\caption{Modulus of $s$-wave partial amplitudes, as a function of the CM energy, for the $W_LW_L\rightarrow HH$ subprocess in the EChL, for different values of the parameters $a$ and $b$. The dashed yellow line corresponds to the SM prediction ($a=b=1$). Left panel shows positive values of $b$ whereas right panel displays negative values.}
\label{partial}
\end{figure}

The plots in \reffi{xsechl1} also show that, the bigger the deviations of $a$ and $b$ from 1, the bigger the cross section. Thus, some possible BSM physics, given by the EChL parameters departing from their SM values, would yield clear experimental evidence, with much larger cross sections than those predicted by the SM. For instance, at $\sqrt{\hat s}=3\, {\rm TeV}$, and for the considered values of $a$ and $b$ in the upper left panel, the cross section can be several orders of magnitude larger than in the SM. For negative values of $b$ (upper right plot), the departure with respect to the SM prediction can be even larger. For instance, for $a=1$, $b=-1$ the cross section is about a factor 1000 larger than that for $a=b=1$. This is a remarkable enhancement, produced by just changing the sign of $b$. 
Notice that  the variations chosen in these two plots for $b$ with respect to 1 are much bigger than for $a$, simply because the experimental bounds existing on $a$ are more restrictive, while the constraints are looser for $b$.  Considering smaller deviations of these parameters reduces the size of the enhancement in the cross section with respect to the SM prediction,  but it still produces important departures,  as it can be seen in the lower left plot in \reffi{xsechl1}.  For instance,  at $\sqrt{\hat s}=3\, {\rm TeV}$,  a deviation in $b$ of $20\%$ ($10\%$) enlarges the SM cross section by a factor of 10 (4).  If both $a$ and $b$ are slightly deviated from 1, with a variation in $a$ of $10\%$, and in $b$ of $20\%$ ($10\%$) the  corresponding enhancement factor is 40 (20).

This large increase of the cross sections when $a$ and/or  $b$  are not equal to 1 might indicate that unitarity is not guaranteed in the EChL predictions.
% as cross sections are unequivocally related to scattering probabilities, a monotonic growth of the %cross section with the energy might mean the probability eventually becomes larger than 1. %
In order to check if unitarity is preserved, we plot, in \reffi{partial}, the energy dependence of the longitudinally polarized $s$-wave partial amplitude for the same values of $a$ and $b$ as those chosen in the previous figure.
These results show that unitarity is, in fact, violated for some parameter $(a,b)$  values at high energies. Except in the SM case (which, as it is well known, perfectly respects unitarity), the rest of the chosen values of the parameters render partial wave amplitudes larger than 1 at energies of several TeVs, which could be reachable at colliders. As it could be anticipated, the further the departure from the SM, the sooner unitarity is violated.
\begin{figure}[h!]
\centering
\includegraphics[scale=0.4]{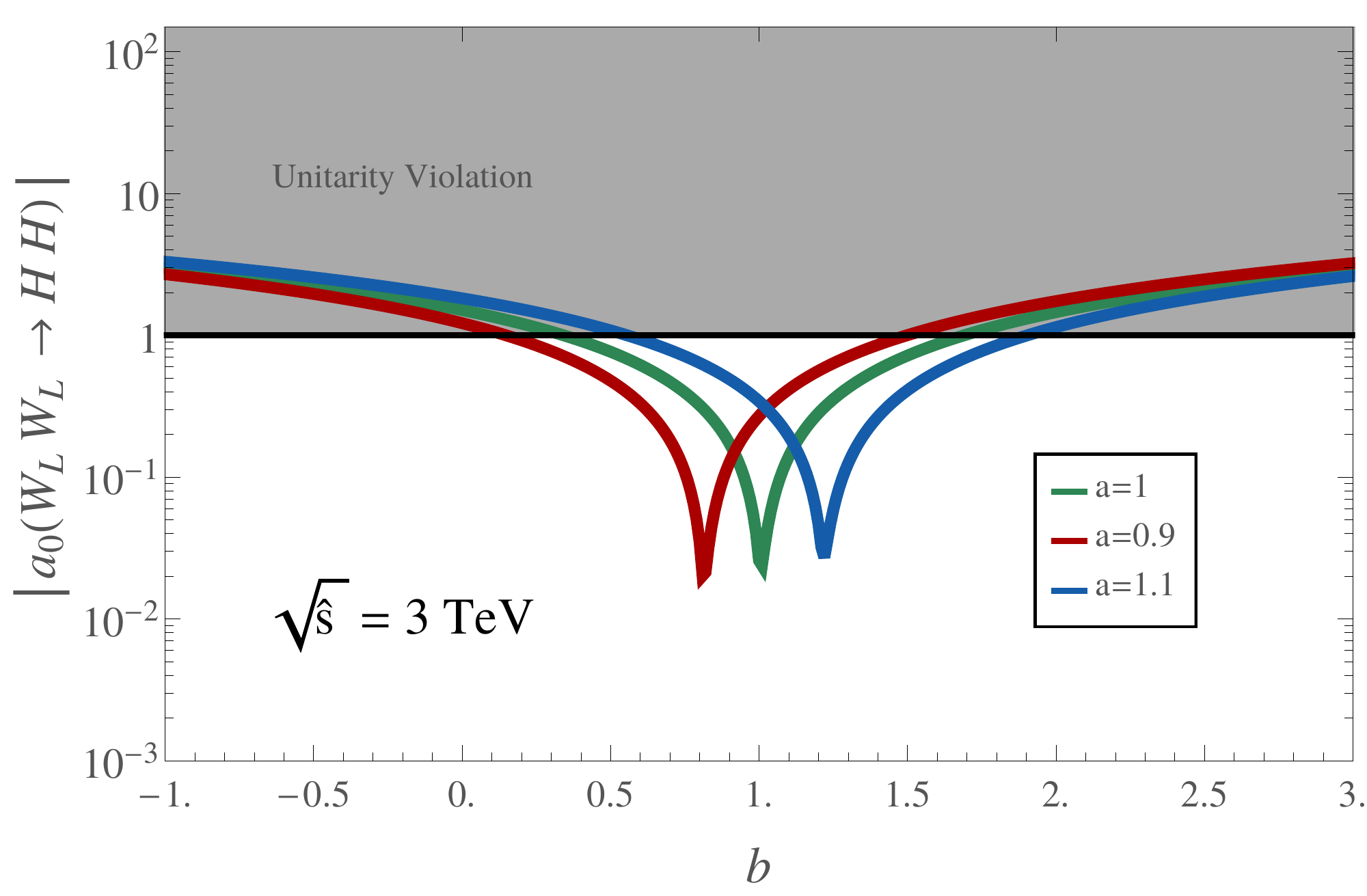}
\includegraphics[scale=0.4]{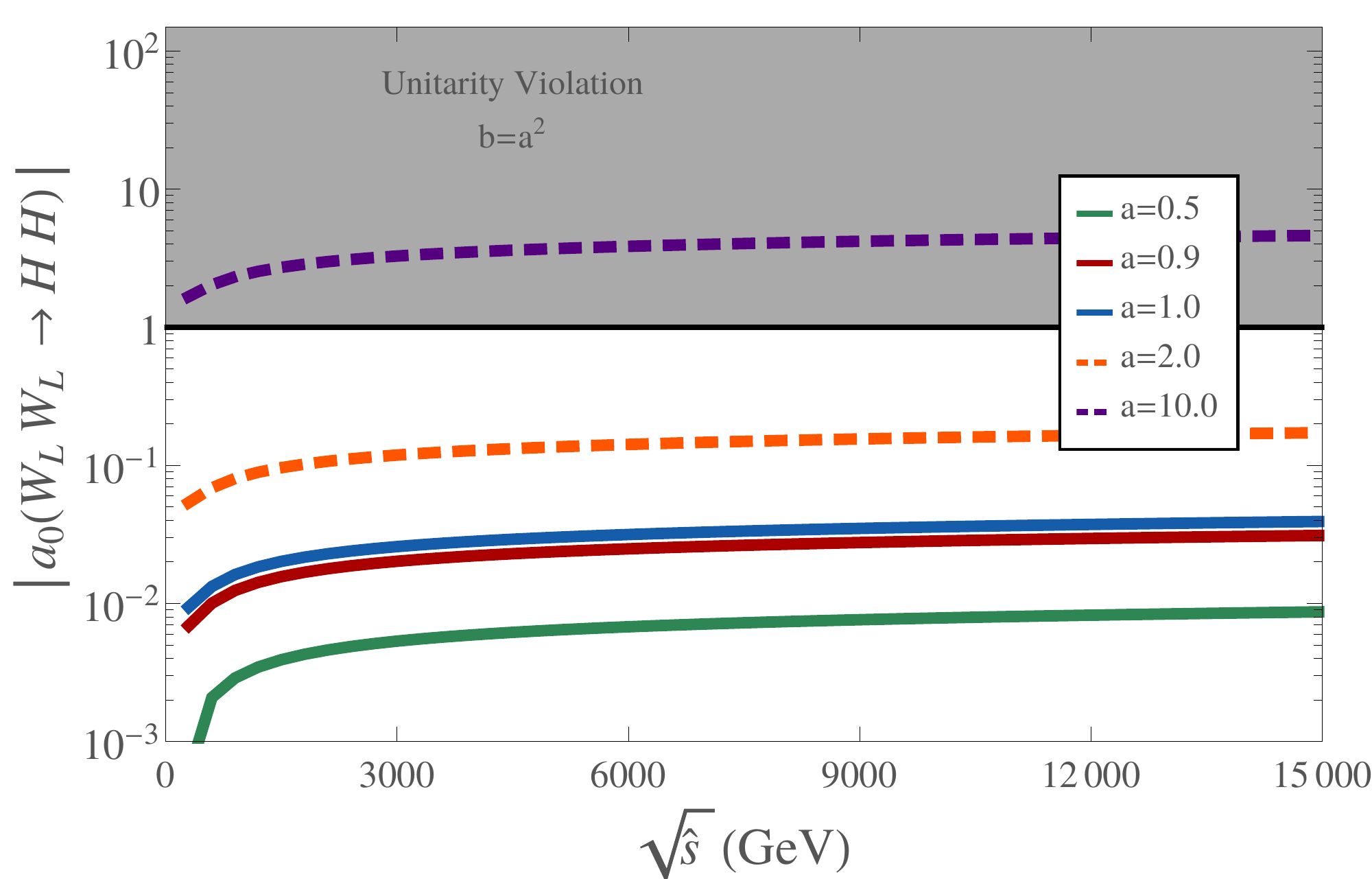}
\caption{Lowest partial wave amplitude for $W_LW_L\rightarrow HH$ within the EChL. Left panel shows the dependance on the $b$ parameter, for different values of $a$ and a fixed energy of  $\sqrt{\hat s}=3$ TeV. Right panel shows the dependance on the energy for several values of $a$, with $b$ determined by the relation $b=a^2$. }
\label{a0echl3}
\end{figure}

Focusing on one particular energy, it is also very illustrative to study the failure of unitarity as a function of the parameters $a$ and $b$. \reffi{a0echl3} (left panel) shows the predictions at $\sqrt{\hat s}=3$ TeV for the lowest partial wave amplitude as a function of $b$ and for several values of the parameter $a$. It can be seen that, at this energy, when $a$ varies from its SM value approximately a 10$\%$ (which, roughly speaking, coincides with experimental bounds), $b$ can depart from 1 a 50$\%$ at most, in order to obtain predictions which preserve unitarity. Thus, for the case of CLIC with the largest planned energy, we learn from this figure that by keeping $a$ within the experimentally allowed interval and by restricting the $b$ parameter in a rather conservative interval of $b \in [0.5, 1.5]$, the predictions will respect unitarity. With this conservative choice, unitarity will consequently be also preserved at other colliders with lower energies. 
%\begin{figure}[h!]
%\centering
%\includegraphics[scale=0.4]{figures/PartialWavesNew.pdf}
%\caption{Predictions from the EChL for the lowest partial wave amplitude as a function of energy, %for various $a$ and $b$ parameters but all related by $a^2=b$. }
%\label{a0-for-a2eqb}
%\end{figure}

Finally, to illustrate the effect of the EChL parameters on the issue of unitarity violation when restricted  by the relation $b=a^2$, we show in \reffi{a0echl3} (right panel) the predictions for the lowest partial wave amplitude as a function of the energy, for various values of $a$ ($b$ is set here to $a^2$). It is clear from this plot that, at high energies ($\sqrt{\hat s}\gg m_W,m_H$), a constant behaviour with the energy, similar to the one in the SM, is obtained in the EChL when the $a$ and $b$ parameters are restricted by $b=a^2$.  We also conclude from this figure that all the displayed predictions respect the unitary bound, expect for the extremely large value of $a=10$, where the constant prediction excedes the unitary limit. Such large values of $a$ are totally out of the experimentally allowed interval, and thus they will not be considered anymore in this work.  

\begin{figure}[h!]
\centering
\includegraphics[width=0.49\textwidth]{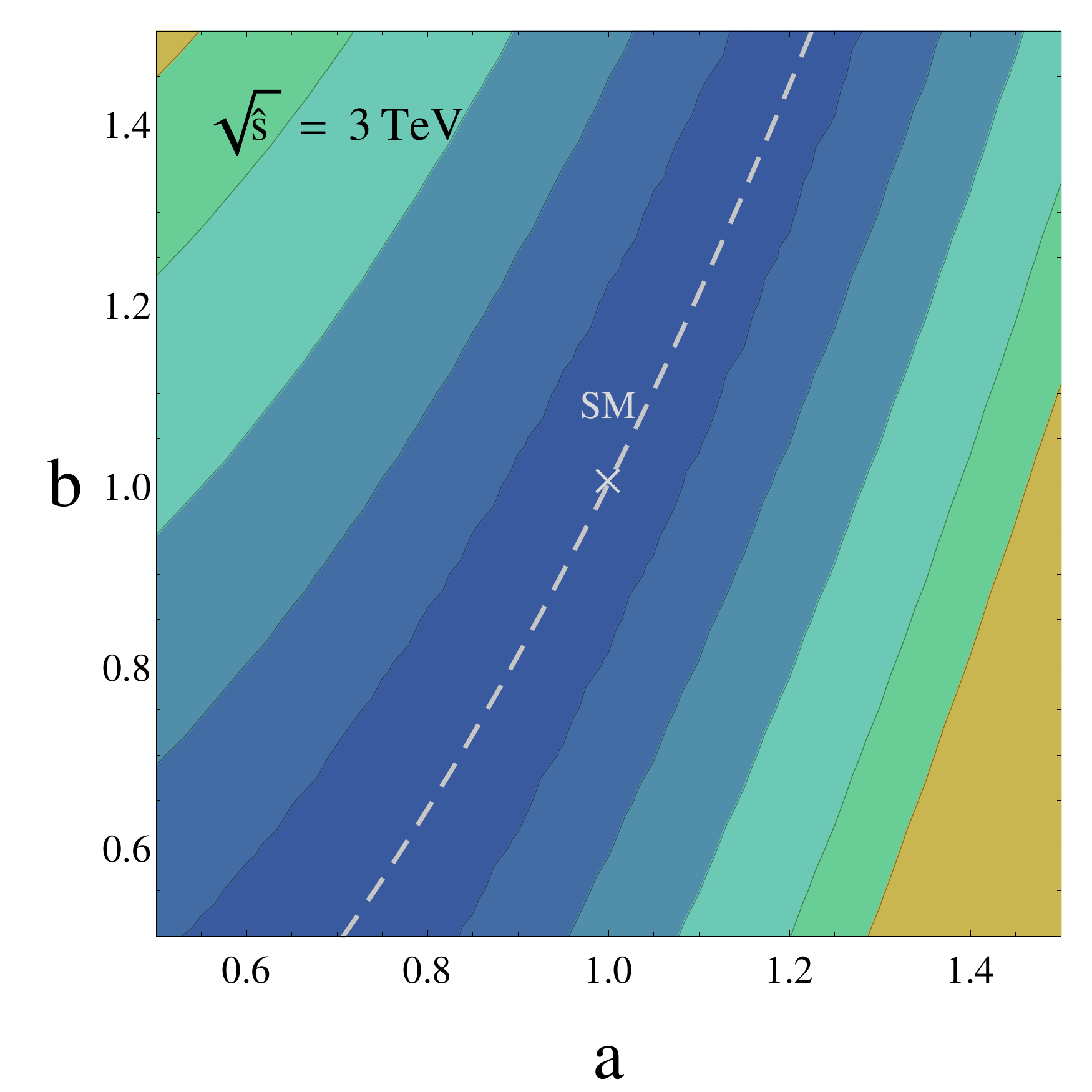}
\includegraphics[width=0.49\textwidth]{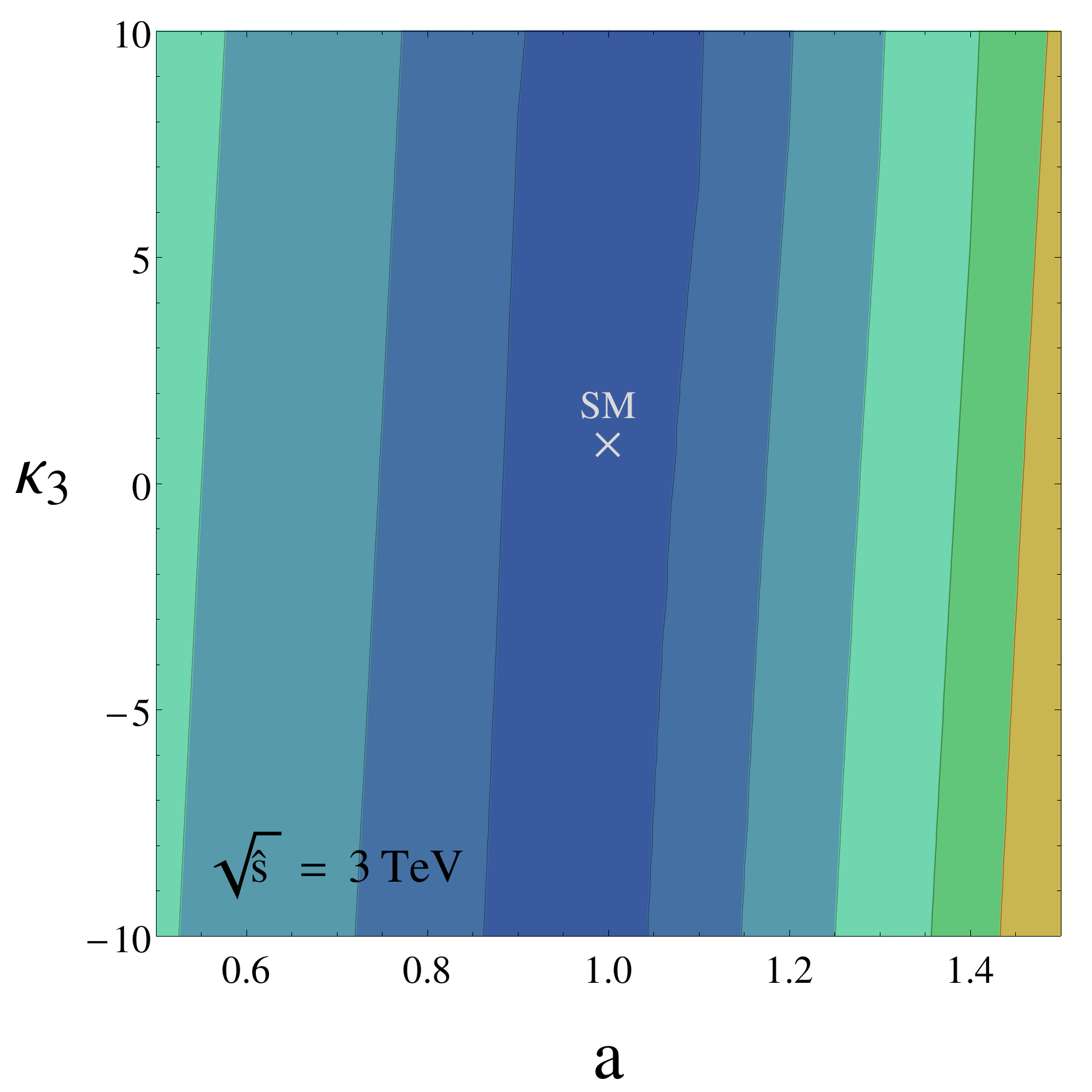}
\includegraphics[width=0.6\textwidth]{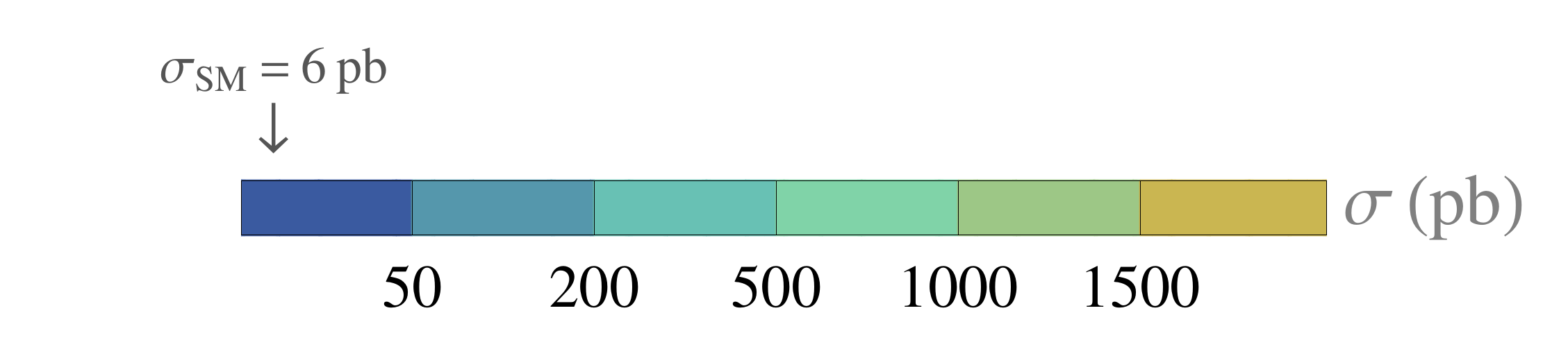}
\caption{EChL predictions for the total cross section of the $W^-W^+\to H H$ subprocess,  at a center of mass energy of 3 TeV,  in the $(a,b)$ plane (left panel) and the $(a,\kappa_3)$ plane (right panel). The dashed line in the left panel corresponds to the direction $b=a^2$.  The white cross represents the SM prediction ($a=b=1$).}
\label{WWHHcontours}
\end{figure}

These results show that the cross section flattens with the energy for the specific choice of $b=a^2$ (with considerably lower rates than for the general case, $b\neq a^2$),  indicating a  particular direction in the parameter space where the cross section is minimum.  To illustrate this correlation more clearly, we show in \reffi{WWHHcontours} (left plot) the contour lines of constant cross section, $\sigma(WW \to HH)$,  in the $(a,b)$ plane.  We have also included the contour line corresponding to  $b=a^2$ for comparison (with the SM point marked on top of it).   We can see  in this plot that the cross section contour lines  are parallel to the $b=a^2$ direction,  and how the cross section reaches the minimum values along this line,  growing very rapidly when separating from this line.  It is also worth mentioning that this correlation we find between $a$ and $b$ is not present in other combinations of the studied EChL parameters.  In particular, we have also explored the cross section in the ($a,\kappa_3$) plane, not finding any interesting correlation between these two parameters.  This is shown in \reffi{WWHHcontours} (right plot),  where the contour lines are practically parallel to the vertical axis; it can also be seen that the effect of $a$ is dominant,  overwhelming that of $\kappa_3$.  
Thus,  from now on we will consider separately the two sets of parameters, $a$ and $b$ on one hand,  $\kappa_3$ and
$\kappa_4$ on the other.  This classification is basically motivated by the type of interactions affected by each set of anomalous couplings,  $H-W$ and Higgs self-interactions,  respectively.

All in all, the results in this subsection clearly show that the EChL predictions are way different than those of the SM when varying the parameters $a$ and $b$ and provide quite large cross sections for the $W^-W^+ \to HH$ subprocess.  We will see in the following that this can lead to measurable departures from the SM predictions at future $e^+e^-$ colliders. And this is true even after including the restrictions on the parameters $a$ and $b$ such that unitarity is respected in all the predictions.
%%%%
\subsection{Effects of $\kappa_3$ and $\kappa_4$ in $W^-W^+ \to HH(H)$}
\label{k3k4inWWS}
In this section we analyse the WWS subprocesses and the BSM deviations with respect to the SM predictions arising from the anomalous triple and quartic Higgs self-couplings, parametrised by $\kappa_3$ and $\kappa_4$ respectively.  For this  study we explore channels of both double and triple Higgs production and, for simplicity, we set  $a,b=1$. Since double Higgs production is not sensitive at all to $\kappa_4$, we focus here mainly on the triple Higgs production case, which is the novel one,  and use the double Higgs case rather as a reference case (better known) regarding the sensitivity to $\kappa_3$. For a more devoted study on the effect of $\kappa_3$ in double Higgs production see, for instance, \citere{Arganda:2018ftn}.

\begin{figure}[h]
\centering
\includegraphics[width=0.999\textwidth]{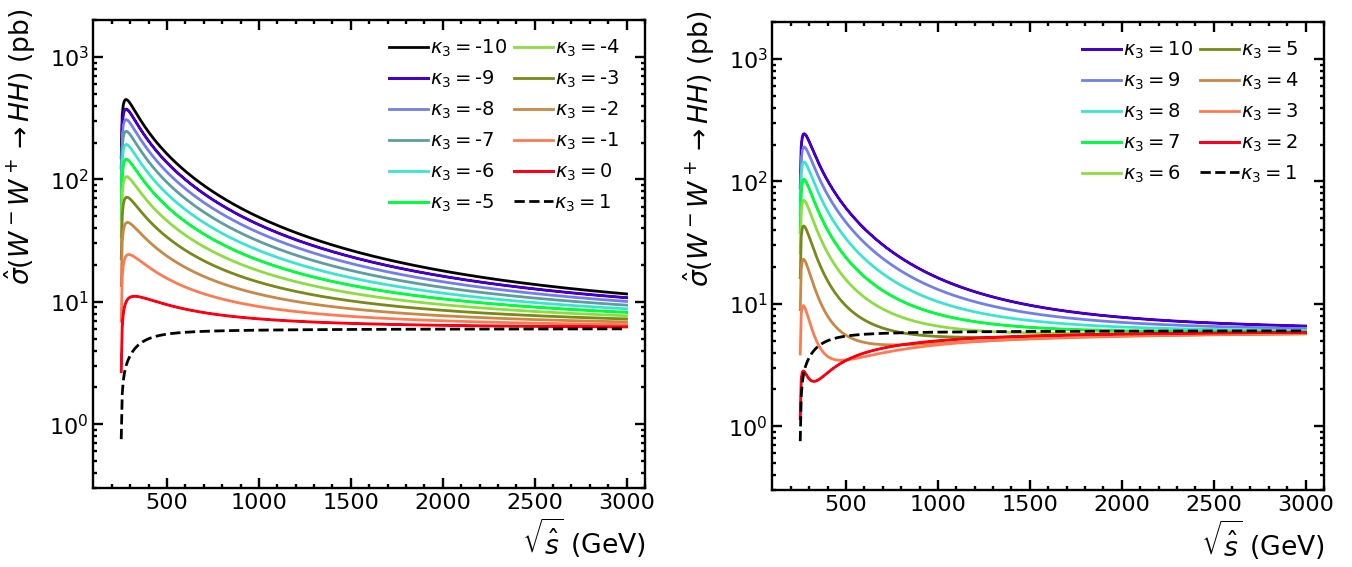}
\caption{Total cross section of the $W^-W^+\rightarrow HH$ subprocess as a function of the CM energy $\sqrt{\hat{s}}$ for different values of the parameter $\kappa_3$, with $\kappa_4$ fixed to 1, compared to the SM prediction (dashed line). Negative (positive) values of $\kappa_3$ are shown in the left (right) panel.}
\label{HHk3vssqrts}
\end{figure}

\begin{figure}[h!]
\centering
\includegraphics[width=0.999\textwidth]{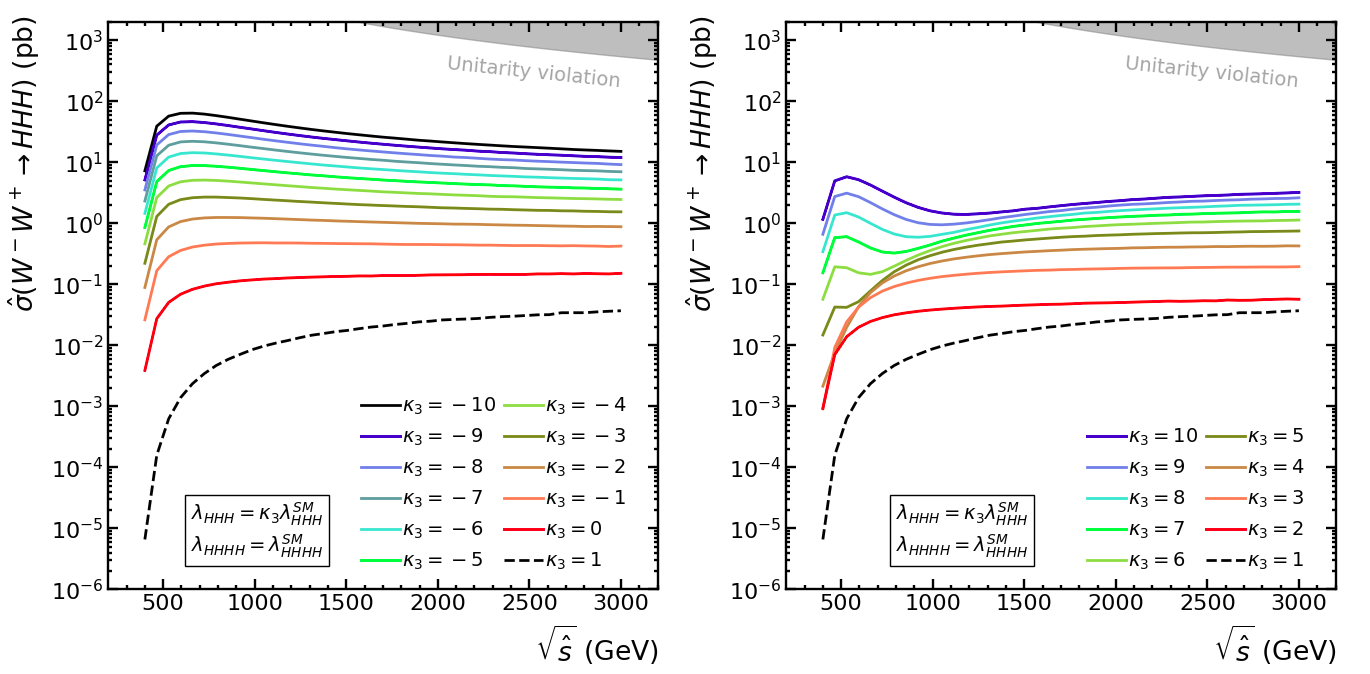}
\caption{Total cross section of the $W^-W^+\rightarrow HHH$ subprocess as a function of the CM energy $\sqrt{\hat{s}}$ for different values of the parameter $\kappa_3$, with $\kappa_4$ fixed to 1, compared to the SM prediction (dashed line). Negative (positive) values of $\kappa_3$ are shown in the left (right) panel.  The unitarity violating region is the shaded area displayed at the right upper corner.}
\label{HHHk3vssqrts}
\end{figure}

\begin{figure}[h!]
\centering
\includegraphics[width=0.999\textwidth]{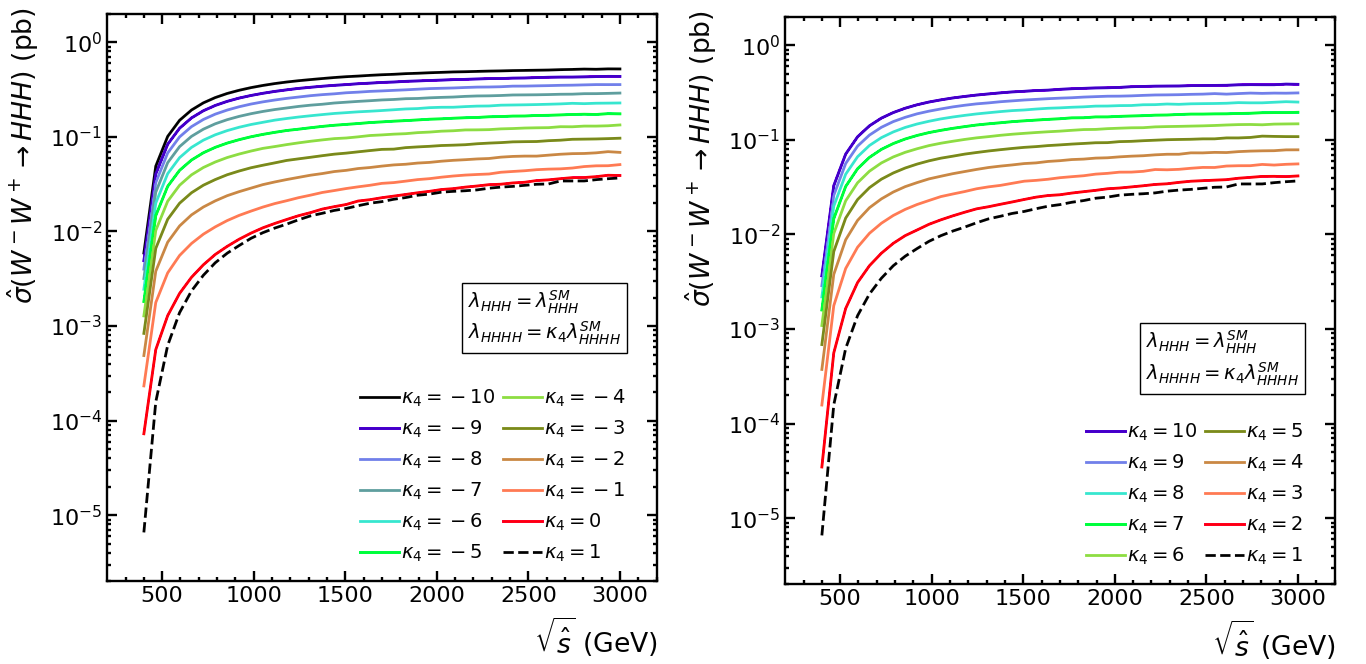}
\caption{Total cross section of the $W^-W^+\rightarrow HHH$ subprocess as a function of the CM energy $\sqrt{\hat{s}}$ for different values of the parameter $\kappa_4$, with $\kappa_3$ fixed to 1, compared to the SM prediction (dashed line). Negative (positive) values of $\kappa_4$ are shown in the left (right) panel.}
\label{HHHk4vssqrts}
\end{figure}

\begin{figure}[h!]
\centering
\includegraphics[width=0.5\textwidth]{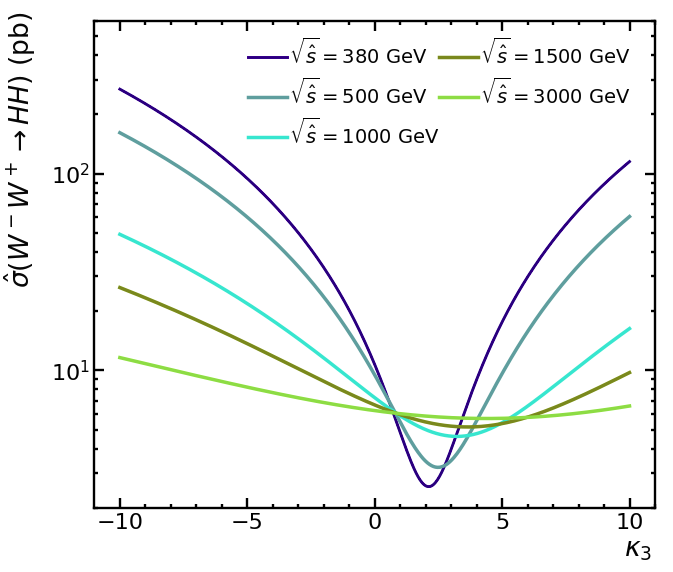}
\caption{Total cross section of the $W^-W^+\rightarrow HH$ subprocess as a function of $\kappa_3$, with $\kappa_4$ set to 1, for different values of the CM energy $\sqrt{\hat{s}}$.}
\label{HHvsk3}
\end{figure}

\begin{figure}[h!]
\centering
\includegraphics[width=0.9\textwidth]{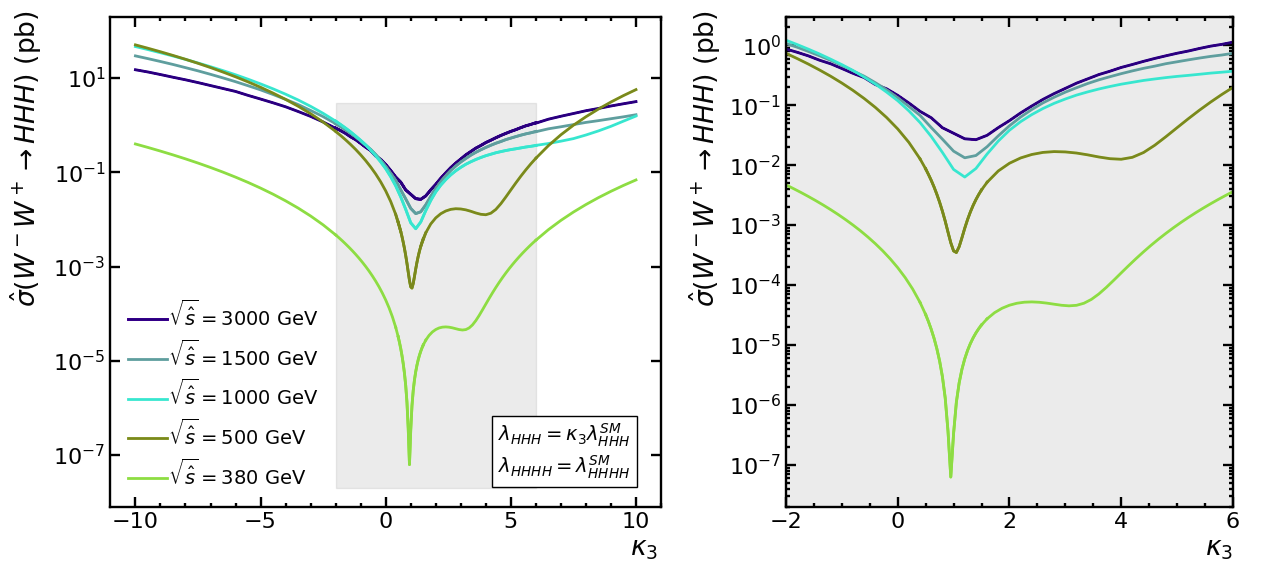}
\caption{Total cross section of the $W^-W^+\rightarrow HHH$ subprocess as a function of $\kappa_3$ (with $\kappa_4$ fixed to 1) for different values of the CM energy $\sqrt{\hat{s}}$. Right panel shows a closer view of the minima.}
\label{HHHvsk3}
\end{figure}

\begin{figure}[h!]
\centering
\includegraphics[width=0.5\textwidth]{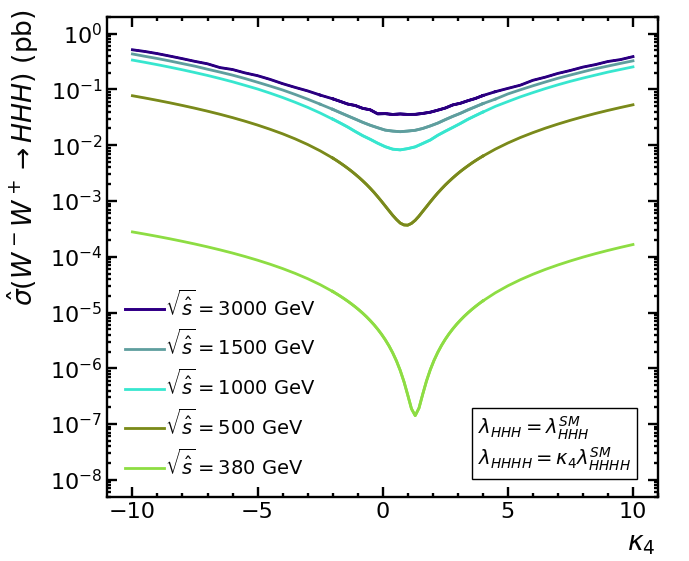}
\caption{Total cross section of the $W^-W^+\rightarrow HHH$ subprocess as a function of $\kappa_4$ (with $\kappa_3$ fixed to 1) for different values of the CM energy $\sqrt{\hat{s}}$.}
\label{HHHvsk4}
\end{figure}

The results of our calculations are collected in Figs. \ref{HHk3vssqrts} to \ref{HHHk4vssqrts}, in which we show the behaviour of the cross section with the energy for different values of $\kappa_3$ and $\kappa_4$, and Figs. \ref{HHvsk3} to \ref{HHHvsk4}, in which we display the cross section variation with $\kappa_3$ and $\kappa_4$, independently, at fixed energies. Notice again that since double Higgs production does not involve the quartic coupling (at least at tree level), the value of 
$\kappa_4$ is not relevant for the predictions in that channel.
%%%%%%%%%%%

We can extract some first conclusions by looking at these plots. Starting from \reffi{HHk3vssqrts}, that shows the double Higgs production case, we observe that modifying $\kappa_3$ in the interval [-10,10]  leads to a notable enhancement in the total cross section with respect to the SM prediction, here represented by the dashed line ($\kappa_3=1$). The maximum deviation occurs slightly above the threshold energy $2m_H$, and it is larger for negative values of $\kappa_3$. In the most extreme case, $\kappa_3=-10$, the BSM prediction can deviate up to two orders of magnitude with respect to the SM value in this close to threshold energy region. Notice also that there are no regions disallowed by unitarity in this figure. It can be checked from \citere{Arganda:2018ftn} that for $\kappa_3 \in [-10,10]$ and for the energies considered here, $\sqrt{\hat{s}}\leq 3$ TeV,  unitarity is preserved in all the predictions for double Higgs production.

\begin{figure}[h!]
\centering
\includegraphics[width=0.999\textwidth]{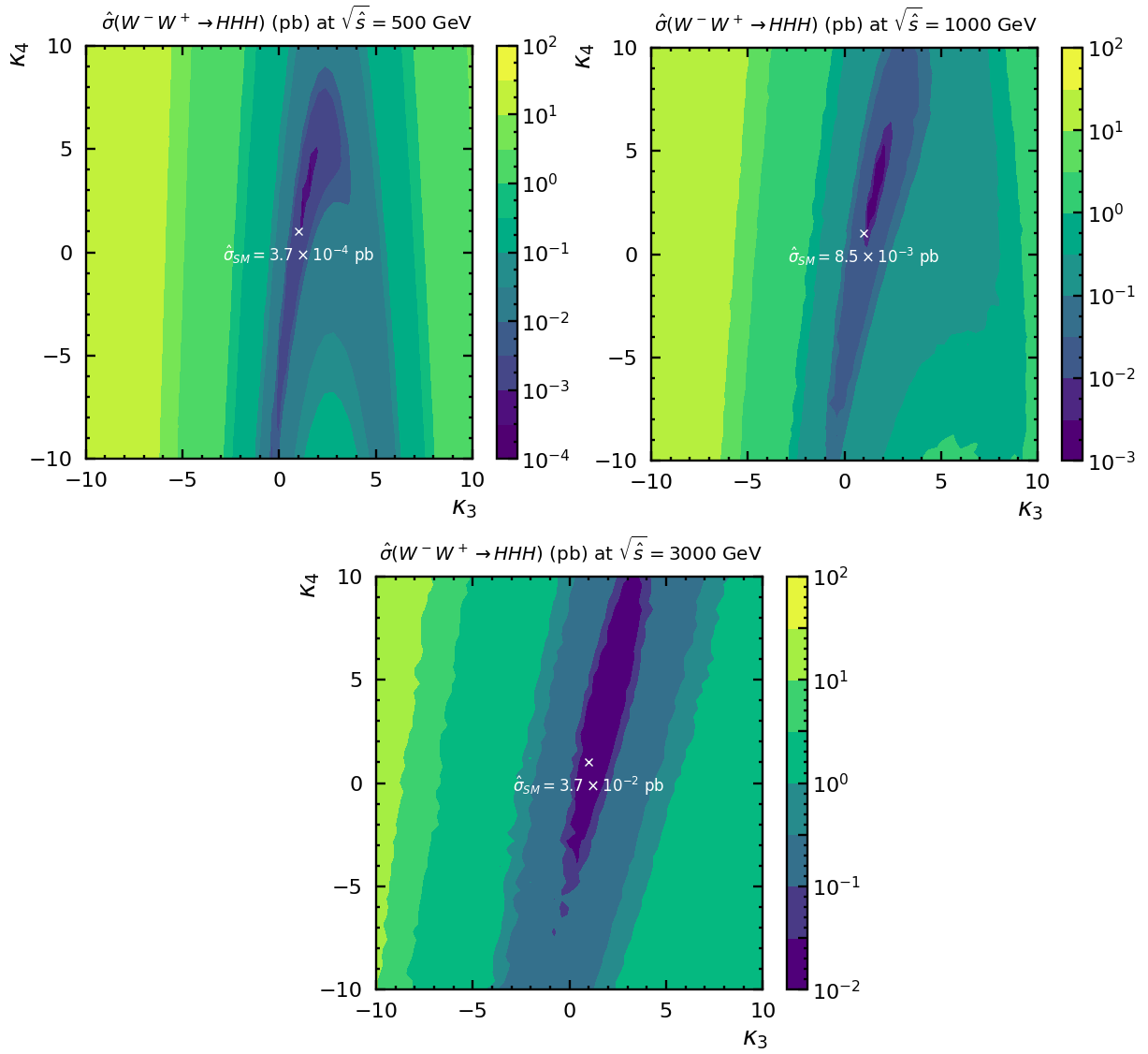}
\caption{Contour levels for the total cross section of the $W^-W^+\rightarrow HHH$ subprocess represented in the $(\kappa_3,\kappa_4)$ plane for different values of the CM energy $\sqrt{\hat{s}}$.}
\label{WWtoHHH_scan2D}
\end{figure} 

 \reffi{HHHk3vssqrts} shows the case of triple Higgs production. Comparing this figure with the previous \reffi{HHk3vssqrts}, we can see that the consequences of modifying $\kappa_3$, with $\kappa_4$ fixed to 1,  are similar in triple and double Higgs production, but now the maximum appears slightly above the new threshold energy, $3m_H$. The deviations with respect to the SM prediction are larger again for negative values of $\kappa_3$, and can reach values of up to five orders of magnitude larger than the SM in that close to threshold energy region. On the other hand, we learn from \reffi{HHHk4vssqrts} that varying the value of $\kappa_4$ within the interval [-10,10], with $\kappa_3$ fixed to 1, does not modify the shape of the cross section significantly. However, it can also increase its value by one or even two orders of magnitude with respect to the SM prediction in the most extreme cases. The dependence on the sign of $\kappa_4$ in triple Higgs production is very mild, in contrast with the strong dependence on the sign of $\kappa_3$ that has been found in both channels.

Coming back to $W^-W^+\rightarrow HH$,  \reffi{HHvsk3} shows the variation of the cross section with $\kappa_3$ at a fixed CM energy. Here we see that there is a minimum in the region $\kappa_3\in[2,5]$. The greater the energy, the larger the value of $\kappa_3$ that minimizes the cross section. Deviations from the SM can reach two orders of magnitude in the most extreme case, that is, for $\kappa_3=-10$ at the lowest energy. The sensitivity to variations of the $\kappa_3$ parameter decreases with increasing energy; note that far from the region of the minimum, the behaviour of the cross sections is inverted, being higher at lower energies. This is the reason for the appearance of the peak near the threshold in \reffi{HHk3vssqrts}, which does not appear in the SM case. 

When reproducing this same plot for $W^-W^+\rightarrow HHH$ as a function of $\kappa_3$ (see  \reffi{HHHvsk3}) we find a very different picture. First of all, in the left plot in \reffi{HHHvsk3} we see that the maxima at the extreme values of $\kappa_3=\pm 10$  are displaced in energy and they are now reached around $\sqrt{\hat{s}}=500$ GeV.  This is in contrast to the double Higgs production case, where these maxima are produced at the minimum energy of $\sqrt{\hat{s}}=380$ GeV. Second, the shape of the curve, especially at low energies, exhibits two minima , in contrast to just one minimum in $HH$. One of them is near the SM value $\kappa_3=1$, and the other one appears at a positive $\kappa_3$ and is displaced to higher values of this parameter as energy grows. The maximum deviations with respect to the SM are large in any case, varying from two orders of magnitude at $\sqrt{\hat{s}}=3000$ GeV to even five orders of magnitude at $\sqrt{\hat{s}}=380$ GeV. One of the most interesting findings is that the cross section  $\sigma(W^-W^+\rightarrow HHH)$ is significantly more sensitive to variations in the $\kappa_3$ parameter than $\sigma(W^-W^+\rightarrow HH)$, mainly at low energies. However, triple Higgs production has the disadvantage that it leads to smaller cross sections than double Higgs production, due to the obvious phase space suppression.

Finally, \reffi{HHHvsk4} shows the dependence with the $\kappa_4$ parameter, fixing 
$\kappa_3=1$, for various values of the CM energy.  We see that, in this case, the predictions of the cross sections only exhibit one minimum for each energy. Besides, the maximum deviations from the SM vary from one order of magnitude at $\sqrt{\hat{s}}=3000$ GeV up to three orders of magnitude at $\sqrt{\hat{s}}=380$ GeV.

Since triple Higgs production depends on both parameters, $\kappa_3$ and $\kappa_4$,  it is also interesting to check what happens if we vary the two of them at the same time. In  \reffi{WWtoHHH_scan2D} we represent the $W^-W^+\rightarrow HHH$ cross section at three fixed energies, $\sqrt{\hat{s}}=$ 500, 1000 and 3000 GeV,  in the $(\kappa_3,\kappa_4)$ plane. The additional information that we can extract looking at these plots is that variations in the cross section are strongly dominated by deviations along the $\kappa_3$ direction, and the higher values are reached when $\kappa_3$ approaches $-10$. The dependence with $\kappa_4$ is softer, and the maximum cross section can be reached either at negative or positive values of $\kappa_4$. For example, in the negative $\kappa_3$ region, the cross section tends to be higher around $\kappa_4=10$, while in the positive $\kappa_3$ region the trend can be this same or the opposite one, depending on the energy. Note that at high energies the combined effect of modifying both parameters at the same time can lead to a notable increase of the cross section, especially if they have opposite signs. For instance, at $\sqrt{\hat{s}}=$ 3000 GeV and for $(\kappa_3,\kappa_4)$ near the corner $(-10,10)$, the cross section exhibits large values of ${\cal O}(100)$ pb, to be compared with the SM prediction of $3.7 \times 10^{-2}$ pb.  We also notice that the lowest values (darker region) are approximately arranged along a line which does not coincide with the $\kappa_3=\kappa_4$ direction. This means that modifying the Higgs self-couplings without altering their ratio, that is, $\lambda_{HHH}/\lambda_{HHHH} = \lambda_{HHH}^\text{SM}/\lambda_{HHHH}^\text{SM}=1$, can also produce an enhancement in the cross section. The size of this darker region is related to the depth of the minimum, and is smaller for lower energies. Note that the SM prediction is contained in this region, which is why we do not expect to measure this triple Higgs production process if the self-couplings are close to their SM values. In any case, what we are studying here is the subprocess, which cannot be seen in a real experiment, so to complete our analysis, in the next section, we will study the full $e^+e^-$ process to understand how sensitive it is to deviations in these parameters.

Regarding unitarity, we find that the $W^-W^+\rightarrow HHH$ subprocess does not violate unitarity for the energies and parameter values considered here, i.e. $\kappa_{3,4}\in[-10,10]$ and $\sqrt{\hat{s}}<3$ TeV. We also confirm the good unitarity behaviour in double Higgs production with respect to $\kappa_3$, for these same settings. 
It is worth mentioning that for triple Higgs production,  the unitarity check has been performed with a different criterion than for double Higgs production. This difference is because
 to study the unitarity bounds for a $2\rightarrow n$ process, one cannot employ the usual partial waves expansion, since the amplitude depends on a higher number of variables, other than $s$, $t$ and $u$. Following for instance \citere{Belyaev:2018fky}, one way of checking unitarity in a $2\rightarrow n$ process is to insert a complete set of intermediate states into its left-hand side, separating the elastic and the inelastic parts:
\begin{align}
    2\text{Im}\left[ \mathcal{M}_\text{el}(2\rightarrow2)\right] = \int_{\Pi_2} |\mathcal{M}_\text{el}(2\rightarrow2)|^2 + \sum_n \int_{\Pi_n} |\mathcal{M}_\text{inel}(2\rightarrow n)|^2.
\end{align}
Here $\Pi_n$ refers to the $n$-body phase space. In this case, the following bound for $\sigma_\text{inel}(2\rightarrow n)$ is obtained after introducing the partial wave expansion for $\mathcal{M}_\text{el}(2\rightarrow 2)$:
\begin{align}
    \sigma_\text{inel}(2\rightarrow n) \leq \frac{4\pi}{s}.
\end{align}
This is the constraint that we have imposed to check the unitarity of our predictions for $\sigma(W^-W^+\rightarrow HHH)$. According to this criterion, all our predictions for BSM physics coming from deviations in $\kappa_3$ and $\kappa_4$ are fully unitary for all the energies considered in this work. Concretely,  in the worst-case scenario, i.e. for the largest energies considered here ($\sqrt{\hat{s}}=3000$ GeV), preservation of unitarity requires that:
\begin{align}
    \hat{\sigma}(W^-W^+\rightarrow HHH) \lesssim 540 \text{ pb},
\end{align}
which is not reached even in the most extreme cases in \reffi{WWtoHHH_scan2D}. 
%
%This result could be expected from the beginning, taking into account that the Higgs boson self-couplings are not responsible for very strong cancellations in these WWS amplitudes, as it is the case of the Higgs couplings to vector bosons.% 
Thus, in contrast 
to $a$ and $b$, which suffer of strong restrictions from unitarity, the parameters $\kappa_3$ and $\kappa_4$ do not.
% --------------------------------------------------------------------------
%%%%%%%%%%
\section{Deviations from BSM Higgs couplings in $e^+e^-$}
\label{TestProcess}
In this section we present the results for the effects of the BSM Higgs couplings within the EChL parametrized by $a$, $b$, 
$\kappa_3$ and $\kappa_4$ in double and triple Higgs production at the real scattering processes $e^+e^-\rightarrow HH\nu_e\bar{\nu}_e$ and $e^+e^-\rightarrow HHH\nu_e\bar{\nu}_e$, respectively. We will explore these effects at the planned $e^+e^-$ colliders, and compare the BSM behaviour with the SM predictions. Again, to simplify this study, we take $\kappa_{3,4}=1$ when exploring the sensitivity to $a$ and $b$, and viceversa. As in the SM case, the full computation of the  BSM rates is done with MG5 (we neglect the Yukawa couplings to the electrons) and includes 8 contributing diagrams in $e^+e^-\rightarrow HH\nu_e\bar{\nu}_e$, 4 mediated by WWS and 4 by $Z$, and 50 contributing diagrams  in $e^+e^-\rightarrow HHH\nu_e\bar{\nu}_e$, 25 mediated by WWS and 25 by $Z$. We do not draw all of them here, for shortness, but the ones mediated by WWS are easily extracted from those of the corresponding subprocess, which have been collected in the appendices. 
\subsection{Deviations from  $a$ and $b$ in $e^+e^-\rightarrow HH\nu_e\bar{\nu}_e$}
The predictions for the cross section $\sigma(e^+e^-\rightarrow HH\nu_e\bar{\nu_e})$ within the EChL  for different values of the parameters $a$ and $b$ and of the $e^+e^-$ CM energies are shown in \refta{ratiosechl}. The values of $a$ and $b$ are chosen in such a way that $\Delta a\,,\Delta b\leq 1$, with $\Delta a(b)\equiv\vert a(b)-1\vert$.  The corresponding predictions for the SM case (i.e. for $a=b=1$) are also included for comparison. We present the results from both methods, the full computation with MG5 and the approximate rates given by the EWA. 
%%%%%%%

\begin{table}[H]
\centering
\begin{tabular}{|c|c|c|c|c|}
\cline{2-5}
\multicolumn{1}{c|}{}& $\mathbf{\sqrt{s}}\enspace$(TeV)&\textbf{0.5}&\textbf{1}&\textbf{3}\\
\hline
\multirow{3}[5]*{SM $a,b=1$} 
&MG5&$1.3 \times10^{-5}$&$8.0 \times 10^{-5}$&$8.0\times 10^{-4}$
\bigstrut\\\cline{2-5}
&EWA&$7.7\times 10^{-6}$&$1.2\times 10^{-4}$&$8.5\times 10^{-4}$
\bigstrut\\\cline{2-5}
&$R_\text{WWS}$&0.3&0.9&0.997
\bigstrut\\
\hline
\hline
\multirow{3}[5]*{$a=1.5$, $b=1$} 
&MG5&$1.2 \times10^{-4}$&$2.4 \times 10^{-3}$&$3.6\times 10^{-2}$
\bigstrut\\\cline{2-5}
&EWA&$1.4\times 10^{-4}$&$2.6\times 10^{-3}$&$3.2\times 10^{-2}$
\bigstrut\\\cline{2-5}
&$R_\text{WWS}$&0.83&0.9921&0.99947
\bigstrut\\
\hline
\hline
\multirow{3}[5]*{$a=0.5$, $b=1$} 
&MG5&$1.5 \times10^{-5}$&$3.0 \times 10^{-4}$&$7.1\times 10^{-3}$
\bigstrut\\\cline{2-5}
&EWA&$7.9\times 10^{-6}$&$2.0\times 10^{-4}$&$5.8\times 10^{-3}$
\bigstrut\\\cline{2-5}
&$R_\text{WWS}$&0.65&0.9975&0.99985
\bigstrut\\
\hline
\hline
\multirow{3}[5]*{$a=1$, $b=1.5$} 
&MG5&$1.7 \times10^{-5}$&$9.0 \times 10^{-5}$&$2.4\times 10^{-3}$
\bigstrut\\\cline{2-5}
&EWA&$2.7\times 10^{-6}$&$6.6\times 10^{-5}$&$2.0\times 10^{-3}$
\bigstrut\\\cline{2-5}
&$R_\text{WWS}$&0.06&0.83&0.9975
\bigstrut\\
\hline
\hline
\multirow{3}[5]*{$a=1$, $b=0.5$} 
&MG5&$1.7 \times10^{-5}$&$3.3 \times 10^{-4}$&$5.5\times 10^{-3}$
\bigstrut\\\cline{2-5}
&EWA&$2.0\times 10^{-5}$&$3.8\times 10^{-4}$&$5.0\times 10^{-3}$
\bigstrut\\\cline{2-5}
&$R_\text{WWS}$&0.66&0.988&0.99944
\bigstrut\\
\hline
\hline
\multirow{3}[5]*{$a=1$, $b=0.9$} 
&MG5&$1.3 \times10^{-5}$&$1.0 \times 10^{-4}$&$1.2\times 10^{-3}$
\bigstrut\\\cline{2-5}
&EWA&$9.5\times 10^{-6}$&$1.6\times 10^{-4}$&$1.2\times 10^{-3}$
\bigstrut\\\cline{2-5}
&$R_\text{WWS}$&0.292&0.932&0.998
\bigstrut\\
\hline
\hline
\multirow{3}[5]*{$a=1$, $b=1.1$} 
&MG5&$1.3 \times10^{-5}$&$2.5 \times 10^{-4}$&$6\times 10^{-4}$
\bigstrut\\\cline{2-5}
&EWA&$6.1\times 10^{-6}$&$9.1\times 10^{-5}$&$6.6\times 10^{-4}$
\bigstrut\\\cline{2-5}
&$R_\text{WWS}$&0.154&0.984&0.996
\bigstrut\\
\hline
\hline
\multirow{3}[5]*{$a=0.9$, $b=1$} 
&MG5&$9.3 \times10^{-6}$&$2.9 \times 10^{-5}$&$4.7\times 10^{-4}$
\bigstrut\\\cline{2-5}
&EWA&$2.5\times 10^{-6}$&$4.1\times 10^{-5}$&$4.9\times 10^{-4}$
\bigstrut\\\cline{2-5}
&$R_\text{WWS}$&0.04&0.741&0.995
\bigstrut\\
\hline
\hline
\multirow{3}[5]*{$a=1.1$, $b=1$} 
&MG5&$2.0 \times10^{-5}$&$2.1 \times 10^{-4}$&$2.6\times 10^{-3}$
\bigstrut\\\cline{2-5}
&EWA&$1.8\times 10^{-5}$&$2.9\times 10^{-4}$&$2.6\times 10^{-3}$
\bigstrut\\\cline{2-5}
&$R_\text{WWS}$&0.4&0.956&0.999
\bigstrut\\
\hline
\end{tabular}
\caption{Predictions of the cross sections $\sigma(e^+e^-\rightarrow HH\nu_e\bar{\nu_e})$  (in pb) within the EChL for different values of the parameters $a$ and $b$ and $e^+e^-$ CM energies. The SM case ($a=b=1$) is also included for comparison. The predictions from both MG5 and the EWA are displayed, as well as the ratio $R_\text{WWS}$, defined  in \refeq{rvbs}, which quantifies the relevance of the WWS within the whole $e^+e^-$ process.}
\label{ratiosechl}
\end{table}

%%%%
From the results in this table we extract several conclusions. First, we see from the predictions with MG5 that the distortions introduced by $a$ and $b$ with respect to the SM can be very large, leading to large enhacements  in the cross sections. If we set one of the parameters to 1 and vary the other one in a certain amount, the enhancement will be larger when the variation is applied on $a$ rather than on $b$ (notice the difference between the cases $a=1, b=1.5$ and $a=1.5,b=1$). This could be expected since the $a$ parameter enters quadratically in the amplitude whereas $b$ enters linearly. We also see in this table that the size of the enhancements obtained for a given $\Delta a$ and $\Delta b$  depends on the energy and grows in general with it. The maximum departures from the SM are found for the largest energies and $\Delta$'s considered here, and can lead to ratios of BSM cross section  over SM cross section of up to ${\cal O}(100)$. Second, from this table we also learn that the EWA  gives quite good predictions for high energies at and above 1 TeV, not only for the SM rates as we have already seen in \refse{EWAvMG}, but also for the BSM rates. Indeed, we see that the accuracy of the EWA is even better for the BSM predictions than for the SM ones, and this occurs because the dominance of the WWS subprocess is more pronounced in the BSM scenarios than in the SM case. This is illustrated by the ratio $R_\text{WWS}$, defined in \refeq{rvbs}, which clearly approaches one at the highest energies, indicating that WWS largely dominates the full BSM cross sections. 
%%%
\begin{figure}[h!]
\centering
\includegraphics[scale=0.35]{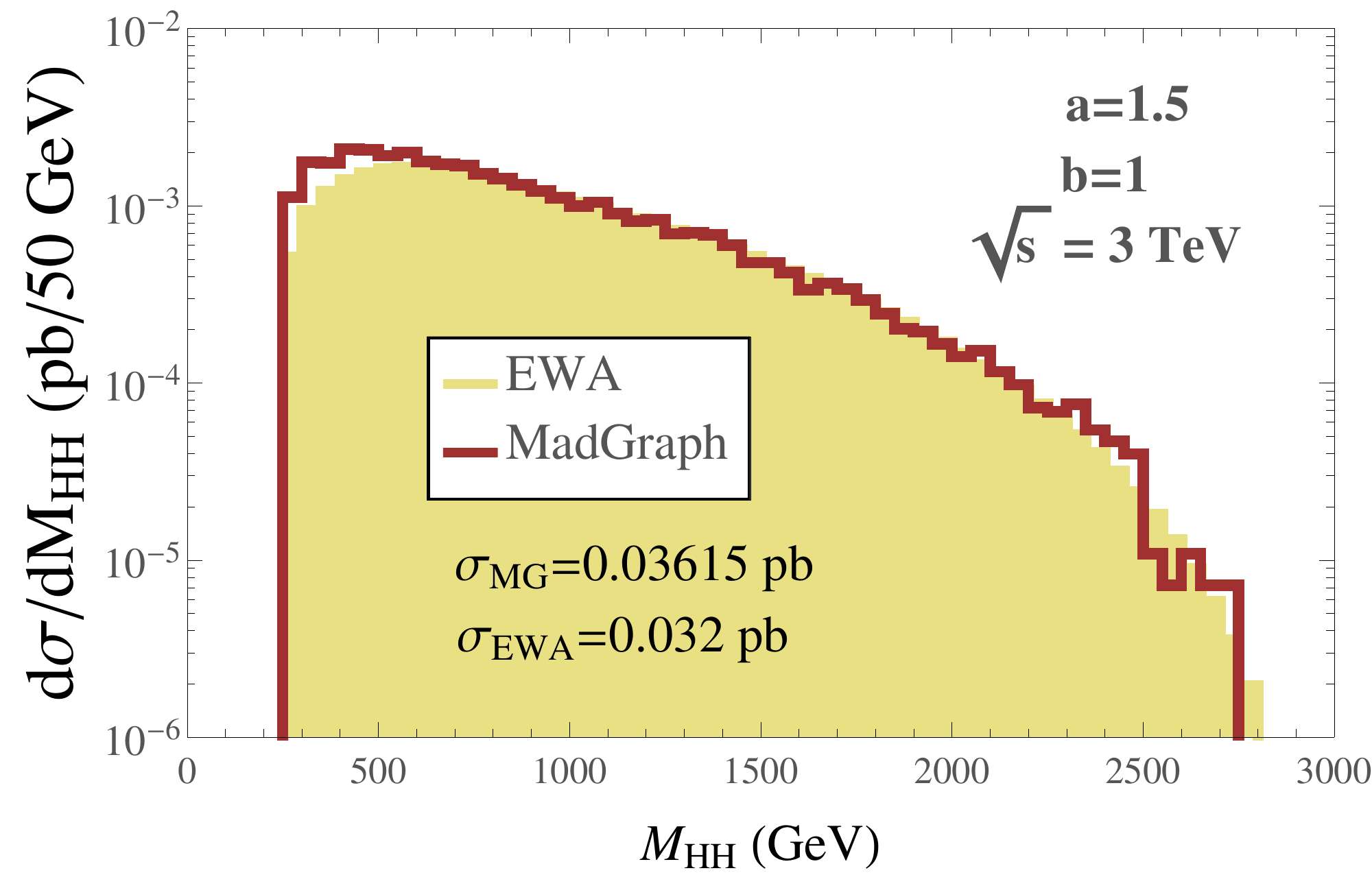}
\includegraphics[scale=0.35]{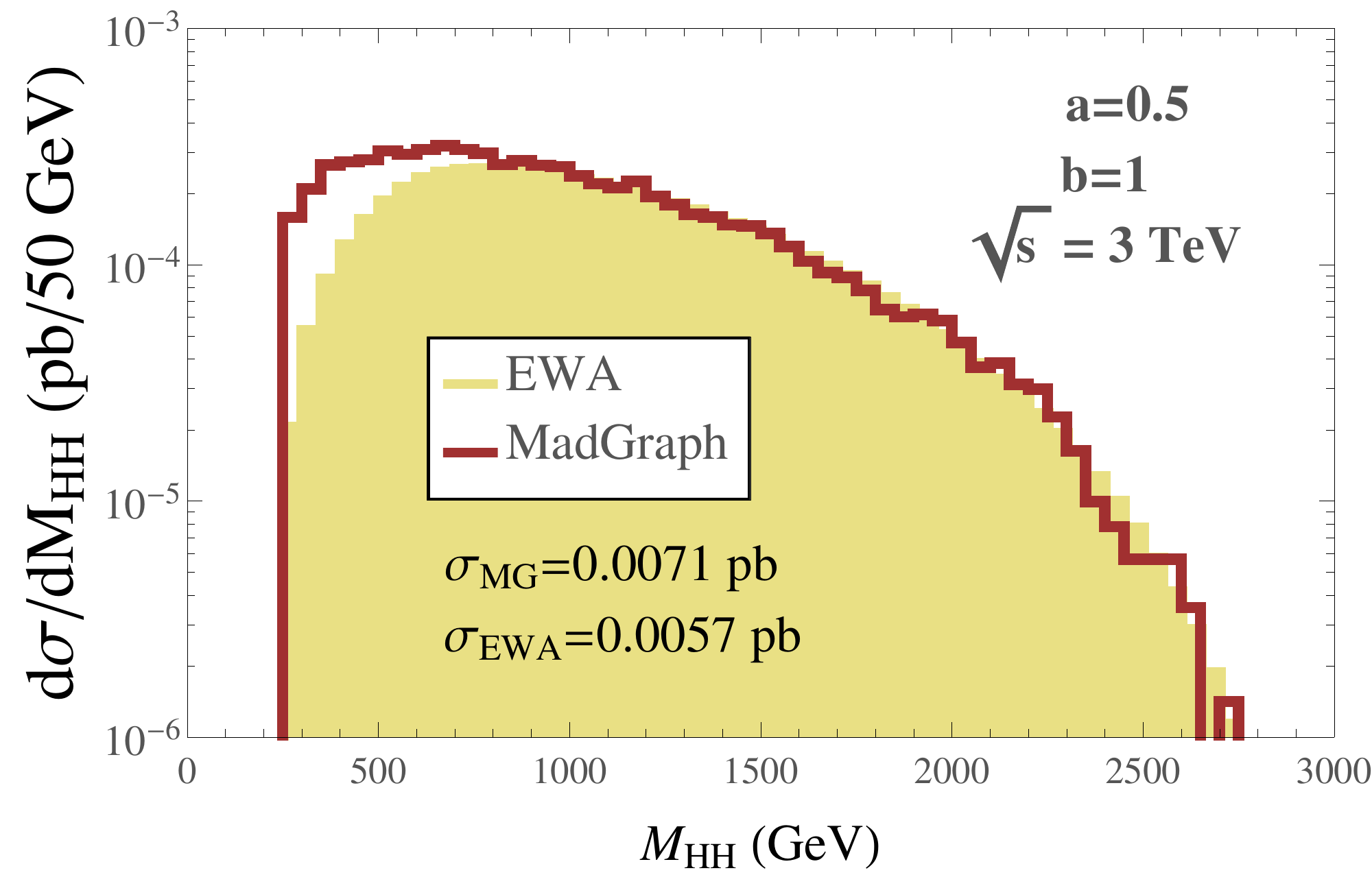}\\
\includegraphics[scale=0.35]{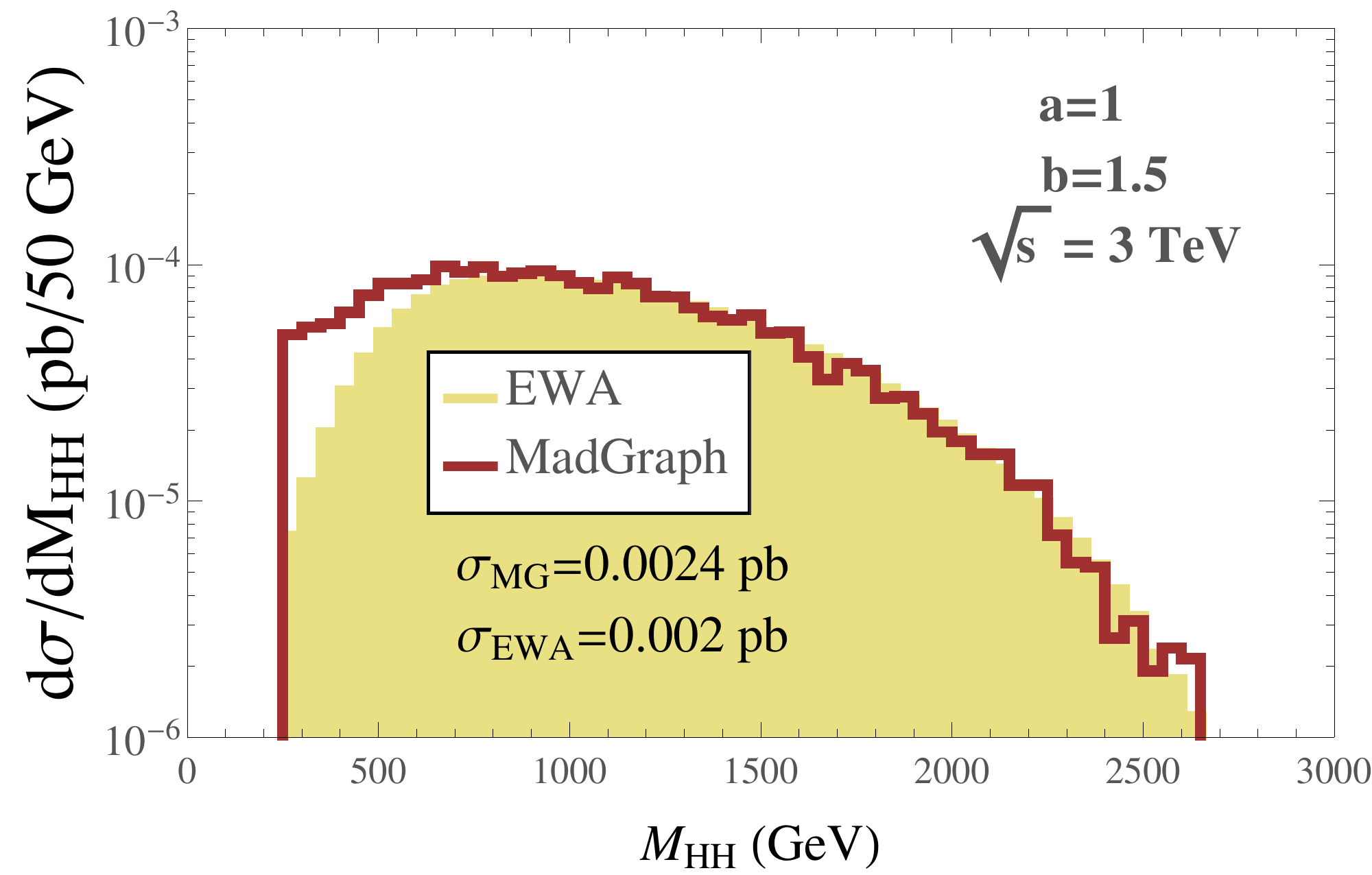}
\includegraphics[scale=0.35]{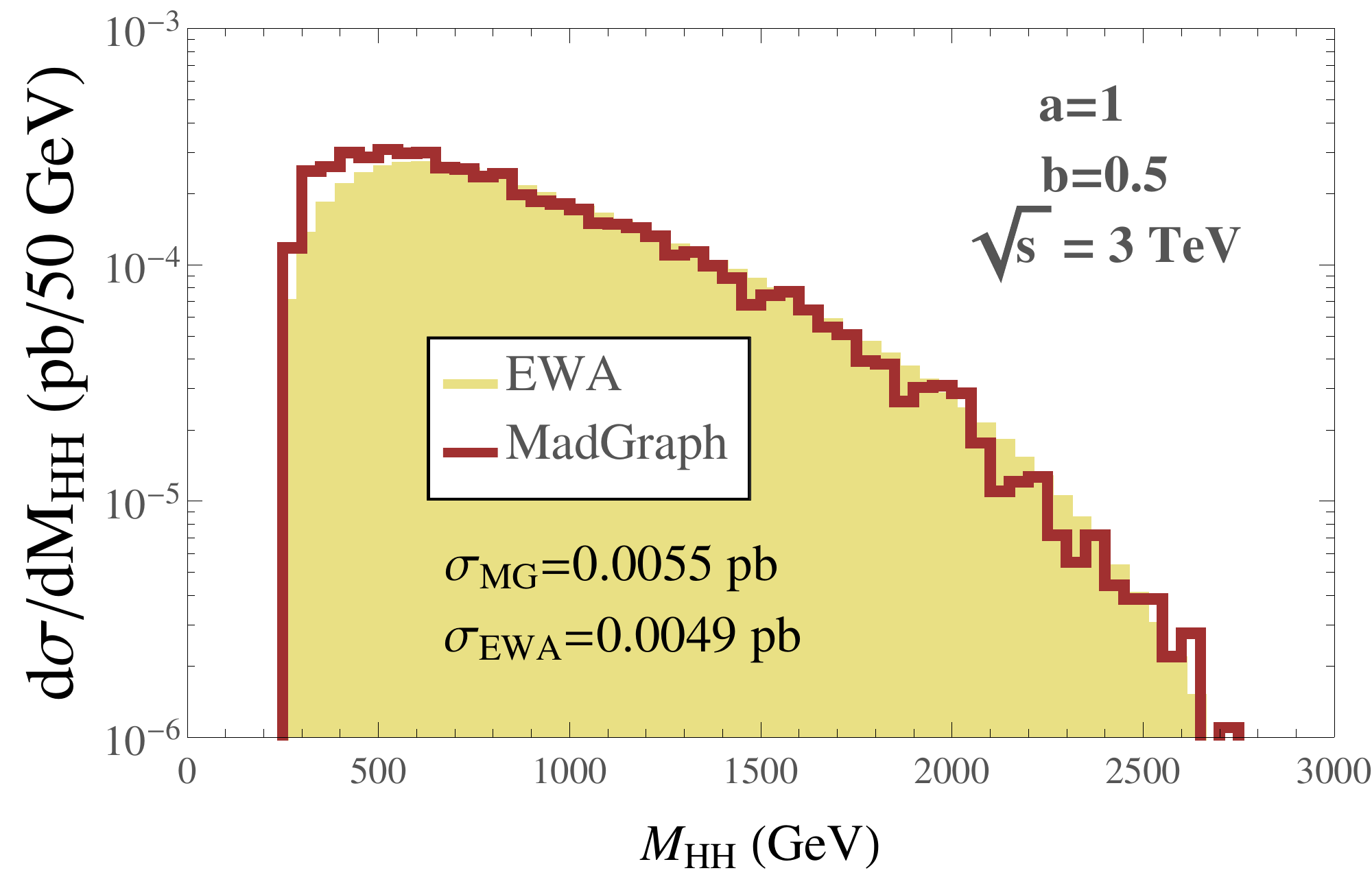}
\caption{Predictions of the cross section distributions for $e^+e^-\rightarrow HH\nu_e\bar{\nu_e}$  with respect to the Higgs pair invariant mass, $M_{HH}$, within the EChL, for different values of the parameters $a$ and $b$. The collider energy is set to 3 TeV. The results from both MG5 and the EWA computations are shown for comparison. }
\label{distewa1}
\end{figure} 
%%%%

This WWS dominance can also be seen in the predictions of the cross sections distributions with the invariant mass $M_{HH}$, as it is shown in \reffi{distewa1}, where the collider energy has been set to 3 TeV. We also see in this figure that the prediction from the EWA is very close to the prediction from MG5 for $M_{HH}$ above 1 TeV.  All in all, we learn from \refta{ratiosechl} and \reffi{distewa1} that the main features found in the predicted cross sections for $e^+e^-\rightarrow HH\nu_e\bar{\nu_e}$ with respect to variations in the $a$ and $b$ parameters follow a similar pattern as those found previously at the $WW \to HH$ subprocess level in \refse{abinWWS}.  The reason for this similarity is again the dominance of the WWS subprocess, which is even more pronounced in the BSM scenarios than in the SM case.  

\begin{figure}[t!]
\centering
\includegraphics[width=0.34\textwidth]{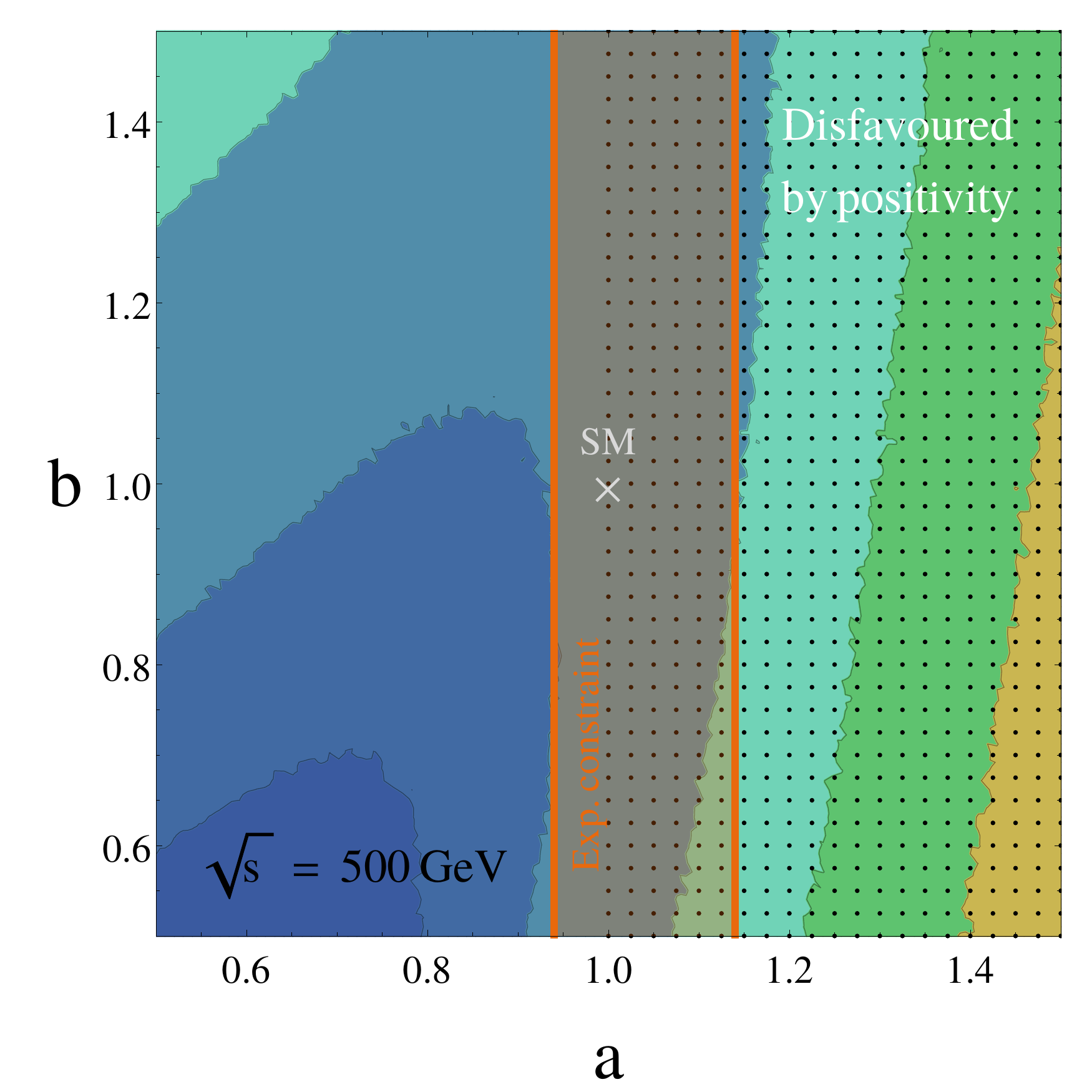}
\hspace{-0.5cm}
\includegraphics[width=0.34\textwidth]{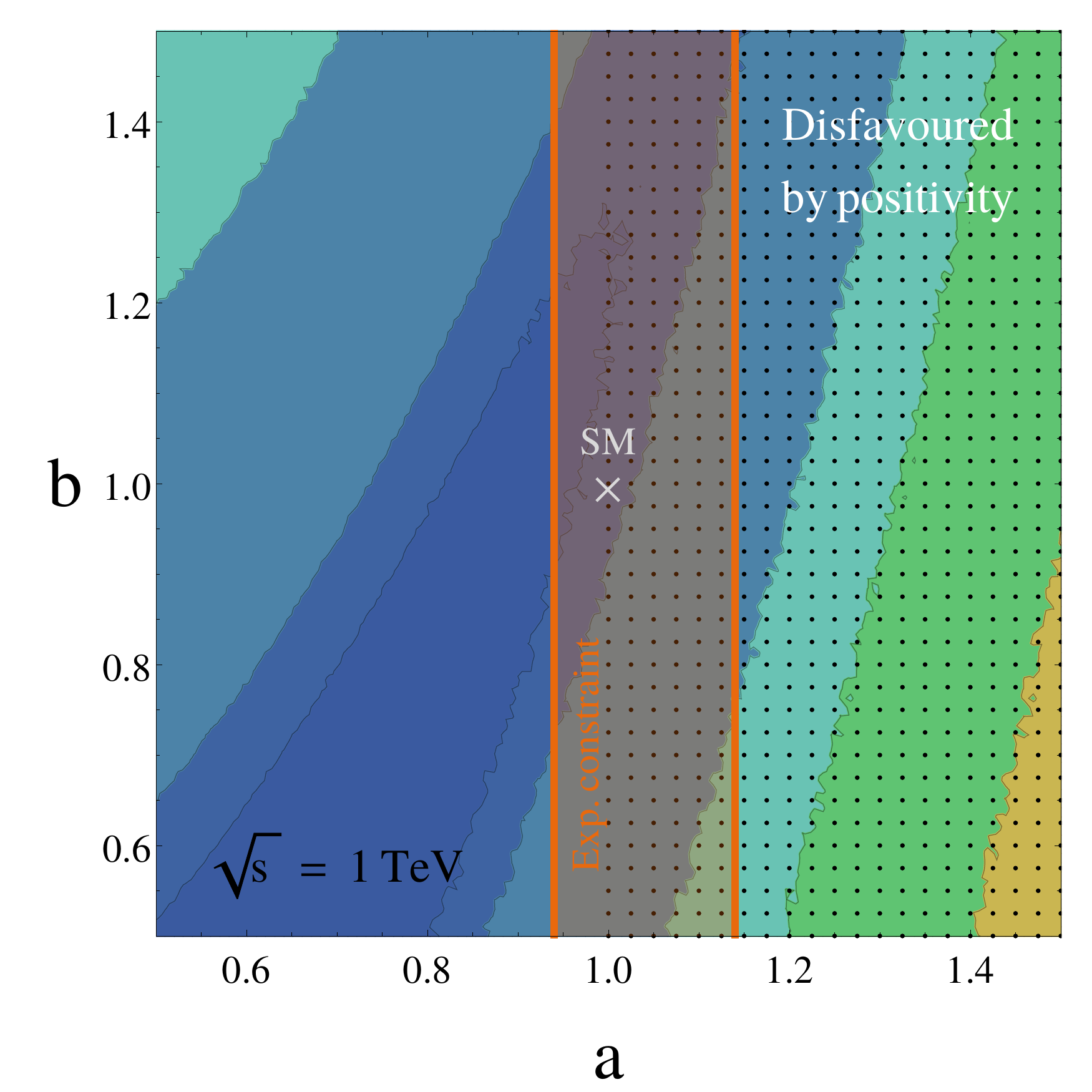}
\hspace{-0.5cm}
\includegraphics[width=0.34\textwidth]{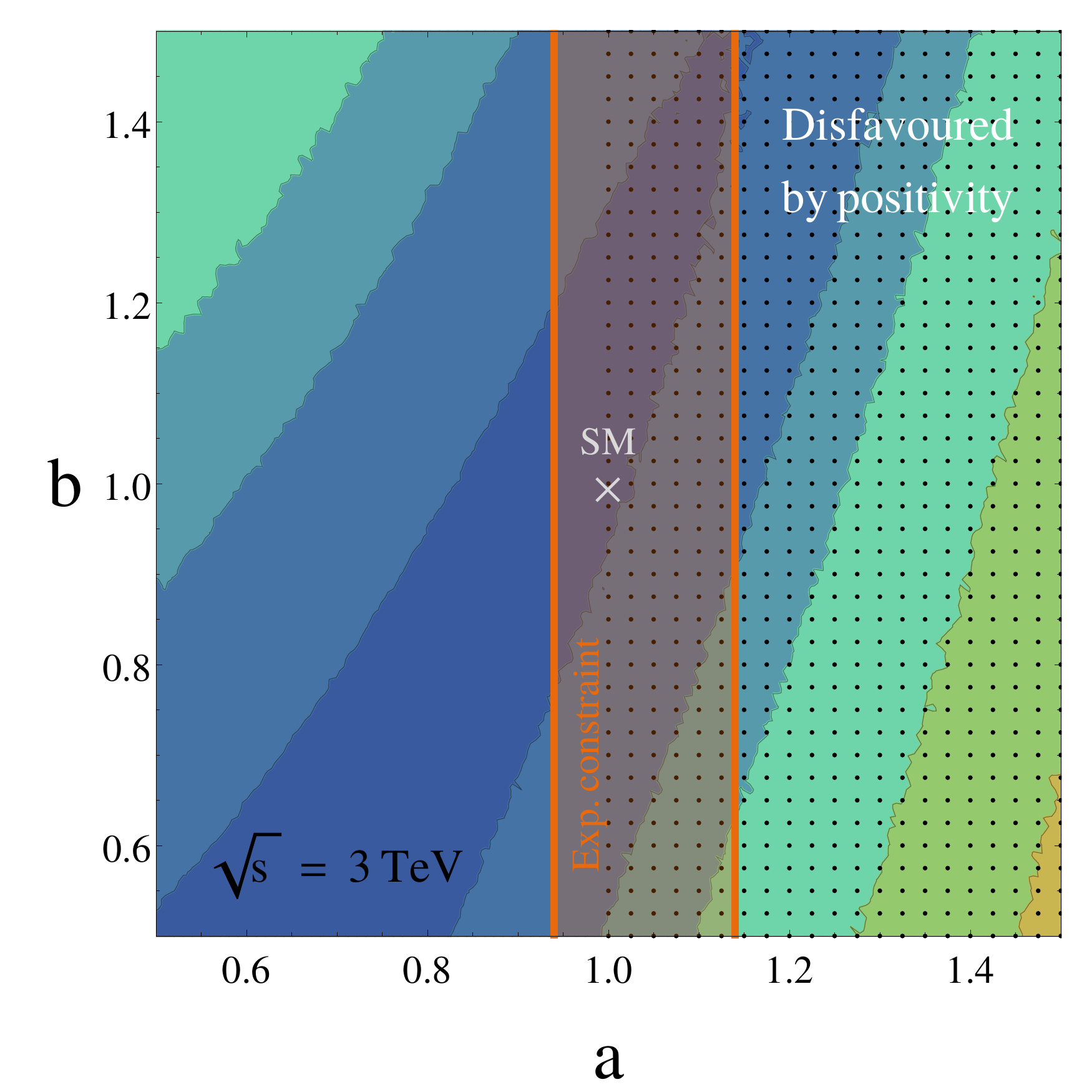}
\includegraphics[width=0.33\textwidth]{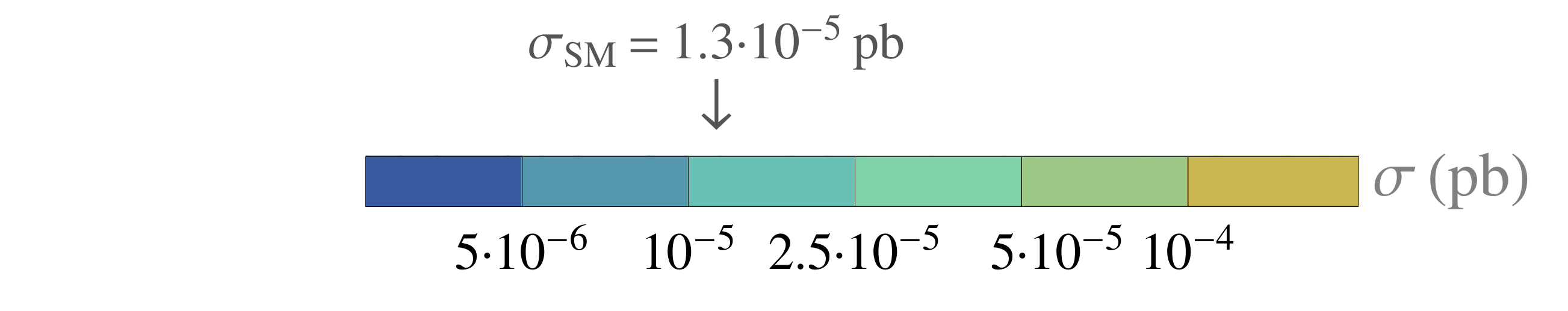}
\hspace{-0.4cm}
\includegraphics[width=0.33\textwidth]{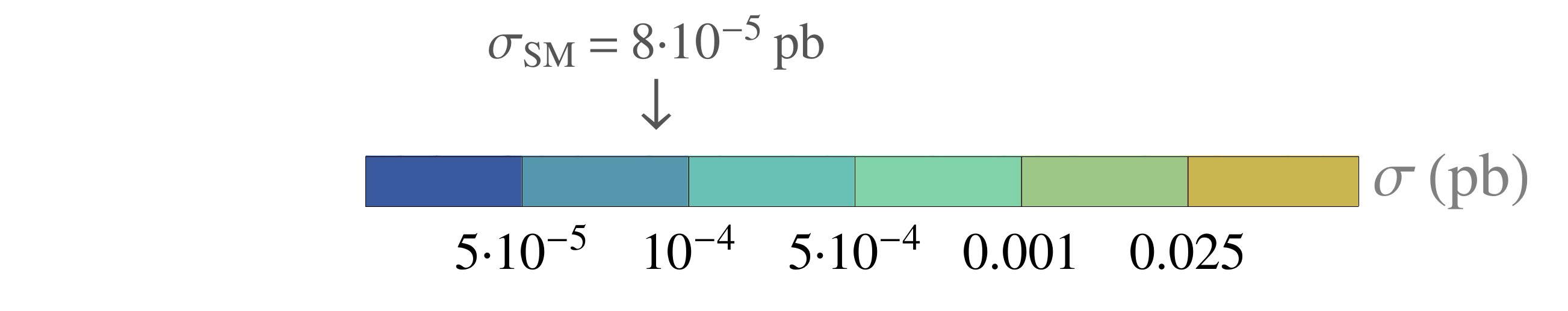}
\hspace{-0.4cm}
\includegraphics[width=0.33\textwidth]{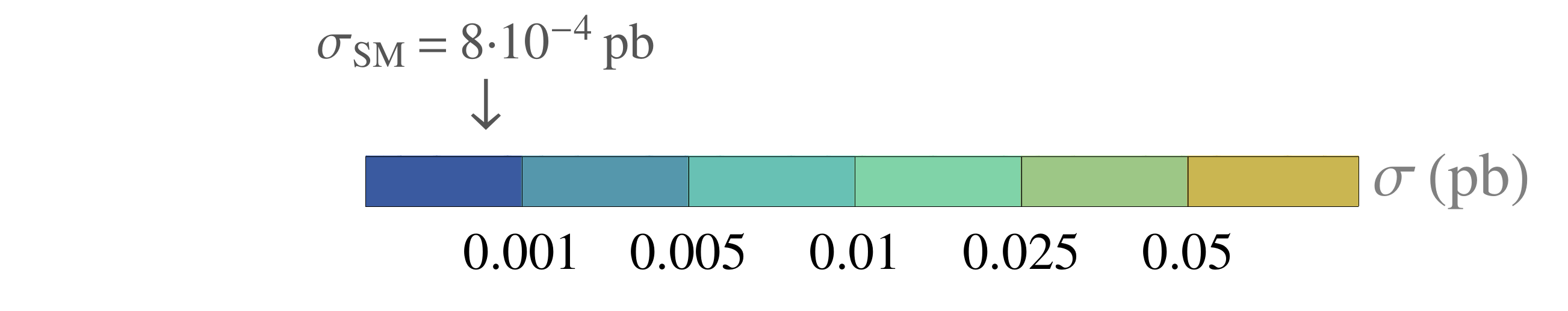}
\caption{EChL predictions for contour lines of $\sigma(e^+e^-\rightarrow HH\nu_e\bar{\nu_e})$ (in pb) in the $(a,b)$ parameter space, at a CM energy of 500 GeV (left panel), 1 TeV (middle panel) and 3 TeV (right panel). The dotted region is disfavoured by positivity, while the orange one shows the experimentally allowed region for $a$. The white cross represents the SM prediction ($a=b=1$).}
\label{contourXS}
\end{figure}

Finally, we present the results by scanning both parameters $a$ and $b$ together in the $(a,b)$ plane and for various collider energies. 
%%%%
\reffi{contourXS} shows the cross section contour lines, obtained with MG5, for $e^+e^-$ collider energies of 500 GeV, 1 TeV and 3 TeV. 
The range for the $b$ parameter in these plots, $b \in [0.5,1.5]$, is chosen such that unitarity is guaranteed for the three considered centre-of-mass energies. Although the range considered for $a$ is larger than the one it is actually constrained to, we wanted to display the same ranges for both parameters in order to obtain a global view. Notice that in these plots we have also marked the region with $a>1$, where the so-called positivity constraint applies. According to Refs. \cite{Delgado:2015kxa} and \cite{Zhang:2018shp}, some values in this region could yield a potential problem due to violation of causality in this EFT.  We include in these plots a dotted region, that is theoretically disfavoured by positivity, and and orange band that displays the experimentally allowed region for $a$, within a 95$\%$ confidence level. We have checked that within this experimentally allowed region for $a$ unitarity is fully preserved.

The following plots finally provide solid conclusions on the sensitivity of the $e^+e^-\rightarrow HH\nu_e\bar{\nu_e}$ process to variations of the $a$ and $b$ EChL parameters. As both $a$ and $b$ are modified simultaneously and in the same range, it is possible to compare their role on the behaviour of the cross section.  Notice that the plots are not symmetrical with respect to the $(a,b)=(1,1)$ point, which means that increasing any of the parameters with respect to 1 in a certain amount is not equivalent to diminishing it in the same amount. This happens for both parameters: the plots do not exhibit any symmetry with respect to the $a=1$ or $b=1$ axes. This means that the cross section is sensitive to the sign of $a-1$ and $b-1$. This was already observed in \reffi{distewa1}, and is now confirmed.  Also, the plots are not symmetrical with respect to the $a=b$ line: equal variations in $a$ and $b$ are not equivalent. Clearly $a$ is the dominant parameter: cross sections grow faster when keeping $b$ constant and varying $a$ than viceversa, which could be expected due to the quadratic dependence on $a$ in the amplitude. 

The most relevant conclusion that we learn from \reffi{contourXS} is that the BSM rates are considerably larger than the SM ones,  especially for the higher energy colliders, even if we limit ourselves to the region of the $(a,b)$ parameter space allowed by unitarity, positivity and present experimental constraints. Thus, it is interesting to perform a more detailed analysis in a collider framework, taking into account the Higgs bosons decays. We will carry out such an analysis in \refse{b-jet}.
 
\subsection{Deviations from  $\kappa_3$ and $\kappa_4$  in $e^+e^-\rightarrow HH(H)\nu_e\bar{\nu}_e$}
\begin{figure}[h!]
\centering
\includegraphics[width=0.999\textwidth]{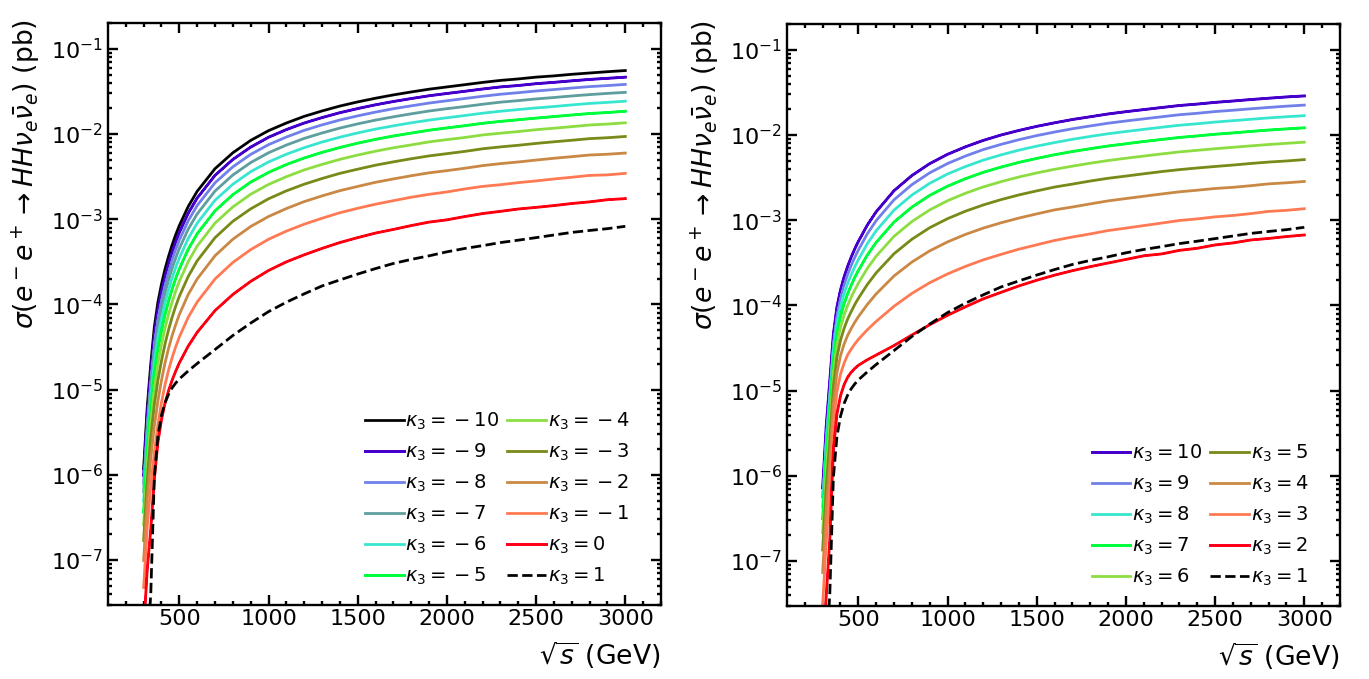}
\caption{Total cross section of $e^+e^-\rightarrow HH\nu_e\bar{\nu}_e$ as a function of the CM energy $\sqrt{s}$ for different values of the parameter $\kappa_3$, with $\kappa_4$ set to 1, compared to the SM prediction (dashed line). Negative (positive) values of $\kappa_3$ are shown in the left (right) panel.}
\label{eeHHk3vssqrts}
\end{figure}

\begin{figure}[h!]
\centering
\includegraphics[width=0.999\textwidth]{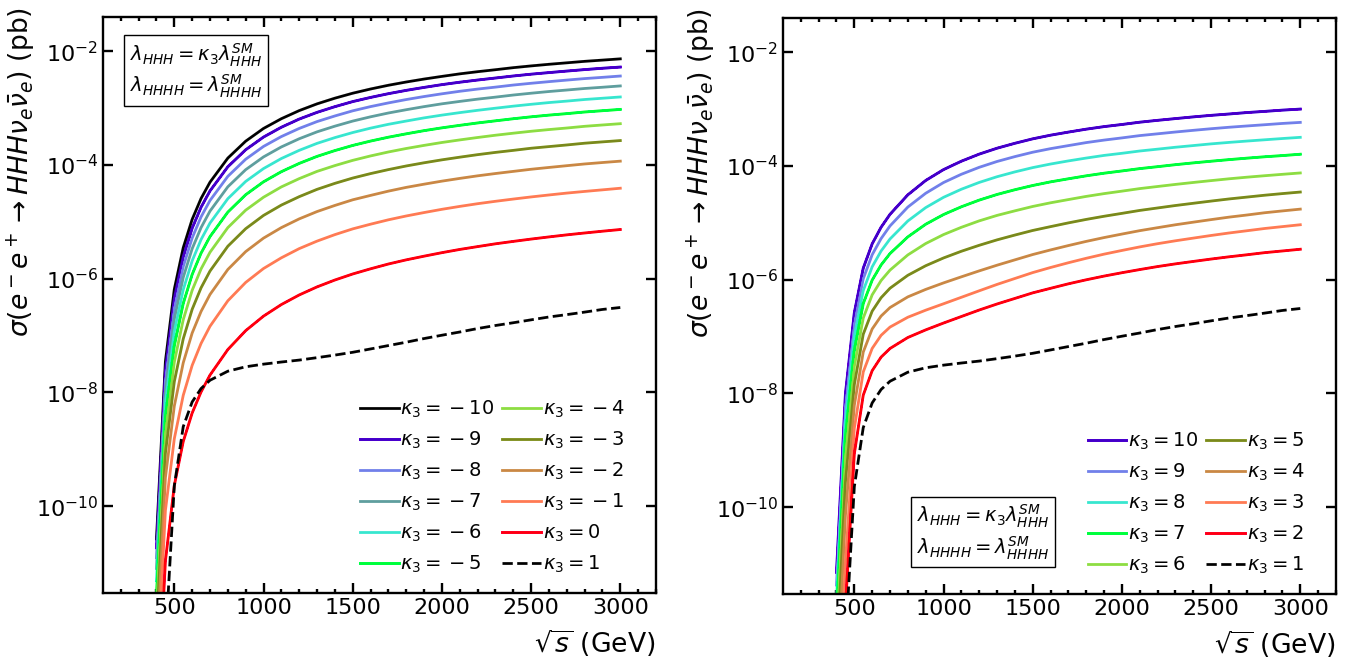}
\caption{Total cross section of $e^+e^-\rightarrow HHH\nu_e\bar{\nu}_e$ as a function of the CM energy $\sqrt{s}$ for different values of the parameter $\kappa_3$, with $\kappa_4$ fixed to 1, compared to the SM prediction (dashed line). Negative (positive) values of $\kappa_3$ are shown in the left (right) panel.}
\label{eeHHHk3vssqrts}
\end{figure}

\begin{figure}[h!]
\centering
\includegraphics[width=0.999\textwidth]{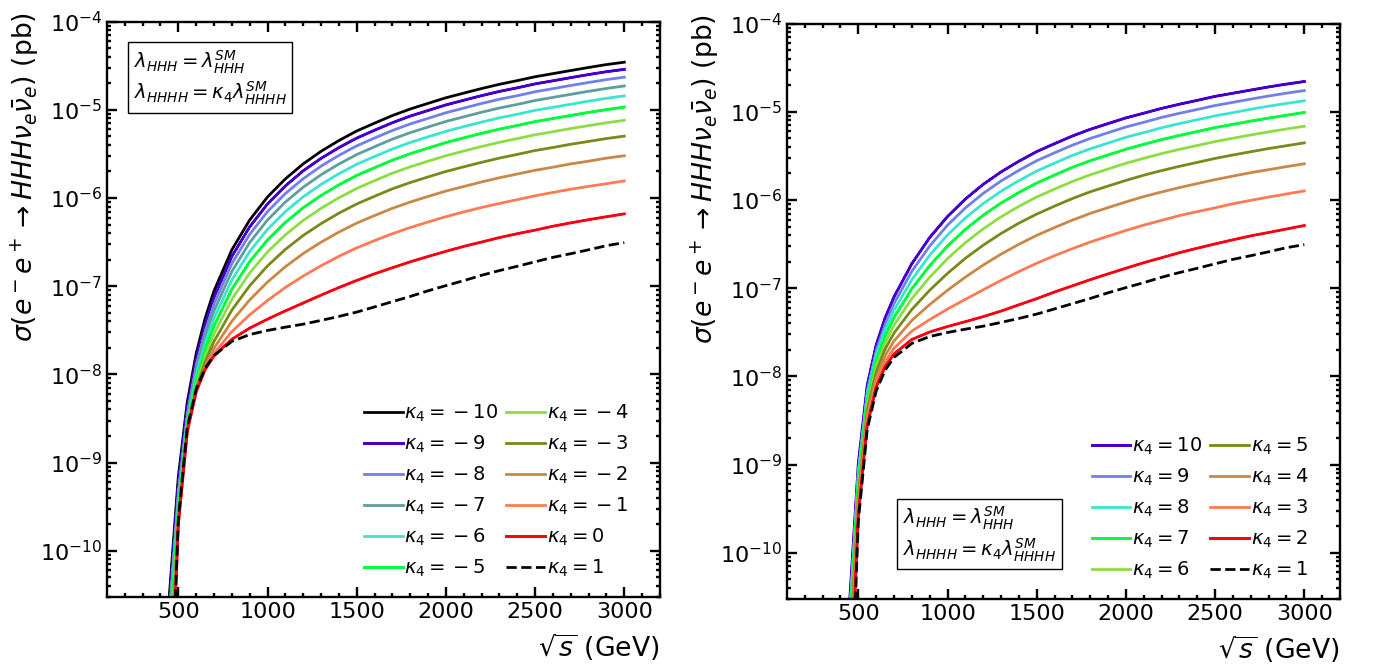}
\caption{Total cross section of $e^+e^-\rightarrow HHH\nu_e\bar{\nu}_e$ as a function of the CM energy $\sqrt{s}$ for different values of the parameter $\kappa_4$, with $\kappa_3$ fixed to 1, compared to the SM prediction (dashed line). Negative (positive) values of $\kappa_3$ are shown in the left (right) panel.}
\label{eeHHHk4vssqrts}
\end{figure}

\begin{figure}[h!]
\centering
\includegraphics[width=0.5\textwidth]{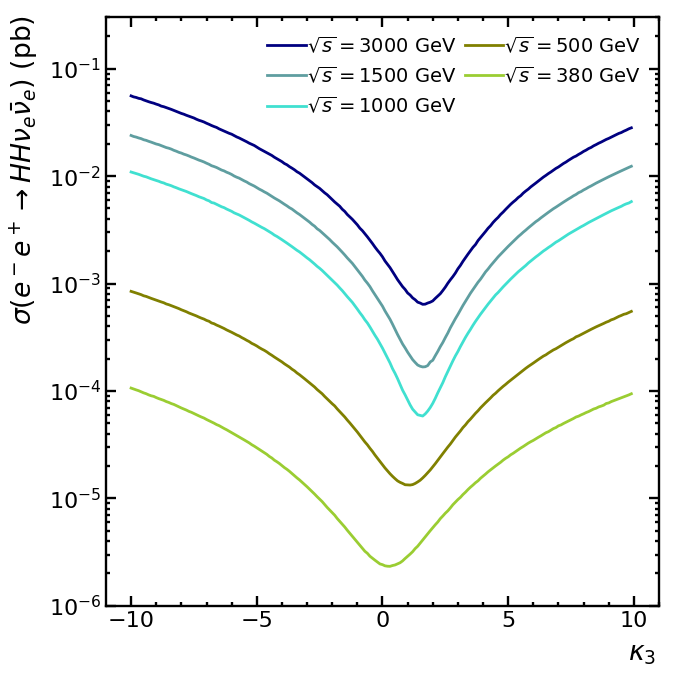}
\caption{Total cross section of $e^+e^-\rightarrow HH\nu_e\bar{\nu}_e$ as a function of $\kappa_3$, with $\kappa_4$ set to 1, for different values of the CM energy $\sqrt{s}$.}
\label{eeHHvsk3}
\end{figure}

\begin{figure}[h!]
\centering
\includegraphics[width=0.999\textwidth]{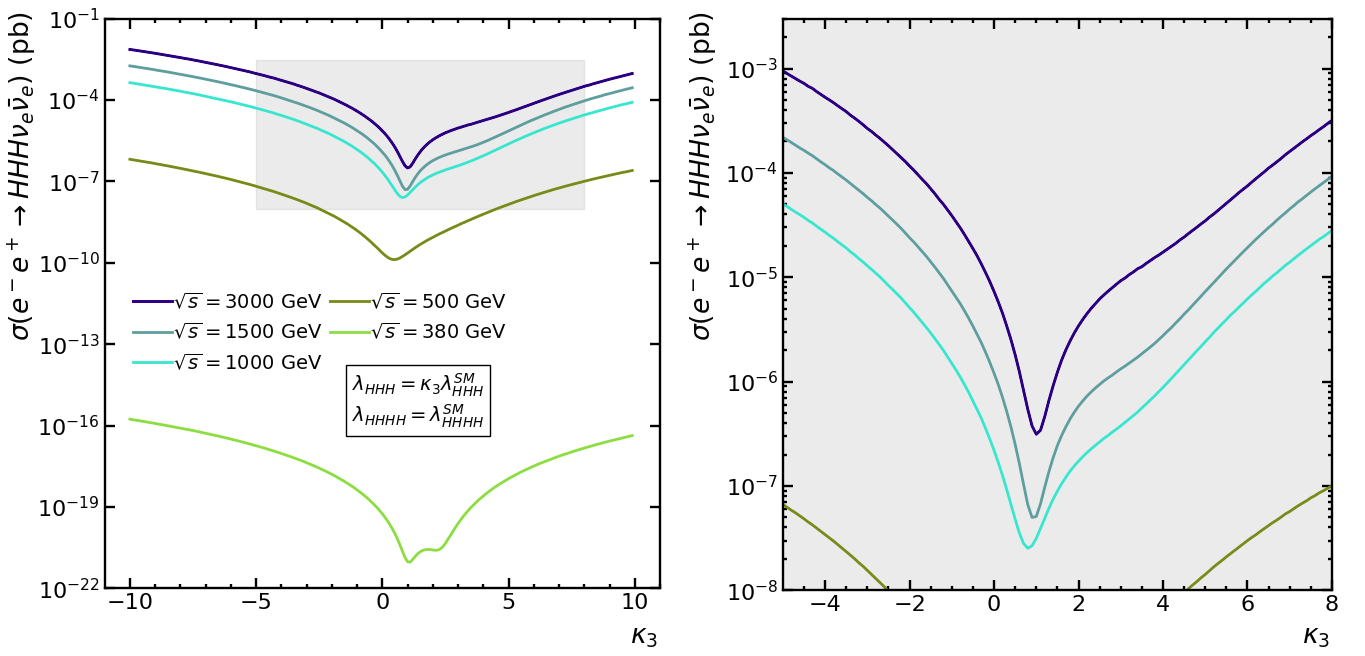}
\caption{Total cross section of $e^+e^-\rightarrow HHH\nu_e\bar{\nu}_e$ as a function of $\kappa_3$ (with $\kappa_4$ fixed to 1) for different values of the CM energy $\sqrt{s}$. Right panel shows a zoom around the dip region.}
\label{eeHHHvsk3}
\end{figure}

\begin{figure}[h!]
\centering
\includegraphics[width=0.999\textwidth]{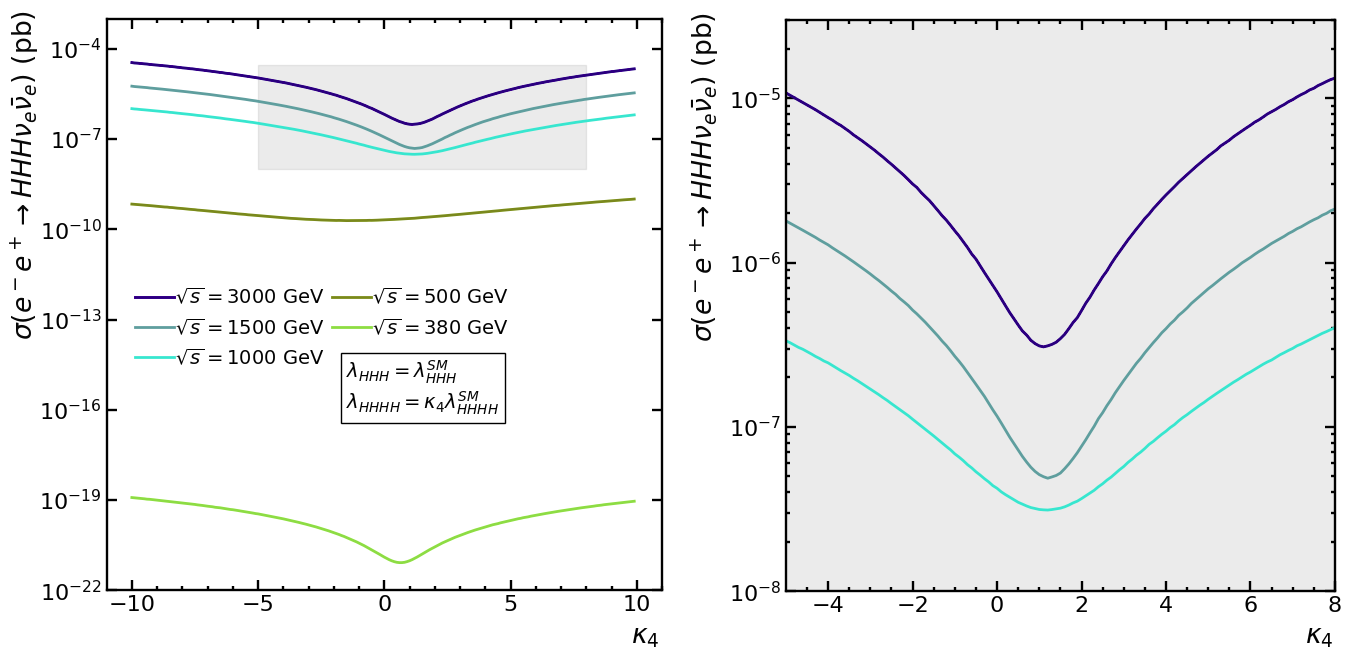}
\caption{Total cross section of $e^+e^-\rightarrow HHH\nu_e\bar{\nu}_e$ as a function of $\kappa_4$ (with $\kappa_3$ fixed to 1) for different values of the CM energy $\sqrt{s}$. The plot on the right shows a zoom around the dip region.}
\label{eeHHHvsk4}
\end{figure}

As we have said, the cross section of $e^+e^-\rightarrow HH \nu_e\bar{\nu}_e$ is only sensitive to $\kappa_3$, whereas that of $e^+e^-\rightarrow HHH \nu_e\bar{\nu}_e$ is sensitive to both parameters $\kappa_3$ and $\kappa_4$. 
The behaviour of the cross section with the energy for $e^+e^-\rightarrow HH\nu_e\bar{\nu}_e$ and for different values of  $\kappa_3$ is shown in \reffi{eeHHk3vssqrts}. The behaviour of the cross section of  $e^+e^-\rightarrow HHH\nu_e\bar{\nu}_e$ for different values of $\kappa_3$ and $\kappa_4$ is shown in \reffi{eeHHHk3vssqrts} and \reffi{eeHHHk4vssqrts}, respectively. In \reffi{eeHHvsk3} the dependence with $\kappa_3$ of $\sigma(e^+e^-\rightarrow HH \nu_e\bar{\nu}_e)$ at various fixed energies is represented. The dependence with $\kappa_3$ and $\kappa_4$ of $\sigma(e^+e^-\rightarrow HHH \nu_e\bar{\nu}_e)$ at various fixed energies is displayed in \reffi{eeHHHvsk3} and \reffi{eeHHHvsk4}, respectively.  
In all cases we include the SM prediction, corresponding to $\kappa_3=\kappa_4=1$,  for comparison. In a first look  to all these plots we see clearly that the main features found at the WWS subprocess level in \refse{k3k4inWWS}, for $W^-W^+ \to HH$ and $W^-W^+ \to HHH$,  are again found here at the collider level, for  $e^+e^- \to HH \nu {\bar \nu}$ and $e^+e^- \to HHH \nu {\bar \nu}$, respectively. In the following we comment in more detail the results in each of these figures.

First, we start with the discussion of  $HH$ production in \reffi{eeHHk3vssqrts}. In these plots we can see how, in general, deviating from $\kappa_3=1$ causes an enhancement of the cross section that is approximately constant with energy. The strongest deviation occurs when $\kappa_3 = -10$, and it differs from the SM prediction by two orders of magnitude. Another thing that can be seen in $HH$ production and will be more significant in $HHH$ is that the bump near the threshold ($2m_H$ in $HH$ and $3m_H$ in $HHH$),  where the associated $Z$ subprocess dominates, disappears when we deviate from the SM (dashed lines).

Second, comparing the  behaviour of $HH$ to the $HHH$ case, which is presented in \reffi{eeHHHk3vssqrts}, we observe something that we had already noticed in the previous section: triple Higgs production is extremely sensitive to variations in the $\kappa_3$ parameter, much more than double Higgs production, reaching BSM deviations of even five orders of magnitude with respect to the SM prediction in the most extreme case ($\kappa_3=-10$). We also learn from  this \reffi{eeHHHk3vssqrts} that in the $HHH$ channel, similarly to the $HH$ channel, the difference with respect to the SM prediction is approximately constant with energy. The bump that is dominated by the associated $Z$ production subprocess  disappears as we separate from $\kappa_3=1$, showing once more that the BSM deviations are clearly dominated by WWS.
\reffi{eeHHHk4vssqrts} is the equivalent to \reffi{eeHHHk3vssqrts}, this time fixing $\kappa_3$ to 1 and varying the $\kappa_4$ parameter. The profile of the deviations due to $\kappa_4$  is very similar to those correspoding to $\kappa_3$, but softer. The maximum deviation now occurs for  $\kappa_4=-10$, this time yielding rates around two orders of magnitude above the SM prediction. As we noted previously, BSM deviations mediated by the $ZHHH$ subprocess due to $\kappa_4\neq 1$ are much smaller than the ones coming from WWS.

Third,  we can now look at the dependence with the $\kappa_3$ and $\kappa_4$ parameters at a fixed CM energy. Starting with the $HH$ case shown in \reffi{eeHHvsk3}, the main difference that we see when comparing this plot with the results for the subprocess in \refse{TestWWS} (see \reffi{HHvsk3}) is that in the $e^+e^-$ case the highest cross section for all $\kappa_3$ values is always achieved at the highest collider energy. 
%All cross sections (this is also true for $HHH$ production) increase with the energy and have no peaks at lower energies independently of the value of $\kappa_3$.% 
The second observation is that there is not a large difference in the sensitivity to $\kappa_3$ depending on the energy, confirming, as stated above, that the deviations with respect to the SM do not depend appreciably on the energy. All the curves in \reffi{eeHHvsk3} for the various energies experiment a variation between one and two orders of magnitude with respect to the SM, having the maximum at the extreme  value of $\kappa_3=-10$, and  they all have just one minimum at the region $\kappa_3\in[0,2]$ (the higher the energy, the larger the value of $\kappa_3$ at the dip). 
 
In \reffi{eeHHHvsk3} we show the corresponding plot for $HHH$ production, where we see the dependence of the cross section with  $\kappa_3$, setting  $\kappa_4=1$, at various CM energies. Here we find again, similarly to the $HH$ case, that the highest cross section for all $\kappa_3$ values is always achieved at the highest collider energy.  Also,  the deviations with respect to the SM prediction are practically insensitive to the energy, and the largest BSM cross section is obtained again at $\kappa_3=-10$, reaching values of up to more than three orders of magnitude higher than the SM value. In contrast to the previous $HH$ case,  the triple Higgs production shows a new feature, namely, the appearance of two minima instead of one. These two minima are obviously correlated with the two minima already observed in \reffi{HHHvsk3} at the $W^-W^+\rightarrow HHH$ subprocess level, and they are more clearly visible at lower collider energies. In particular, for $\sqrt{s}=380$ GeV, there is one minimum around $\kappa_3=1$ and the other one is between $\kappa_3=2$ and $\kappa_3=3$. We also observe a deformation in the $\sqrt{s}=$ 1000, 1500 and 3000 GeV curves, in the region $\kappa_3\in[2,6]$, apart from the minimum around $\kappa_3=1$. This deformation does not appear (at least so clearly) in the curve $\sqrt{s}=500$ GeV, which only has one minimum around $\kappa_3=1$. %
Finally, \reffi{eeHHHvsk4} shows the dependence of $\sigma(e^+e^-\rightarrow HHH\nu_e\bar{\nu}_e)$ with $\kappa_4$, setting $\kappa_3=1$,  for various fixed energies. As we already commented, the deviations with respect to the SM  are softer in this case. In contrast with the previous plot, in this one the curves exhibit only one minimum, which is around $\kappa_4=1$. The only curve in which this does not occur is the one corresponding to $\sqrt{s}=500$ GeV. The cross section at this energy is with difference the least sensible to variations in the $\kappa_4$ parameter.
\begin{figure}[h!]
\centering
\includegraphics[width=0.48\textwidth]{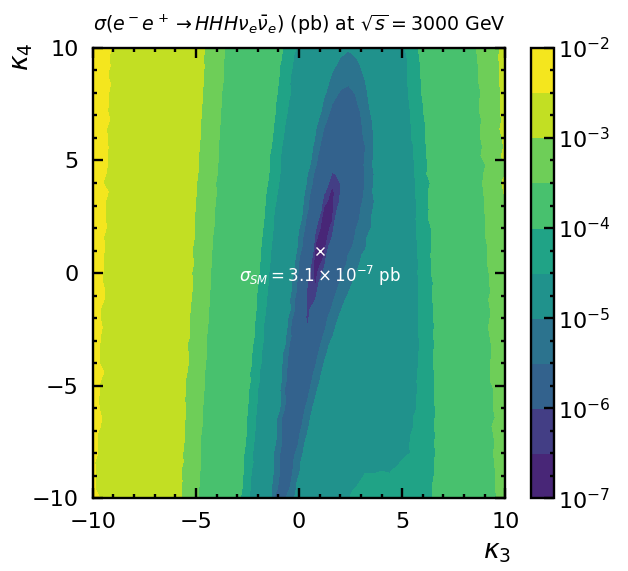}
\includegraphics[width=0.48\textwidth]{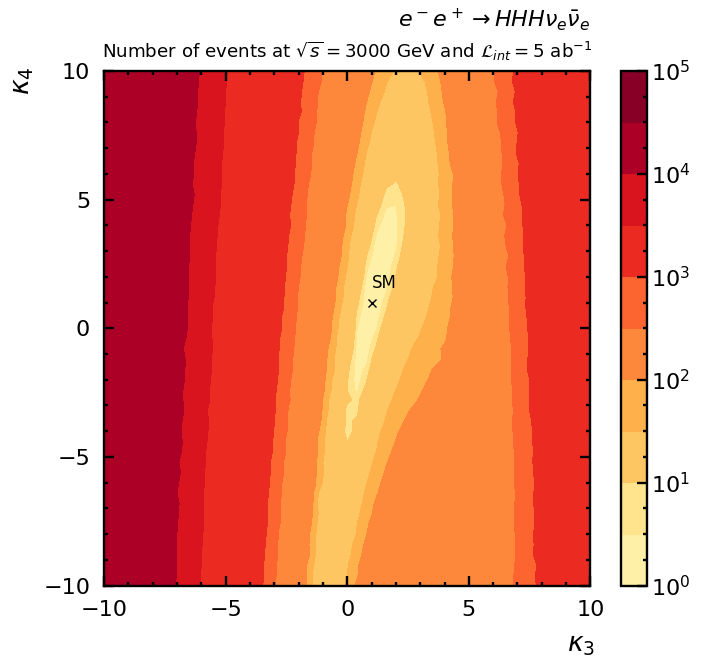}
\caption{Left: Contour levels for the total cross section of the $e^+e^-\rightarrow HHH\nu_e\bar{\nu}_e$ process represented in the $(\kappa_3,\kappa_4)$ plane at a CM energy of 3000 GeV. Right: Corresponding contour levels for the number of $e^+e^-\rightarrow HHH\nu_e\bar{\nu}_e$ events expected at the last stage of CLIC with ${\cal L}_{\rm int}=5$ ab$^{-1}$. }
\label{eetoHHH_3000_2D}
\end{figure}

Now that we have seen the consequences of varying each of the $\kappa_3$ and $\kappa_4$ parameters separately, we study next the combined effect in triple Higgs production of varying both parameters at the same time. For reasons that we will motivate later, we will restrict ourselves in this study to the particular case of $\sqrt{s} = 3000$ GeV. The results of the contour lines for $\sigma(e^+e^- \to HHH \nu_e {\bar \nu_e})$ in the $(\kappa_3,\kappa_4)$ plane are shown in the left plot in \reffi{eetoHHH_3000_2D}.  These results are consistent with the observations made in the previous plots, and confirm clearly that the deviations in the cross section are much stronger in the $\kappa_3$ direction than in the $\kappa_4$ one. In addition, the sensitivity to $\kappa_4$ is larger at the central region of the explored  $\kappa_3$ interval, and the largest rates in this plot are obtained in the upper left corner, i.e. for the extreme values of $(\kappa_3, \kappa_4)=(-10,10)$. Furthermore, if we look back to the subprocess plots in \reffi{WWtoHHH_scan2D}, we notice that, qualitatively, the variations in the cross section of the $e^+e^-$ process at $\sqrt{s}=3000$ GeV behave similarly to those of the $W^-W^+\rightarrow HHH$ subprocess in the region around $\sqrt{\hat{s}}=1000$ GeV. This suggests that the 'effective energy' for WWS in $e^+e^-$ collisions at $\sqrt{s}=3000$ GeV is approximately around $\sqrt{\hat{s}}\sim 1000$ GeV.

Finally, we show in the right plot in \reffi{eetoHHH_3000_2D} the corresponding predictions for the expected number of events in the 
($\kappa_3,\kappa_4$) plane in the most favourable case, namely, at the last stage of CLIC with $\sqrt{s}=3000$ GeV and 
$\mathcal{L}=5\, \mathrm{ab}^{-1}$. As it can be seen in this plot, the expected number of $HHH\nu {\bar \nu}$ events from BSM Higgs couplings can be sizeable in a large region of the ($\kappa_3,\kappa_4$) plane. Indeed, it is much higher than in the SM case, which as we have seen produces negligible event rates. This is true even if we stay within the experimental limits, $\kappa_3\in[-2.3,10.3]$ at a 95\% CL (remember again that there are no present constraints on $\kappa_4$).

Now that we have characterized both the $W^-W^+\to HH(H)$ subprocesses and the whole $e^+e^-\to HH(H)\nu\bar{\nu}$ processes, we are close to being able to explore the final sensitivity to the EChL parameters. However, we still need to study these processes in the framework of a collider, studying the real final state particles, namely, after the Higgs bosons decays. In the following section we will perform such an analysis, motivating why 3 TeV is the optimal energy to obtain the best BSM signals in double and especially triple Higgs production.

%Now that we have characterized both the subprocess and the process for triple Higgs production, we are one step closer to be able to make a prediction, but there is still one ingredient remaining: the collider. In the last part of this section we will consider different possibilities and motivate why our choice of $\sqrt{s}=3000$ GeV is the optimal to obtain (if possible) a signal of $HHH$ production.

%%%%%%%%%%%%%%%%%%%%%%%%%%%%%%%%%%%%%%%%%%%%%%%%%%%%%%%%%%%%
\section{Sensitivity to BSM couplings in multiple $b$-jet events } 
\label{b-jet}
We will focus our forthcoming analysis in the two future linear colliders that are currently under study, the ILC and CLIC (see \refta{eecolliders}). As we already mentioned in the introduction, they are both $e^+e^-$ colliders, and will operate at energies between a few hundreds of GeV and 3 TeV. Each of these energy stages serves a different purpose, being the higher energy configurations the ones oriented to measuring the SM triple Higgs self-coupling via $HH$ production. In principle, none of them is expected to yield measurable signals of $HHH$ production, assuming SM rates. Therefore, our studies of BSM signals via $HHH$ production are very singular in the sense that they will not have the SM as a competitor, since it produces negligible rates (less than 1 event in all cases).
 On the other hand, to obtain 
 testable results in these $e^+e^-$ colliders via the processes of our interest, $e^+e^- \to HH(H) \nu \bar{\nu}$, it is necessary to analyze final states where the Higgs bosons have decayed. Here we will choose the Higgs main decay channel, $H\rightarrow b \bar{b}$, yielding final states with multiple $b$-jets, arising from the final $b$ quarks, plus missing energy, associated to the final neutrinos. Thus, we explore the sensitivity to the EChL parameters is this type of multiple $b$-jet events: 1)  $a$ and $b$ in events with missing transverse energy and four $b$-jets from the decays of the two final $H$'s; and 2) $\kappa_3$ and $\kappa_4$ in events with missing transverse energy and six $b$-jets from the decays of the three final $H$'s. For the present study we ignore potential backgrounds to multiple $b$-jets production accompanied by missing energy, which a priori are expected to be negligible in this $e^+e^-$ context. A more refined analysis, including realistic backgrounds and considering the peculiarities of the planned detectors, is clearly beyond the scope of the present work and is left for future studies. 
 %%%%%
\subsection{Sensitivity to $a$ and $b$ in events with 4 $b$-jets and missing transverse energy} 
We explore here the sensitivity to $a$ and $b$ in $e^+e^- \to HH \nu {\bar \nu} \to b {\bar b} b {\bar b}\nu \bar \nu$ events. For this study we perform a Monte Carlo simulation employing MG5: the same computation that was already worked out, but including now the decays of the Higgs bosons. Regarding the particular collider project,  we consider here three of them: 1) The ILC with $\sqrt{s}=$ 500 GeV and $\mathcal{L}=$ 4 ab$^{-1}$;  2) The ILC with $\sqrt{s}=$ 1 TeV and $\mathcal{L}=$ 8 ab$^{-1}$; and 3) The CLIC with $\sqrt{s}=$ 3 TeV and  $\mathcal{L}=$  5 ab$^{-1}$.

In order to characterise the BSM signal arising from the EChL anomalous couplings $(a,b)$, we have first to define the final state with particles that can be detected in the experiments.  The events that we study here contain four $b$-jets from the hadronization of the final $b$ quarks, which are produced in the Higgs decays, and missing energy corresponding to the final $\nu$ and $\bar \nu$ escaping detection. This characteristic missing energy from the neutrinos and antineutrinos could allow to differentiate the BSM signal from potential SM backgrounds, like those of QCD origin. These channels exhibit multiple jets in the final state due to the hadronization of quarks and gluons, but typically show no relevant missing energy. Therefore, to explore the BSM signal we will implement some cuts on the relevant variables of the final particles. In particular, the $b$-jets will be required to have a minimum transverse momentum in order to be detected, and also the missing transverse energy will be required to be above a minimum value.  Besides, a cut in the polar angle $\theta$, or equivalently, a maximum pseudorapidity $\eta^j\equiv -\log\left(\tan\frac{\theta}{2} \right) $ for the jets, is required. Finally, in order to be detectable, the jets need to exhibit a certain angular separation among them. This is equivalent to stablishing a minimum value for the variable $\Delta R_{jj}\equiv \sqrt{(\Delta \eta_{jj})^2+(\Delta \phi_{jj})^2}$, where $\Delta\eta_{jj}$ and $\Delta\phi_{jj}$ are the separations in pseudorapidity and azimuthal angle of the two jets $jj$ , respectively. The values chosen for all these cuts in this work, which are similar to those taken in references \cite{Contino:2013gna} and \cite{Abramowicz:2016zbo}, are summarised as follows:
\begin{equation}
\label{cuts}
%\begin{split}
p_T^j>20\enspace \mathrm{GeV} \,\,\,\,;\,\,
\vert\eta^j\vert<2 \,\,\,\,;\,\,
\Delta R_{jj}>0.4 \,\,\,\,;\,\,E\!\!\!\!/_T>20\enspace \mathrm{GeV}
%\end{split}
\end{equation}
The decay of the two Higgs bosons will lead to a reduction factor in the event rates of $0.58^2$, due to the branching ratios (BRs) of the decays to $b$ quarks. Once the cuts on the final state are implemented, the signal event rates will obviously suffer a further reduction. To estimate the effects of these cuts in \refeq{cuts} we have evaluated the acceptance, ${\cal A}$,  given by the ratio of  the predicted $b \bar b b \bar b \nu \bar \nu$ rates after applying these cuts divided by the rates before cuts (the latter are the ones already shown in \reffi{contourXS}). We have evaluated ${\cal A}$  in the $(a,b)$ plane, considering the region delimited by $a$ and $b$ in the interval [0.5,1.5],  as in \reffi{contourXS}, and we have done it for the three energies chosen here, of 500 GeV, 1000 GeV and 3000 GeV. We have found that  ${\cal A}$ diminishes when increasing the collider energy. For 500 GeV it varies approximately in the range  0.65-0.72, for 1 TeV in the range 0.55-0.65, and for 3 TeV in the range 0.3-0.5. There is a qualitative difference between the case of 500 GeV and the other two collider stages, with 1 TeV and 3 TeV respectively. The acceptance for 500 GeV  is maximal, above 0.7,  at the region close to the SM point $(a,b)=(1,1)$, whereas in the 1 TeV and 3 TeV cases it is minimal in this region, with values around 0.55 and 0.3 respectively. This qualitative difference can be understood from the fact that the kinematical configuration of the final state particles in two cases of 1 and 3 TeV is driven from the dominant WWS subprocess, which is not the case for the 500 GeV collider, where the $Z$-mediated diagrams dominate.  
%It should also be noticed that additional cuts like 
%requiring the invariant mass of the two $b \bar b$ jet pairs, $M_{b \bar b}$, to be close to the Higgs %mass value could also be applied, without affecting much the signal rates. These kind of cuts could  %exclude very efficiently other potential backgrounds  like $ZZ\nu \bar \nu$ production with the $Z$ %gauge bosons decaying to $b \bar b$ pairs. But we do not include additional cuts here, only those in %\refeq{cuts}.

In order to conclude on the potential accesibility to these $a$ and $b$ parameters at future  $e^+e^-$ colliders, we also have to take into account the efficiency in the detection of the four $b$-jets. Thus, the computed rates must be reduced by an extra factor of $\epsilon^4$, where $\epsilon$ is the $b$-tagging efficiency factor. We will assume here $\epsilon$ of an 80$\%$, which is the value commonly used in the literature, see for instance \cite{Bishara:2016kjn} and \cite{Contino:2013gna}.  
Taking into account all the cuts shown above, the BRs of the decays, and the efficiency reduction factors, we have finally computed the $b \bar b b \bar b \nu \bar \nu$ event rates of our BSM signal for each choice of the anomalous couplings $a$ and $b$ and for each selected collider setup.  
\begin{figure}[h!]
\centering
\includegraphics[width=0.34\textwidth]{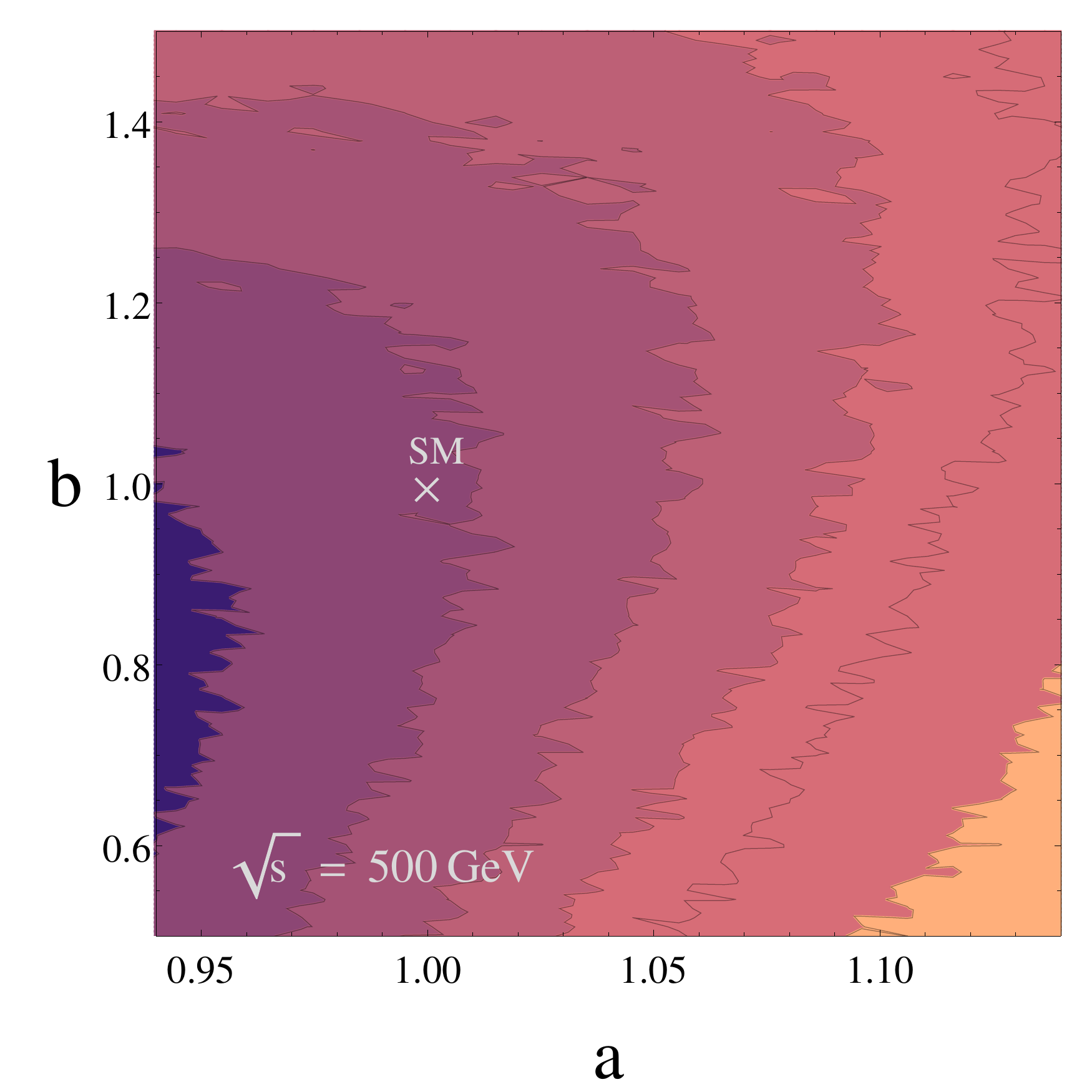}
\hspace{-0.5cm}
\includegraphics[width=0.34\textwidth]{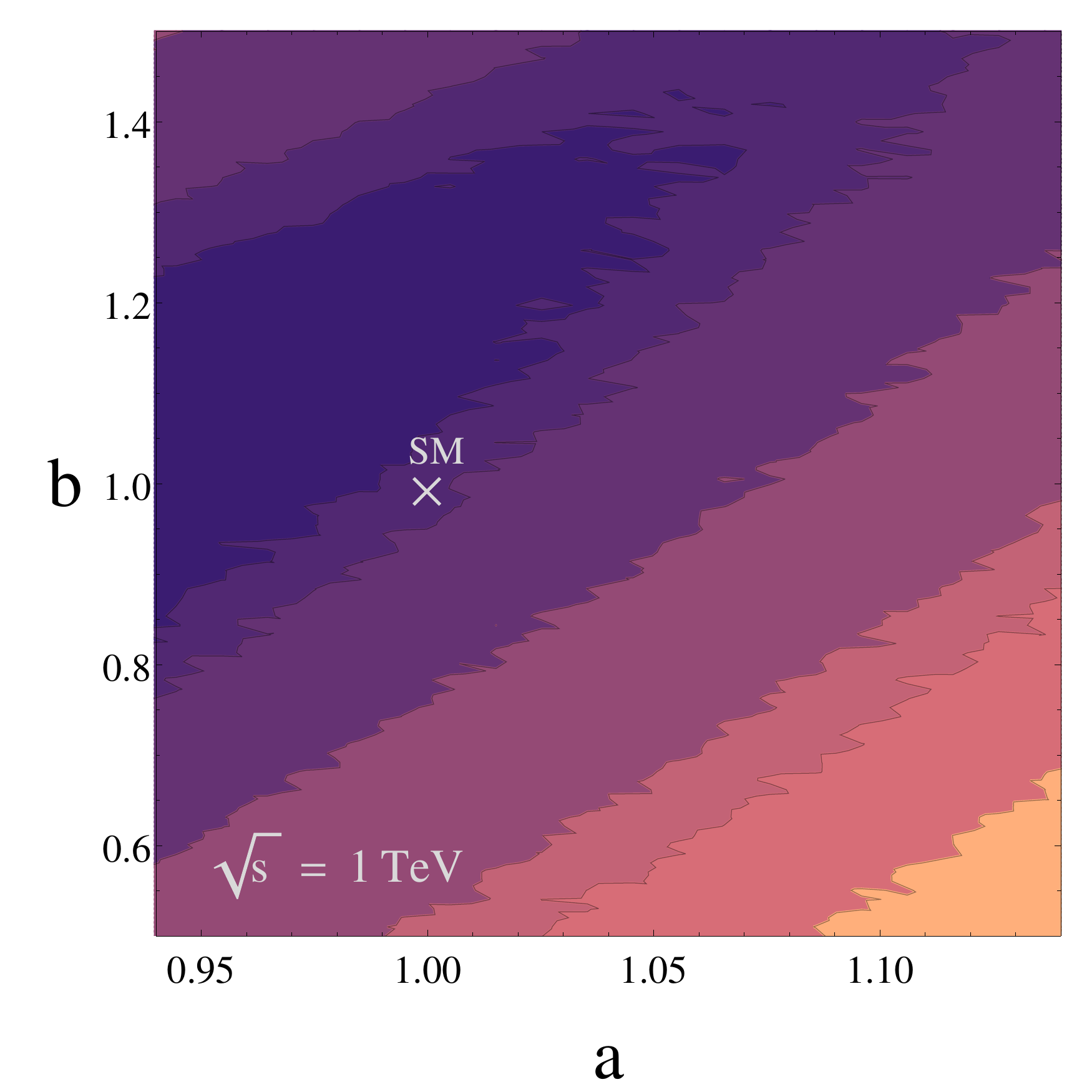}
\hspace{-0.5cm}
\includegraphics[width=0.34\textwidth]{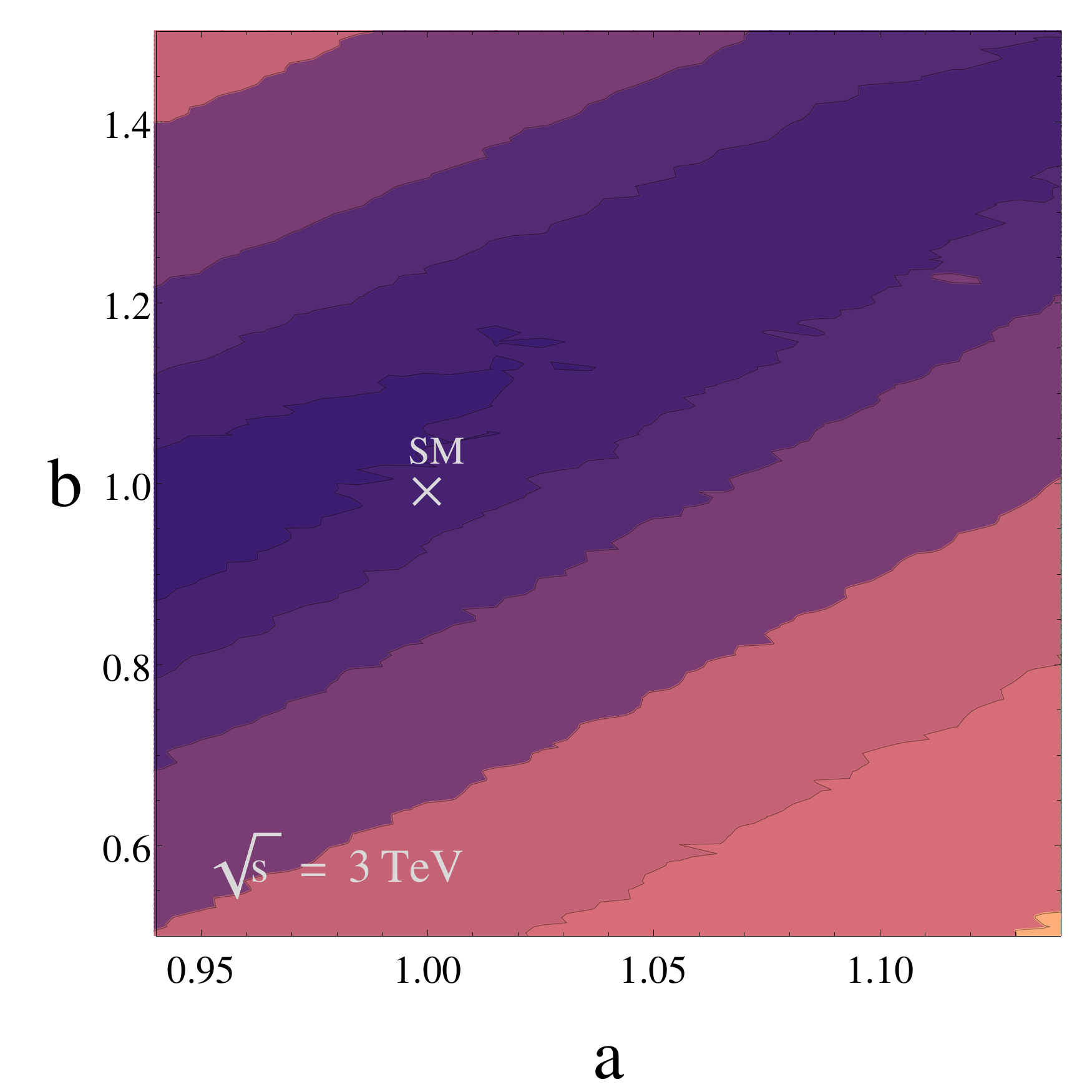}
\includegraphics[width=0.34\textwidth]{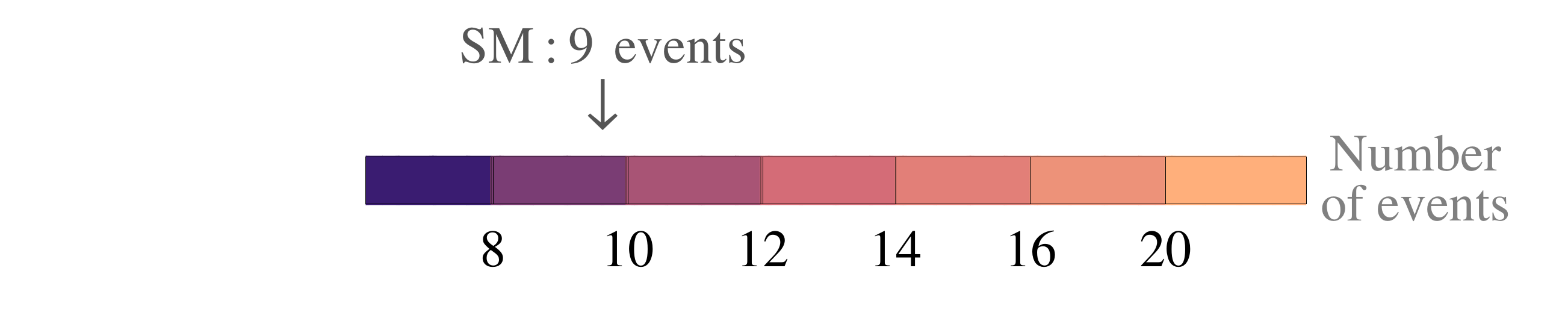}
\hspace{-0.5cm}
\includegraphics[width=0.34\textwidth]{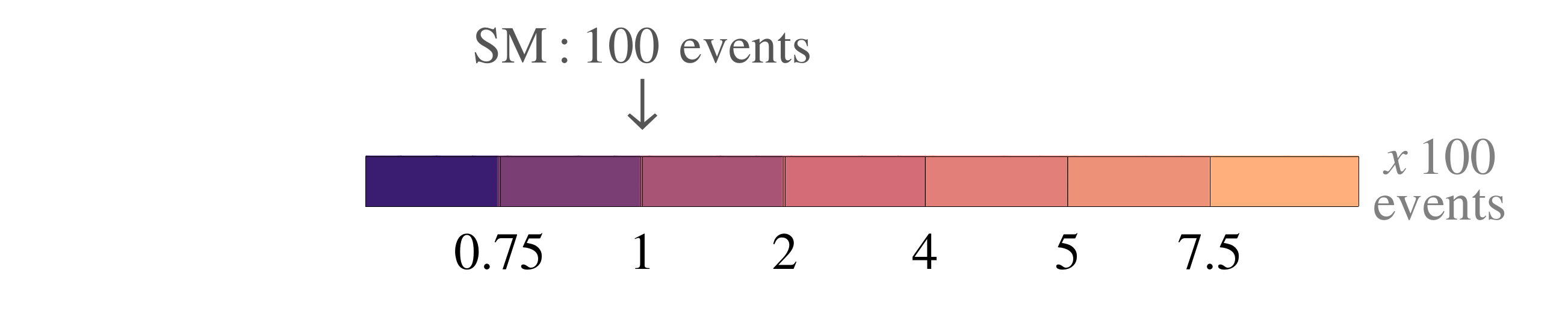}
\hspace{-0.5cm}
\includegraphics[width=0.34\textwidth]{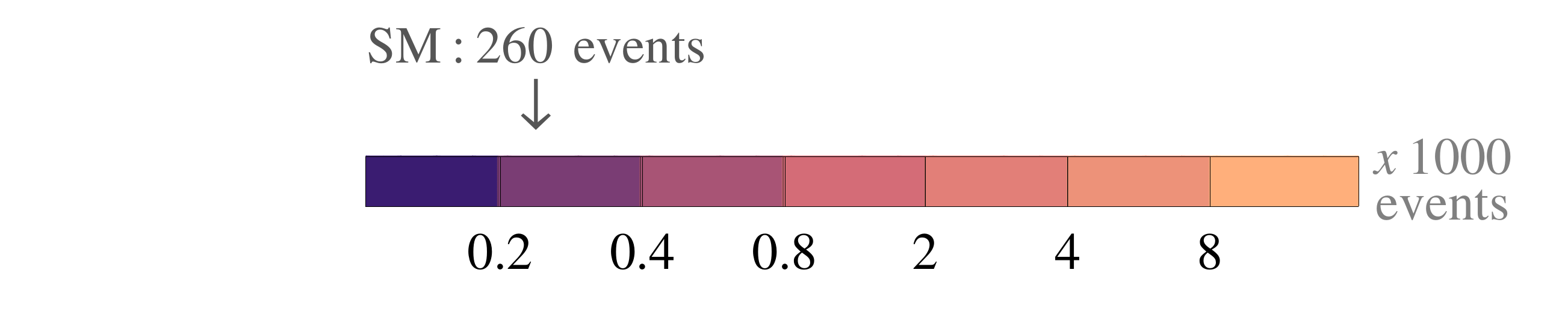}
\caption{Contour lines for the expected number of events of the $e^+e^-\to HH\nu_e\bar{\nu}_e\to b\bar{b}b\bar{b}\nu_e\bar{\nu}_e$ process in the $(a,b)$ plane. The cuts in \refeq{cuts} and the efficiency factors have already been implemented, see text for details. Several collider stages are considered: ILC with 500 GeV and 4 ab$^{-1}$ (left panel), ILC with 1 TeV and 8 ab$^{-1}$ (middle panel), and CLIC with 3 TeV and 5 ab$^{-1}$ (right panel). The range chosen for $a$ is the present experimentally allowed region. The white cross represents the SM prediction ($a=b=1$).}
\label{eventszoom}
\end{figure}

We display, in  \reffi{eventszoom}, the contour lines for these predicted number of events in the $(a,b)$ plane for:  ILC (500 GeV) with 4 ab$^{-1}$ (left plot),  ILC (1TeV) with 8 ab$^{-1}$ (middle plot), and CLIC (3TeV) with 5 ab$^{-1}$ (right plot). Notice that we have adjusted the interval displayed for the paramater $a$ so as to coincide with its presently experimentall allowed interval, whereas for the $b$ parameter we have displayed the full interval [0.5,1.5], as in the previous figures.  
From this \reffi{eventszoom} we can already extract some conclusions regarding the BSM event rates compared to the SM ones. In the three plots shown we find areas (depicted in the lighter colors), where the ratio BSM/SM is considerably larger than 1. Regarding the relative events statistics, the lowest collider energy provides the smallest rates, and the highest collider energy the largest ones, as expected. For ILC (500 GeV) with $4$ ab$^{-1}$  we only find a few events, varying from less than 10 in the lower-left part of the plot to around 22 in the lower-right corner. These should be compared with the 9 events predicted in the SM.  
For ILC(1TeV) with $8$ ab$^{-1}$ we find larger rates, around 50-70 events in the region close and around $b=1$ with $a<1$, and up to more than 750 events in the lower-right corner. The SM predicts 100 events in this case. For CLIC (3TeV) with $5$ ab$^{-1}$  the largest rates are found, ranging from around 100 events in the region very close and around $b=1$ with $a<1$ to more than 8000 events in the lower-right corner. Therefore,  CLIC offers better rates than the other two explored projects and seems to be the best option in order to be sensitive to the $(a,b)$ parameters.

\begin{figure}[h!]
\centering
\includegraphics[scale=0.5]{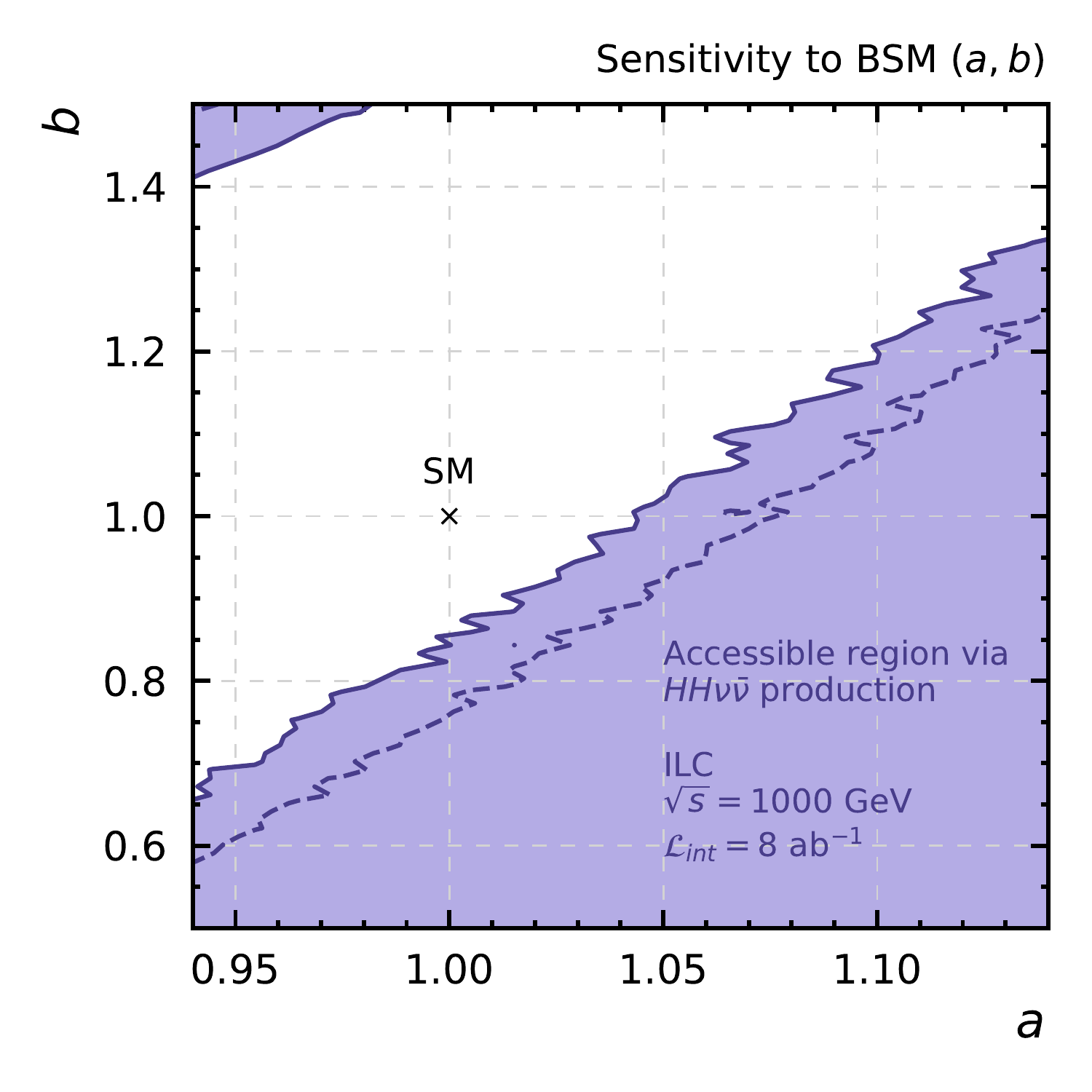}
\includegraphics[scale=0.5]{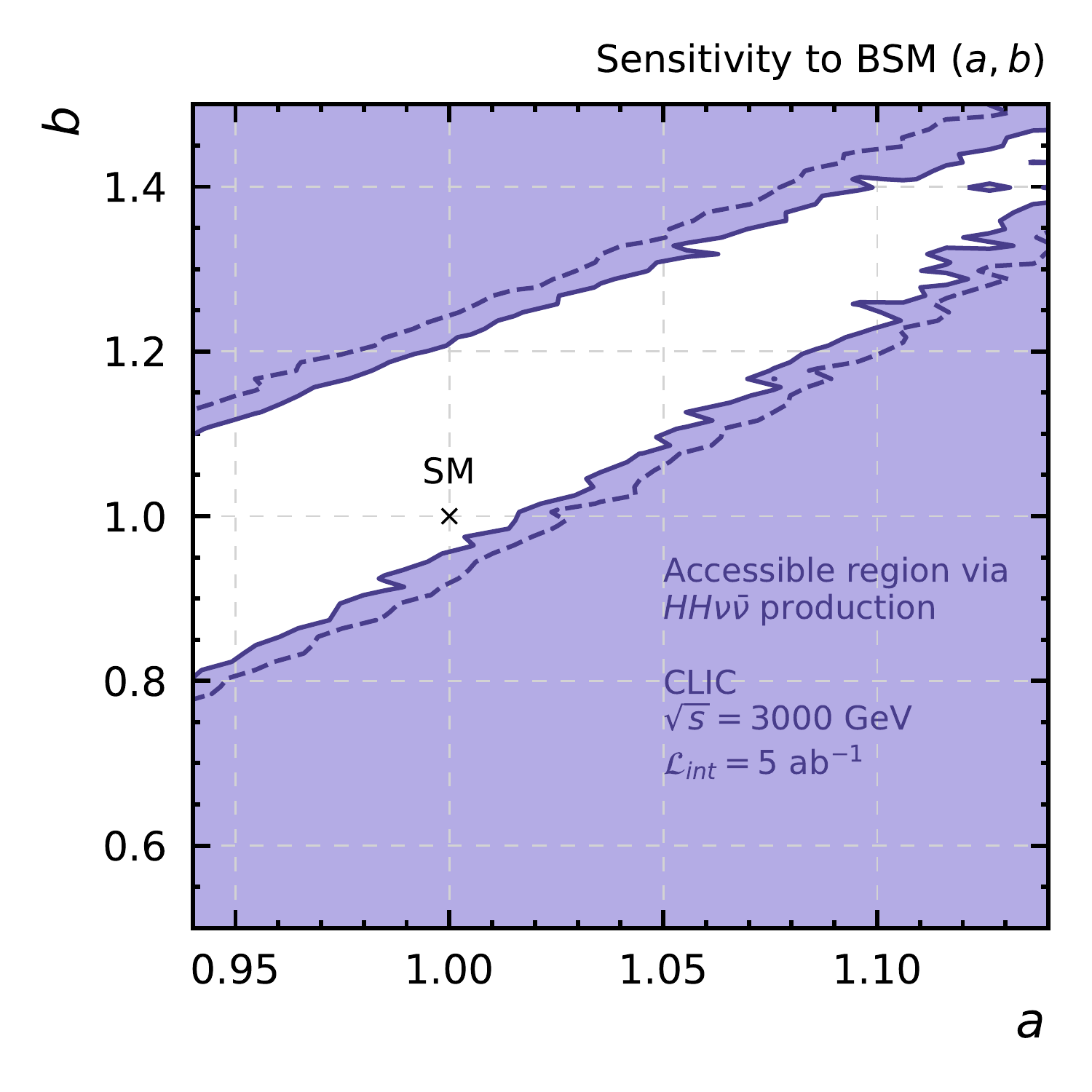}
\caption{Accessible region in the ($a,b$) plane to BSM scenarios via $e^+e^- \to  HH\nu\bar{\nu} \to b {\bar b} b {\bar b} \nu {\bar \nu}$ events, at the ILC with $\sqrt{s}=1000$ GeV and $\mathcal{L}_\text{int}=5$ ab$^{-1}$ (left panel) and at CLIC with $\sqrt{s}=3000$ GeV and $\mathcal{L}_\text{int}=5$ ab$^{-1}$ (right panel). The criterion of accesibility assumed here is by means of the quantity $R\equiv(N_\mathrm{BSM}-N_\mathrm{SM})/\sqrt{N_\mathrm{SM}})$, leading to the purple area bounded by the solid (dashed) contourline for $R\geq 5(10)$. }
\label{Accestoab}
\end{figure}

Finally, to provide a more concrete conclusion on the accessibilty to these two anomalous couplings, we have evaluated the ratio $R\equiv(N_\mathrm{BSM}-N_\mathrm{SM})/\sqrt{N_\mathrm{SM}}$, where $N_\mathrm{(B)SM}$ is the number of events in the (B)SM case. This quantity is a way to measure the size of the deviation of the BSM  signal with respect to the SM prediction. Since there are very low statistics in the 500 GeV case, both for BSM and the SM, we focus on the other two options. We show in \reffi{Accestoab} the accessibility regions in the $(a,b)$ plane for the ILC(1TeV) and  CLIC(3TeV) cases, defined as $R>5$ (purple region bounded by solid contourline) in the most optimistic case,  and as $R>10$ (purple region bounded by dashed contourline) in the more conservative case. It is clear from this plot that both options, ILC(1TeV) and  CLIC(3TeV),  will offer a good accessibility to measure $a$ and $b$ beyond their present experimental constraints from LHC. The best option will be clearly CLIC (3TeV) with $5{\rm ab}^{-1}$, where the unaccessible area (in white) shrinks around the SM point, especially in the $b$ direction, which is the worst explored at present. The sizes of these unaccesible areas provide an approximate estimate of the expected improvements on the constraints on these anomalous couplings. Of course, in order to determine more accurately the sensitivity to variations in $a$ and $b$ with respect to their SM values, it would be necessary to analyze the possible backgrounds as well.  This is well beyond the intention of this work and is left for future research.  

Nevertheless,  we believe that a more complete analysis including backgrounds will not change the main conclusions of this work,  since a first,  order of magnitude estimate of the main backgrounds already shows they can be dealt with easily.  In particular,  if we naively ignore the detection effects and assume no cuts at all,  we find cross section rates for the most relvant backgrounds (corresponding to those processes with $Z$ bosons instead of $H$ bosons) which are comparable to our reference SM value.  For instance,  at 3 TeV,  using MG5,  we get the SM cross section reference value of  $\sigma(e^+e^-\to \nu_e\bar{\nu}_eHH; H\to b \bar{b})=0.54$ fb. For the main backgrounds,  we find $\sigma(e^+e^-\to \nu_e\bar{\nu}_eHZ; H,Z\to b \bar{b})=1.28$ fb and $\sigma(e^+e^-\to \nu_e\bar{\nu}_eZZ;Z\to b \bar{b})=1.25$ fb.  Presumably,  these two processes involving $Z$'s could be efficiently reduced by the action of proper cuts,  in particular,  that on the $b \bar b$ invariant mass,  requiring it to be close to the Higgs mass value.  Other possible backgrounds from QCD (i.e.,  ${\cal O}(\alpha_S  \alpha^2)$ at the amplitude level) are much smaller,  since we require neutrinos in the final state.  For instance,  we get $\sigma(e^+e^-\to b \bar{b}b \bar{b} Z;Z\to \nu\bar{\nu})=0.002$ fb.  Being the signal rates for our BSM scenarios with anomalous $(a,b)$ couplings  well above the SM rates,  we expect all these backgrounds to be treatable.

%% %%%%%%%%%%%%%%%%%%%%%%%%%%%%%%%%%%%%%%%%%%%%%%%%%%%%%%%%%%%%%%%%%%%
\subsection{Sensitivity to $\kappa_3$ and $\kappa_4$ in events with 6 $b$-jets and missing transverse energy}
In this section we study the sensitivity to the anomalous couplings $\kappa_3$ and $\kappa_4$ in 
 $e^+e^- \to HHH \nu {\bar \nu} \to b {\bar b} b {\bar b} b {\bar b} \nu \bar \nu$ events. Therefore, we  consider the dominant decays of the final Higgs bosons leading to $b$-jets and take into account the missing energy left by the final neutrinos and antineutrinos. 
 We perform our analysis in the most favourable scenario, which is the last stage of CLIC, at a CM energy of 3000 GeV with an integrated luminosity of 5 ab$^{-1}$. In the previous sections we learnt that it is at that high energies and luminosities where the channel with three Higgs bosons and neutrinos can be most sensitive to variations of the Higgs self-couplings.  As we have said, it is just in triple, and not double, Higgs production where there is a unique sensitivity to the quartic Higgs coupling $\kappa_4$. Indeed, considering at the same time deviations in $\kappa_3$ and $\kappa_4$  in triple Higgs production plus neutrinos can boost the number of events from not more than one in the SM (which does not yield a detectable signal) to tens, hundreds or even thousands in the most extreme cases of BSM scenarios. Thus, in this section we will focus on characterising these BSM signals from the triple Higgs channel, with mainly 6 b-jets and missing transverse energy, at the last stage of CLIC.

First, we investigate the main features of the kinematical configuration in these multi-particle events  for our BSM signal with 
$(\kappa_3,\kappa_4) \neq (1,1)$ and compare it with the SM case with  $(\kappa_3,\kappa_4) = (1,1)$ . For this characterization, we will generate a set of samples for different values of $\kappa_3$ and $\kappa_4$ and study some interesting event distributions. All the events will be generated using MG5 and analyzed using ROOT 6 \cite{Brun:1997pa}. Since the hadronization of the $b$ quarks in the final state is a very expensive task, in order to define the $b$-jets we will use a resolution criterion instead. Following the reasoning in \cite{Contino:2013gna}, we will consider an energy resolution of $\Delta E/E =5\%$ and assume that two quarks with a small separation of $\Delta R_{qq} < 0.4$ cannot be resolved individually. This condition will be applied recursively until we converge to a final list of quarks that we will identify as the $b$-jets.

%%%%%%%%%%
 \begin{figure}[H]
\centering
\includegraphics[width=0.59\textwidth]{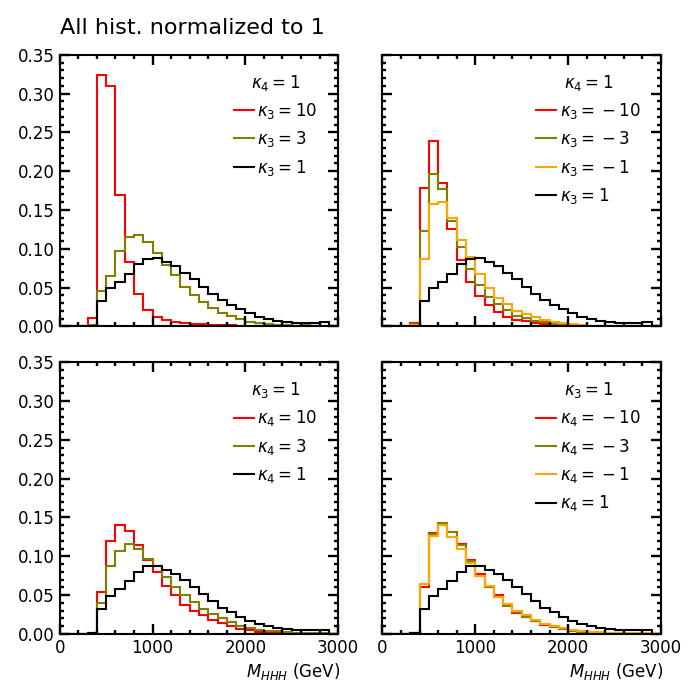}
\caption{Distribution of $e^+e^-\to b\bar{b}b\bar{b}b\bar{b}\nu\bar{\nu}$ events with respect to the invariant mass of the three Higgs bosons, that is, of the six $b$-jets, for several values of the parameters $\kappa_3$ and $\kappa_4$.}
\label{MHHHnocuts}
\end{figure}

\begin{figure}[H]
\centering
\includegraphics[width=0.59\textwidth]{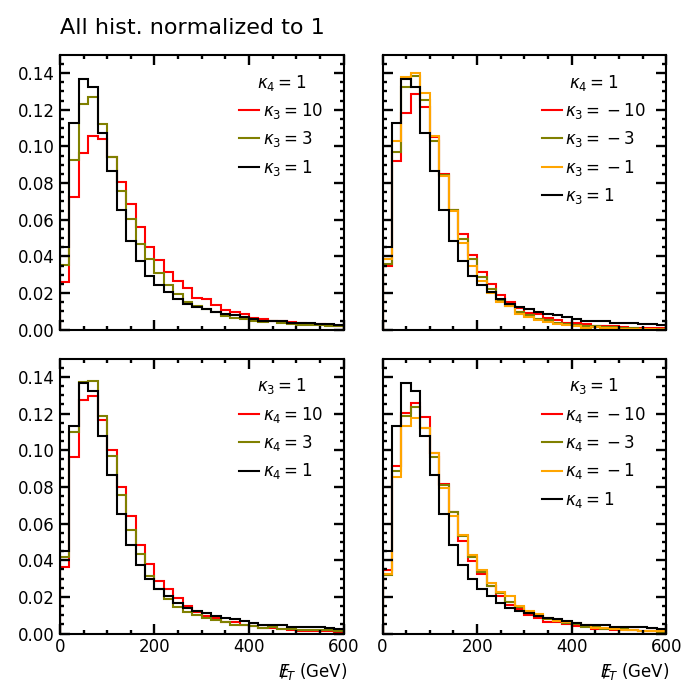}
\caption[Caption for LOF]{Distribution of $e^+e^-\to b\bar{b}b\bar{b}b\bar{b}\nu\bar{\nu}$ events with respect to the transverse missing energy, for several values of the parameters $\kappa_3$ and $\kappa_4$.}
%\footnotemark.}
\label{METnocuts}
\end{figure}

\begin{figure}[H]
\centering
\includegraphics[width=0.58\textwidth]{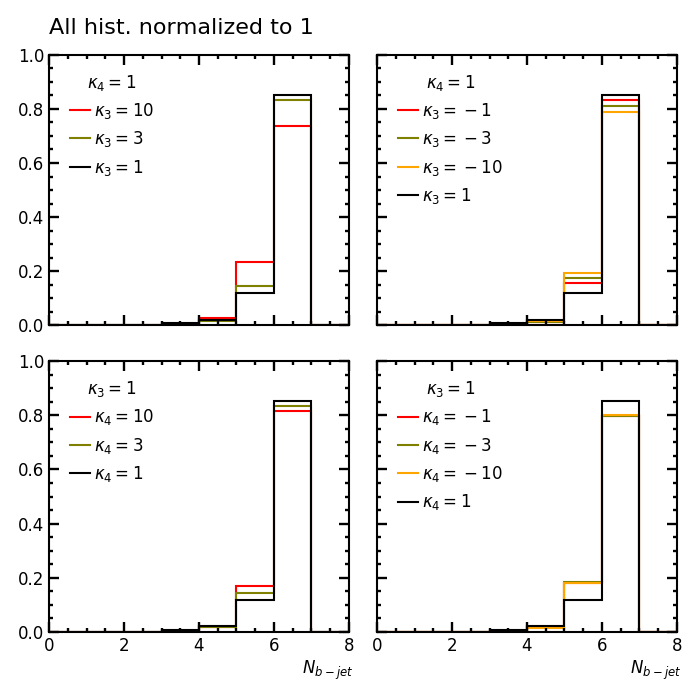}
\caption{Distribution of $e^+e^-\to b\bar{b}b\bar{b}b\bar{b}\nu\bar{\nu}$ events with respect to the number of $b$-jets, for several values of the parameters $\kappa_3$ and $\kappa_4$. Note that the number of $b$-jets is not necessarily six due to the resolution criterion.}
\label{Nbnocuts}
\end{figure}

\begin{figure}[H]
\centering
\includegraphics[width=0.58\textwidth]{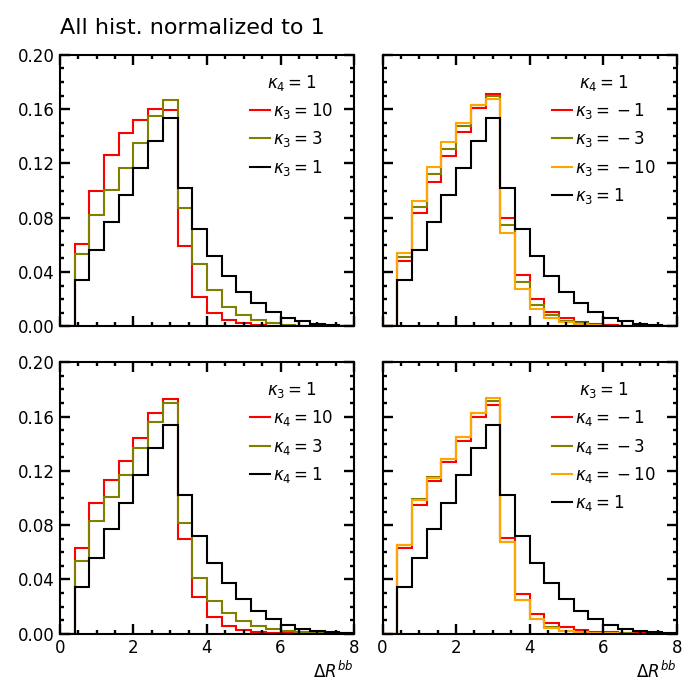}
\caption{Distribution of $e^+e^-\to b\bar{b}b\bar{b}b\bar{b}\nu\bar{\nu}$ events with respect to the angular separation between $b$-jets, for several values of the parameters $\kappa_3$ and $\kappa_4$.}
\label{DRnocuts}
\end{figure}

\begin{figure}[H]
\centering
\includegraphics[width=0.59\textwidth]{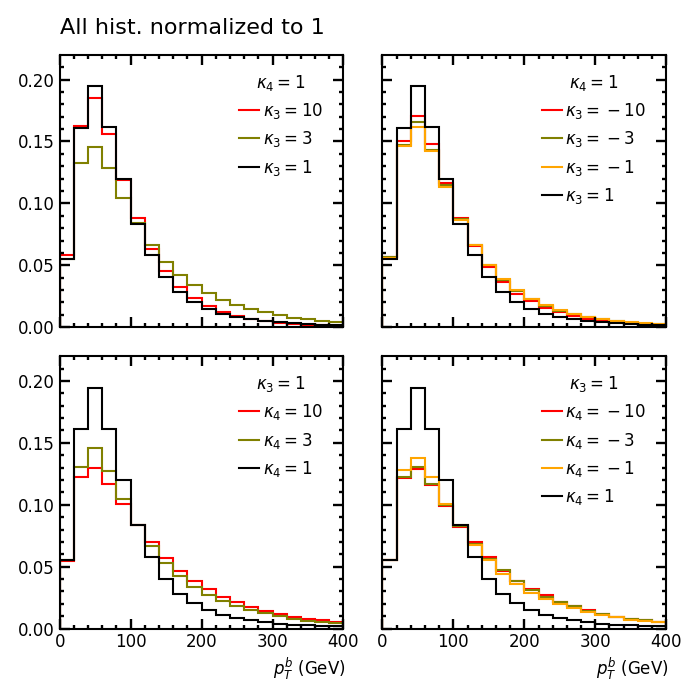}
\caption{Distribution of $e^+e^-\to b\bar{b}b\bar{b}b\bar{b}\nu\bar{\nu}$ events with respect to the transverse momentum of the $b$-jets, for several values of the parameters $\kappa_3$ and $\kappa_4$.}
\label{bPtnocuts}
\end{figure}

\begin{figure}[H]
\centering
\includegraphics[width=0.59\textwidth]{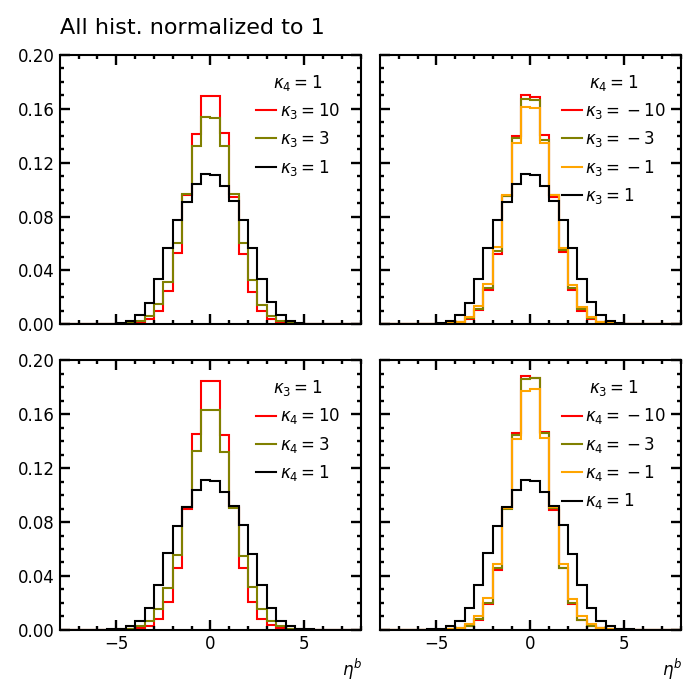}
\caption{Distribution of $e^+e^-\to b\bar{b}b\bar{b}b\bar{b}\nu\bar{\nu}$ events with respect to the pseudorapidity of the $b$-jets, for several values of the parameters $\kappa_3$ and $\kappa_4$.}
\label{bEtanocuts}
\end{figure}

The plots in Figs. \ref{MHHHnocuts} to \ref{bEtanocuts} show the distributions with respect to several kinematic variables for different values of $(\kappa_3,\kappa_4)$. Concretely, we chose: 1) the invariant mass of the three Higgs bosons $M_{HHH}$, equivalent to that of the six $b$-jets (\reffi{MHHHnocuts}); 2) the missing transverse energy $E\!\!\!\!/_T$, equivalent to that of the final $\nu \bar \nu$ pair (\reffi{METnocuts}); 3) the realistic number of $b$-jets in the events,  $N_{b-\mathrm{jet}}$, due to the resolution criterion commented above (\reffi{Nbnocuts}); 4)  the angular separation between two $b$-jets, $\Delta R^{bb}$ (\reffi{DRnocuts}); 5) the transverse momentum of the $b$-jets, $p_T^b$ (\reffi{bPtnocuts}), and 6) the pseudorapidity of the $b$-jets, $\eta^b$ (\reffi{bEtanocuts}). For simplicity we have only plotted deviations of either $\kappa_3$ or $\kappa_4$, with the other one set to 1. Notice also that all the histograms are normalized to unity (and not to the real number of events) since we are interested now in comparing just their shape.

%%%%%%%%%
%%%
We will next comment some general features learnt from these plots. First, we note that the invariant mass distribution of the three Higgs bosons, which is equivalent to the energy of the subprocess, $\sqrt{\hat{s}}$, confirms the observation we made when comparing  \reffi{eetoHHH_3000_2D} with the corresponding plots for the $W^-W^+\rightarrow HHH$ subprocess: we suggested that the 'effective energy' for WWS in $e^+e^-$ collisions at $\sqrt{s}=3\, {\rm TeV}$ is approximately around $\sqrt{\hat{s}}= 1\,{\rm TeV}$.  In fact, in the distribution for the SM case  we see in \reffi{MHHHnocuts} a maximum centered around $M_{HHH}=1000$ GeV, while for other values of $\kappa_3$ and $\kappa_4$ the maximum is generally displaced to lower energies, and the peaks exhibit different shapes. From the $E\!\!\!\!/_T$ and $p_T^b$ plots, we also see that both the neutrino-antineutrino pairs and the $b$-jets tend to be produced with higher transverse momentum in BSM scenarios than in the SM, which is consistent with a smaller value of the pseudorapidity, as we see in the plot of $\eta_b$. As a final remark, we also notice that the realistic number of $b$-jets appears to decrease as we deviate from the SM. This is due to the $b$ quarks being produced with smaller relative angles, as can be seen in the distribution with respect to the variable $\Delta R^{bb}$. The area of this distribution in \reffi{DRnocuts} with larger rates is displaced to lower values of $\Delta R^{bb}$ in the BSM cases than in the SM, meaning that the $b$-jets may not be always identified individually, thus yielding in some cases an apparently lower number of produced $b$-jets. Finally, for the jet analysis, we need to take also into account the $b$-jet identification efficiency.  As in the previous section, we will adopt here a  $b$-tagging efficiency of 80\% \cite{Contino:2013gna}. This corresponds to a misidentification efficiency of 10\% for $c$-jets and 1\% for light flavour jets.
%%%%%
%%%%
%%%%

Overall, considering the features learnt from all these distributions and the needed requirement of not loosing too much signal, we will impose the following set of cuts, which are similar to those in \cite{Contino:2013gna}, but also requiring a minimum value of the missing transverse energy: \begin{equation}
    |\eta_\text{jet}| < 2.72\,\,\,\,;\,\,
    N_\text{jet} \geq 6, \,\,\,\,;\,\,
    p_T^\text{jet} \geq 20 \text{ GeV} \,\,\,\,;\,\,
    N_{b-\mathrm{jet}} \geq 5 \,\,\,\,;\,\,
    E\!\!\!\!/_T \geq 20 \text{ GeV}.
\label{cuts2}    
\end{equation}
Some comments are in order. First, regarding the cut in pseudorapidity  we use here the value reported in \cite{Arominski:2018uuz} of $|\eta|_\text{max}=2.72$ for the particular case of CLIC. Notice that this is slightly less restrictive than the one chosen in the previous section of $|\eta^j|<2$. Since the rates for triple Higgs production are much lower than for double Higgs production, we relax this cut to avoid the loss of too many signal events. 
Second, we require our events to include at least six jets. We also impose that these jets have a minimum transverse momentum of 20 GeV, since very low $p_T$ jets can be difficult to detect. Next, we want at least five jets to be identified as $b$-jets. The reason why we do not require 6 tagged $b$-jets lies on the efficiency.  Assuming that all jets are equal, i.e., that they are not sorted or classified in any way, the probability of identifying five out of the six as $b$-jets is:
\begin{align}
    \varepsilon_5 = 6\times0.8^5\times0.2 + 0.8^6 = 0.66,
\end{align}
while if we tag the six of them:
\begin{align}
    \varepsilon_6 = 0.8^6 = 0.26,
\end{align}
so allowing five tagged jets is way more efficient. It is also important to note that since all $b$-jets come from on-shell Higgs bosons, the three pairs should reconstruct the Higgs invariant mass. Although the resolution will not be optimal since we are treating with jets, this could be an additional cut to reject backgrounds. We have computed the value of the acceptance, $\mathcal{A}$, for the BSM events after applying all these cuts, and found that for the studied values of $\kappa_3$ and $\kappa_4$ it is between 0.44 and 0.51. In the SM case, with $\kappa_3=\kappa_4=1$,  the acceptance is lower and drops to 0.34. Thus, in summary, to finally compute the number of predicted BSM events, we use the following reduction factors, considering the BRs to $b$ quarks, the $b$-tagging efficiencies, and the acceptance, $\mathcal{A}$,  after applying all the cuts in \refeq{cuts2}:
\begin{align*}
    N_\text{events} = N \times 0.58^3 \times (6\times0.8^5\times0.2 + 0.8^6) \times 0.48,
\end{align*}
where we use an approximation for $\mathcal{A}$  setting it to  its average value of 0.48 in all cases. 

\begin{figure}[H]
    \centering
    \includegraphics[width=0.999\textwidth]{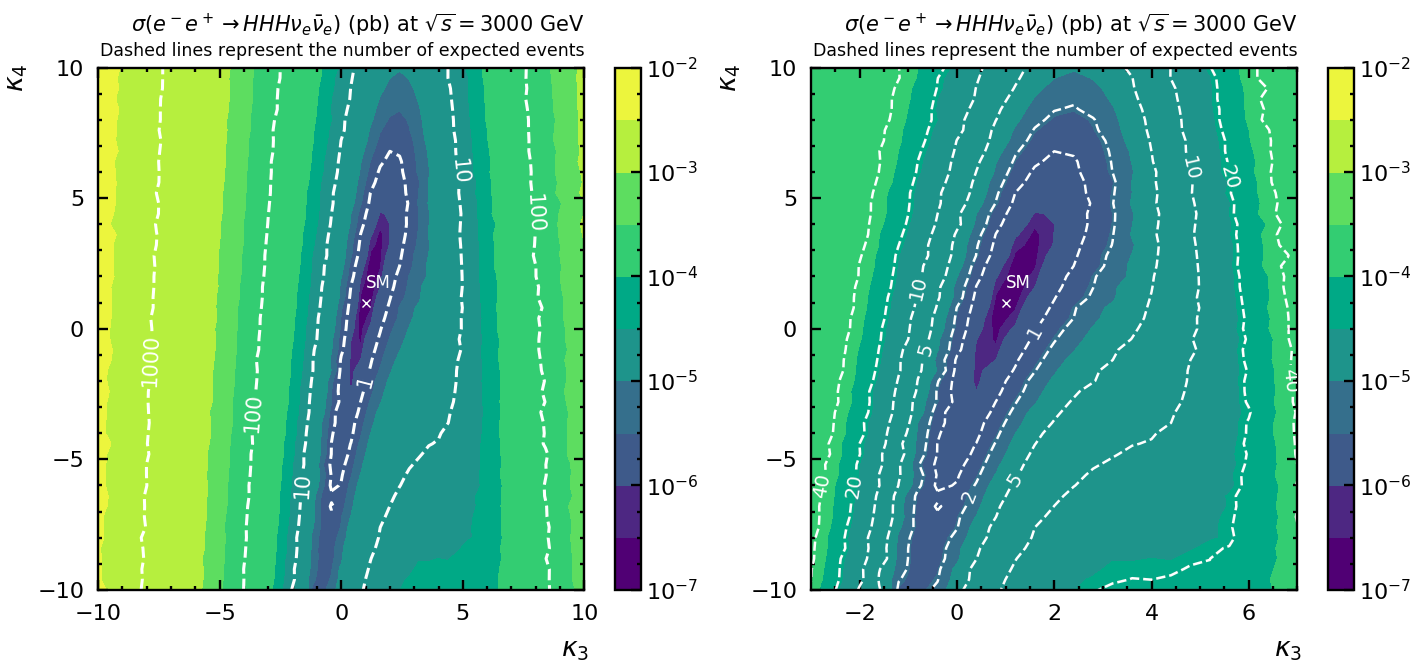}
    \caption{Contour lines for the expected number of $e^+e^-\rightarrow b\bar{b}b\bar{b}b\bar{b}\hspace{0.5mm}\nu_e\bar{\nu}_e$ events (white dashed lines) after applying all the cuts and efficiency factors, represented in the $(\kappa_3,\kappa_4)$ plane. The corresponding contours for the $e^+e^-\rightarrow HHH\nu_e\bar{\nu}_e$ cross section are also included for comparison. Right shows a narrower range in the $\kappa_3$ axis.}
    \label{Nobs}
\end{figure}

The final results for the predicted event rates in the $(\kappa_3,\kappa_4)$ plane, after all the cuts and the efficiencies are taken into account, are presented in \reffi{Nobs}.
As it can be clearly seen in this \reffi{Nobs},  the $b\bar{b}b\bar{b}b\bar{b}\hspace{0.5mm}\nu\bar{\nu}$ event rates are very low in the region of the $(\kappa_3, \kappa_4) $ plane close to the SM point, leading to less than one event, therefore yielding unobservable signals. Separating from this area, the BSM event rates increase reaching values above 10, 100 and even 1000 in the extreme cases with $\kappa_3$ near  -10. We believe that having this signal statistics as a starting point  is very promising and motivates a more complete study including backgrounds, which we are neglecting here.  As we already stated, a detailed analysis taking into account all the backgrounds and the characteristics of the particular detectors at CLIC is beyond the scope of this work. 

However,  as in the case of double Higgs production,  we also believe that the dominant backgrounds to triple Higgs production could be easely treated.  Our naive estimate of the main backgrounds, where the Higgs bosons are replaced by $Z$'s,  provides rates which,  although above the SM reference value,  we still believe to be reasonable.  For instance,  at 3 TeV,  using MG5 and without applying any cuts,  we obtain for the SM cross section $\sigma(e^+e^-\to \nu_e\bar{\nu}_eHHH; H\to b \bar{b})=6\cdot10^{-5}$ fb.  For the main backgrounds,  we find $\sigma(e^+e^-\to \nu_e\bar{\nu}_eHHZ;H,Z\to b \bar{b})=5\cdot10^{-4}$ fb, 
$\sigma(e^+e^-\to \nu_e\bar{\nu}_eHZZ;H,Z\to b \bar{b})=2\cdot10^{-3}$ fb, and $\sigma(e^+e^-\to \nu_e\bar{\nu}_eZZZ; Z\to b \bar{b})=3\cdot10^{-3}$ fb.  Again,  our signal rates from BSM scenarios with anomalous $(\kappa_3,\kappa_4)$ are clearly above the SM ones.  Besides,  we expect that the proper cuts,  in particular the one on the invariant mass of the $b \bar b$ pairs being close to the Higgs mass,  will be able to reduce the backgrounds down to negligible rates.

\begin{figure}[H]
\centering
\includegraphics[width=0.55\textwidth]{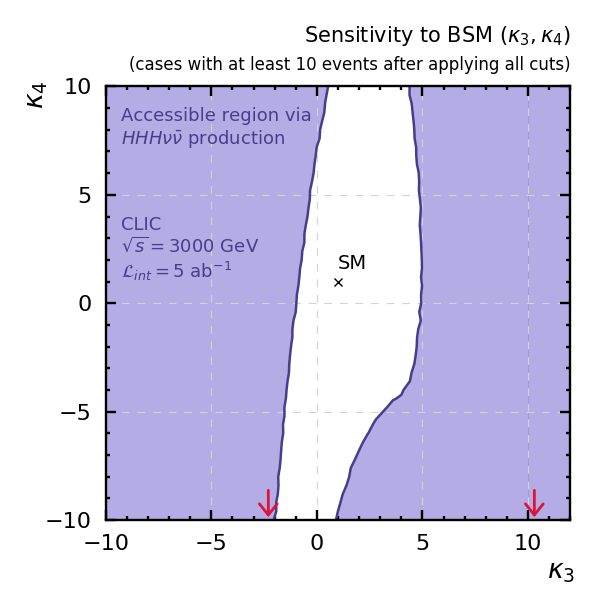}
\caption{Accessible region in the ($\kappa_3,\kappa_4$) plane to BSM scenarios via $e^+e^- \to  HHH\nu\bar{\nu} \to b {\bar b}  b {\bar b} b {\bar b} \nu {\bar \nu}$ events at CLIC, with $\sqrt{s}=3000$ GeV and $\mathcal{L}_\text{int}=5$ ab$^{-1}$. The criterion of accesibility assumed here is requiring more than 10 signal events (purple area). The red arrows mark the limits of the present bound from ATLAS \cite{ATLAS:2019pbo} for $\kappa_3$, given in \refeq{ATLASconstraintEq}.}
\label{result}
\end{figure}

Finally, we wish to conclude with our estimate of the accessible region to the BSM anomalous couplings in the $(\kappa_3, \kappa_4)$ plane. This is summarised  in \reffi{result}, where we have set the 'naive' criterion for accessibilily to a given $(\kappa_3, \kappa_4)$ (in absence of background) by requiring a minimum value of 10 events.  Thus,  we find that the purple area in this \reffi{result} leads to more than 10 events (reaching large values of hundreds and even thousands at the extreme values) and will give access to values of the anomalous self-couplings which are at present unconstrained. In particular, smaller values of $|\kappa_3|$ inside the presently allowed interval $\kappa_3\in[-2.3,10.3]$ (marked by the red arrows in this plot) could be tested at CLIC. Regarding the $\kappa_4$ parameter, it is clear from \reffi{result} that this $HHH\nu \bar \nu$ channel opens the possibility to test the yet unexplored BSM quartic Higgs self-coupling at CLIC. Concretely, if we assume BSM scenarios that are outside the unreachable white region at the center of this plot (which contains the SM point), there seems to be access to  $\kappa_4 \in [-10,10]$.

\section{Conclusions}
In this work we have explored the sensitivity at future $e^+e^-$ colliders to BSM Higgs physics induced by the $WWH$, $WWHH$, $HHH$ and $HHHH$ interaction vertices within the context of the non-linear effective field theory given by the EChL. These interactions provide the BSM Higgs anomalous couplings of our interest, which depend respectively on the EChL  parameters $a$, $b$, $\kappa_3$ and $\kappa_4$. In this non-linear EFT context, these coefficients are independent and uncorrelated by any symmetry argument, because the Higgs field is a singlet in the EChL. This is in contrast to the Higgs couplings in the SM, where some relations among the interaction vertices, such as $V_{WWH}^{\rm SM}= v V_{WWHH}^{\rm SM} $ and $V_{HHH}^{\rm SM}= v V_{HHHH}^{\rm SM}$, are derived from the Higgs being a component of a doublet. 

We have shown that the double and triple Higgs production channels, in particular $e^+e^- \to HH \nu \bar{\nu}$ and $e^+e^- \to HHH \nu \bar{\nu}$, provide the proper window to explore efficiently these BSM Higgs couplings. Whereas double Higgs production can access to $a$, $b$ and $\kappa_3$, the triple Higgs channel is the only one that can access to $\kappa_4$, a parameter which is at present totally unconstrained. Our study here then provides a first try to test the BSM quartic Higgs self-coupling, which is so far evading all experimental searches, at $e^+e^-$ colliders.    

We have also understood why these two particular channels are so efficient in the searches for BSM signals arising from the anomalous Higgs couplings. The main reason is the fact that, at the TeV scale and above, they are dominated by WWS mediated diagrams. It is in these configurations where the largest enhancement due to BSM physics occurs.  We also conclude that the main features found in the WWS subprocesses, $W^-W^+ \to HH(H)$,  are basically reproduced in the corresponding  $e^+e^-\to HH(H) \nu \bar{ \nu}$  processes.  In our study of the dominance of the WWS subprocesses, we also showed that the improved EWA works remarkably well for the double Higgs production case, but it fails in the predictions for triple Higgs production. This is a new result, and convenient to be aware of for future BSM  searches. The conclusion in this concern is that the EWA can be safely used for $HH$ but it  should not be used for the $HHH$ case.   

Our final study of a more realistic experimental scenario, with multiple $b$-jets and missing transverse energy in the final state, shows that the sensitivity to these parameters will indeed be improved considerably at the future $e^+e^-$ colliders. The final results in \reffi{eventszoom} and \reffi{result} summarise our main conclusion on the accessibility region for these EChL parameters, $a$ and $b$ in \reffi{eventszoom}, and $\kappa_3$ and $\kappa_4$ in \reffi{result}. These plots also suggest that the best tests of these parameters  will be provided by CLIC.  In particular, triple Higgs production seems to provide access to $\kappa_4$ in the latest stage of this collider, with a luminosity of $5\, {\rm fb}^{-1}$. Of course, in order to obtain a more accurate conclusion, a more complete study should be performed, including realistic backgrounds and taking into account the detector properties.  
\label{conclu}

%%%%%%%%%%%%%%%%%%%%%%%%%%%%%%%%%%%%%%%%%%%%%%%%%%%%%%%%%%%%%%%%%%%%%%%%%%%%%%

\subsection*{Acknowledgements}

\begingroup %\small
%%%
We wish to thank Claudia Garcia-Garcia for her participation in the early stages of the $\sigma(W^-W^+ \to HH)$ computation. 
This work is partially supported by the European Union through the ITN ELUSIVES H2020-MSCA-ITN-2015//674896 and the RISE INVISIBLESPLUS H2020-MSCA-RISE-2015//690575, by the 
  `Spanish Agencia Estatal de Investigaci\'on'' (AEI) and the EU
``Fondo Europeo de Desarrollo Regional'' (FEDER) through the project
FPA2016-78022-P and from the grant IFT
Centro de Excelencia Severo Ochoa SEV-2016-0597. We also acknowledge partial financial support from the National project  with reference PID2019-108892RB-I00/AEI/10.13039/501100011033.
\endgroup

%%%%%%%%%%%%%%%%%%%%%%%%%%%%%%%%%%%%%%%%%%%%%%%%%%%%%%%%%%%%%%%%%%%%%%%%%%%%%%%

%%%%%%%%%%%%%%%%%%%%%%%%%%APPENDIX%%%%%%%%%%%%%%%%%%%%%%%%%%%%%%%%%%%%%%%%%%%%%%%%%%%%%

%%%%%%%%
\begin{appendices}
\newpage
\section{Diagrams contributing to  $W^-W^+\rightarrow HH$}
\label{appendix:I}
\begin{figure}[h!]
\centering
\includegraphics[scale=0.65]{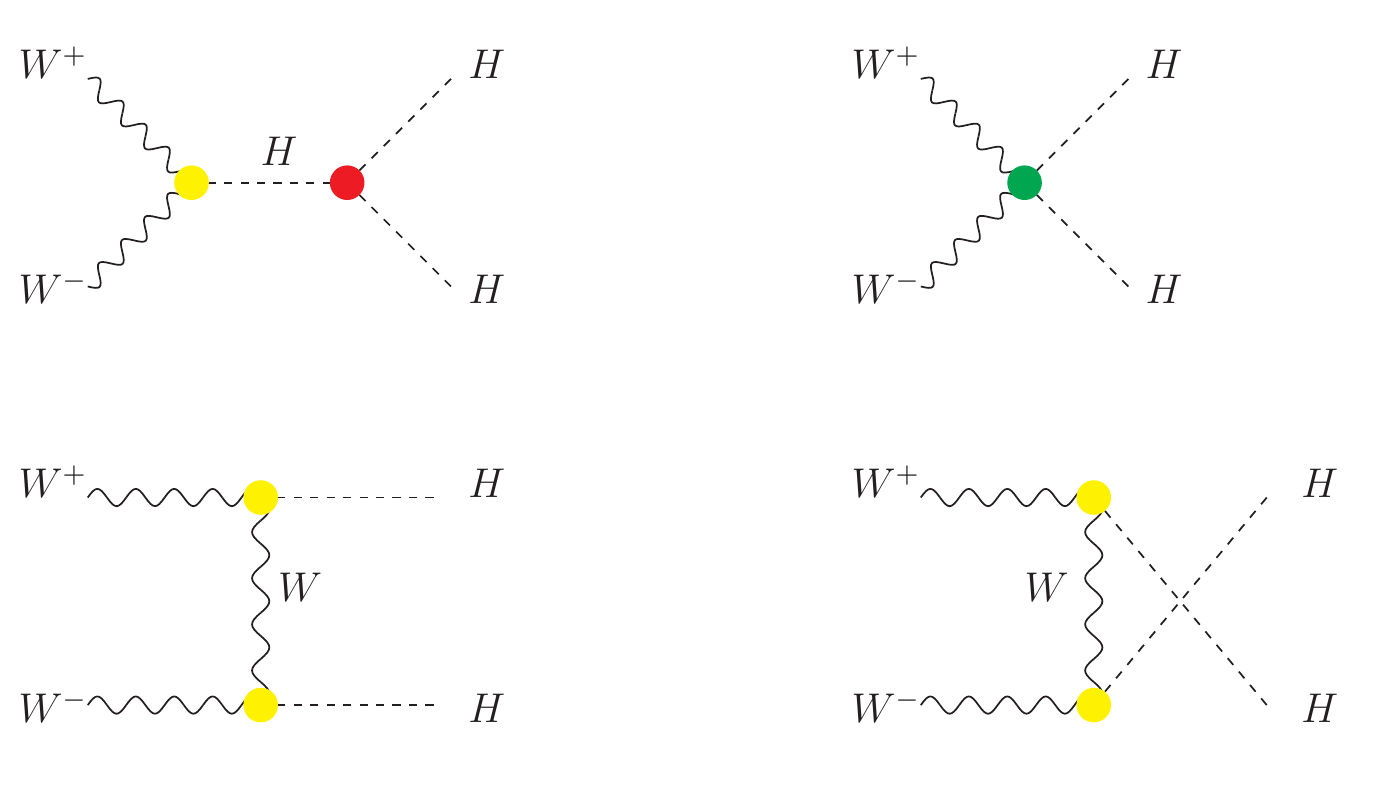}
\caption{Diagrams contributing to the $W^-W^+\rightarrow HH$ subprocess in the unitary gauge.}
\label{subdiagramas}
\end{figure}

There are 4 diagrams contributing to the scattering amplitude of the  $W^-W^+\rightarrow HH$ subprocess in the unitary gauge: $s$ channel, contact channel, $t$ channel and $u$ channel. These are drawn in \reffi{subdiagramas}. The anomalous couplings parametrized by $a$, $b$ and 
$\kappa_3$ (depicted by yellow, green and red dots respectively) enter in the vertices of these channels: $a \cdot \kappa_3 $ in $s$ channel,  $b$ in contact channel, $a^2$ in $t$ and $u$ channels. We will omit the analytical result of the corresponding amplitudes for shortness.
\newpage
%%%%%
\section{Diagrams contributing to  $W^-W^+\rightarrow HHH$}
\label{appendix:II}
\begin{figure}[h!]
\centering
\includegraphics[width=0.95\textwidth]{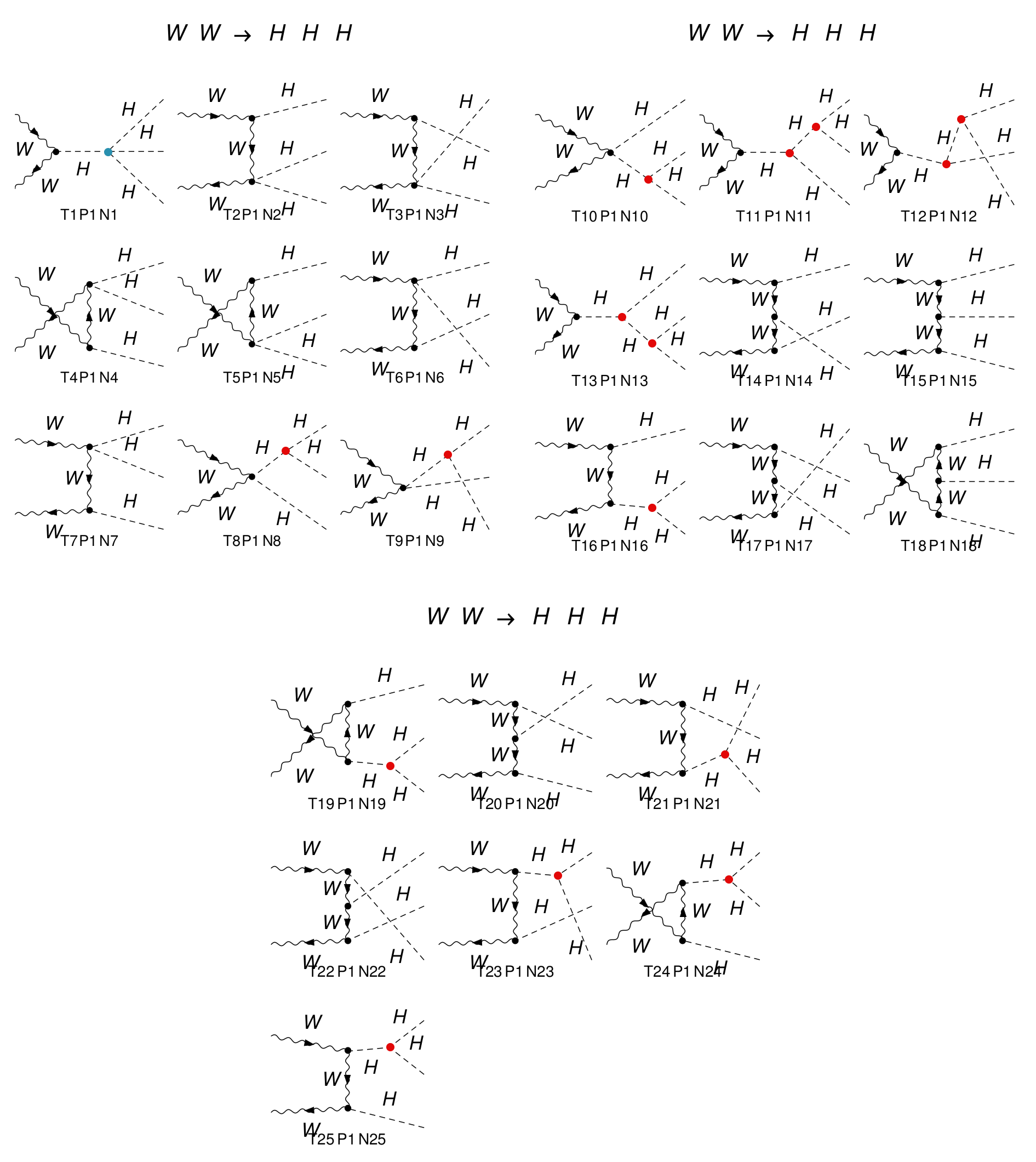}
\caption{Diagrams contributing to the $W^-W^+\rightarrow HHH$ subprocess in the unitary gauge.}
\label{subdiagHHH}
\end{figure}

There are  25 diagrams contributing to the scattering amplitude of the $W^-W^+\rightarrow HHH$ subprocess in the unitary gauge. These have been generated using \textsc{FeynArts-3.10} and are displayed in \reffi{subdiagHHH}.  The red and blue dots represent the triple and quartic Higgs self-interactions, respectively. We will omit the analytical result of the corresponding amplitudes for shortness.
\vspace{5pt}
\end{appendices}
%%%%%%%%%%%%%%%%%%%%%%%%%%REFERENCES%%%%%%%%%%%%%%%%%%%%%%%%%%%%
\newpage

%\bibliography{references}
%\bibliographystyle{BiblioStyle}
\providecommand{\href}[2]{#2}\begingroup\raggedright\endgroup

\end{document}

%% file: paperdef.tex
\newcommand\ReDiag{\mathop{%
  \raise .5pt\hbox{[}%
  \widetilde{\mathrm{Re}}%
  \raise .5pt\hbox{]}}}
\newcommand\ReOffDiag{\mathop{%
  \raise .5pt\hbox{$\llbracket$}%
  \widetilde{\mathrm{Re}}%
  \raise .5pt\hbox{$\rrbracket$}}}

\newcommand\refeq[1]{Eq.~(\ref{#1})}

\newcommand\refta[1]{Tab.~\ref{#1}}
\newcommand\refse[1]{Sect.~\ref{#1}}

\newcommand\citere[1]{Ref.~\cite{#1}}

\def\reffi#1{\mbox{Fig.~\ref{#1}}}

\definecolor{Orange}{named}{orange}
\definecolor{Purple}{named}{purple}
\definecolor{Lightblue}{cmyk}{0.9,0.1,0.1,0.3}
\definecolor{dgelborange}{cmyk}{0.,0.3,0.5, 0.}
\definecolor{Lila}{rgb}{0.5,0.,1}
\definecolor{Darkgreen}{rgb}{0.,.7,0.2}